\makeatletter
\newcommand{\DisableACMJournalFooterInAppendix}{%
  \fancypagestyle{appendixpagestyle}{%
    \fancyhf{}%
    \renewcommand{\headrulewidth}{\z@}%
    \renewcommand{\footrulewidth}{\z@}%

    \fancyhead[LE]{\ACM@linecountL\@headfootfont\@acmArticlePage}%
    \fancyhead[RO]{\@headfootfont\@acmArticlePage}%
    \fancyhead[RE]{\@headfootfont\@shortauthors}%
    \fancyhead[LO]{\ACM@linecountL\@headfootfont\shorttitle}%

  }%

  \let\ps@standardpagestyle\ps@appendixpagestyle
}
\makeatother

\documentclass[acmtog]{acmart}

\acmJournal{TOG}
\acmYear{2025}
\acmMonth{12}          %
\acmVolume{44}
\acmNumber{6}
\acmArticle{5}
\acmDOI{10.1145/3757377.3763830}

\setcopyright{acmlicensed}

\newif\ifdraft
\drafttrue
\ifdraft
\newcommand{\dcc}[1]{{\color{red}[\textbf{DC:} #1]}}
\newcommand{\avc}[1]{{\color{purple}[\textbf{AV:} #1]}}
\newcommand{\odc}[1]{{\color{blue}[\textbf{OD:} #1]}}
\newcommand{\syc}[1]{{\color{orange}[\textbf{SY:} #1]}}

\else
\newcommand{\dcc}[1]{}
\newcommand{\avc}[1]{}
\newcommand{\odc}[1]{}
\newcommand{\syc}[1]{}

\fi

\usepackage{multirow}
\usepackage{overpic} %
\usepackage{xcolor}
\usepackage{makecell} %
\usepackage{colortbl}
\usepackage{ifthen}
\usepackage{color}
\usepackage{subcaption} %

\newboolean{archive}
\newboolean{submission}  %

\definecolor{Salmon}{rgb}{1, 0.55, 0.41} %
\definecolor{LightGreen}{rgb}{0.88, 1, 0.88}
\definecolor{LightRed}{rgb}{1, 0.88, 0.88}

\newcolumntype{Y}{>{\raggedleft\arraybackslash}X}

\setboolean{archive}{true}  %
\setboolean{submission}{true}  %
\newboolean{acm}
\setboolean{acm}{false}  %

\usepackage[ruled]{algorithm2e} %
\usepackage{appendix}

\SetAlFnt{\small}
\SetAlCapFnt{\small}
\SetAlCapNameFnt{\small}
\SetAlCapHSkip{0pt}

\begin{document}

\title{Navigating with Annealing Guidance Scale in Diffusion Space}

\author{Shai Yehezkel$^*$}
\email{shai444@gmail.com}
\orcid{0009-0006-8156-583X}
\affiliation{%
    \institution{Tel Aviv University}
    \country{Israel}
}

\author{Omer Dahary$^*$}
\email{omer11a@gmail.com}
\orcid{0000-0003-0448-9301}
\affiliation{%
    \institution{Tel Aviv University}
    \country{Israel}
}

\author{Andrey Voynov}
\email{an.voynov@gmail.com}
\orcid{0009-0000-2997-9601}
\affiliation{%
    \institution{Google DeepMind}
    \country{Israel}
}
\author{Daniel Cohen-Or}
\email{cohenor@gmail.com}
\orcid{0000-0001-6777-7445}
\affiliation{%
    \institution{Tel Aviv University}
    \country{Israel}
}

\begin{abstract}

Denoising diffusion models excel at generating high-quality images conditioned on text prompts,  
yet their effectiveness heavily relies on careful guidance during the sampling process.  
Classifier-Free Guidance (CFG) provides a widely used mechanism for steering generation by setting the guidance scale,  
which balances image
quality and prompt alignment. 
However, the choice of the guidance scale has a critical impact on the convergence toward a visually appealing and prompt-adherent image.
In this work, we propose an annealing guidance scheduler
which dynamically adjusts the guidance scale over time based on the conditional noisy signal. 
By learning a scheduling policy, our method addresses the temperamental behavior of CFG.  
Empirical results demonstrate that our guidance scheduler significantly enhances image
quality and alignment with the text prompt,  
advancing the performance of text-to-image generation. 
Notably, our novel scheduler requires no additional activations or memory consumption, 
and can seamlessly replace the common classifier-free guidance, 
offering an improved trade-off between prompt %
alignment and quality. Code is available at our
project page: https://annealing-guidance.github.io/annealing-guidance/.

\vspace{-3pt}
\end{abstract}

\begin{teaserfigure}
    \centering
    \setlength{\tabcolsep}{1pt} %
    \renewcommand{\arraystretch}{0.0}
    \begin{tabular}{@{}c@{\hskip 1pt}c@{\hskip 3pt}c@{\hskip 1pt}c@{\hskip 3pt}c@{\hskip 1pt}c@{}}
        \multicolumn{1}{c}{\small CFG} & \multicolumn{1}{c}{\small \textbf{Ours}} &
        \multicolumn{1}{c}{\small CFG} & \multicolumn{1}{c}{\small \textbf{Ours}} &
        \multicolumn{1}{c}{\small CFG} & \multicolumn{1}{c}{\small \textbf{Ours}} \\[2pt]
        \includegraphics[width=0.157\textwidth]{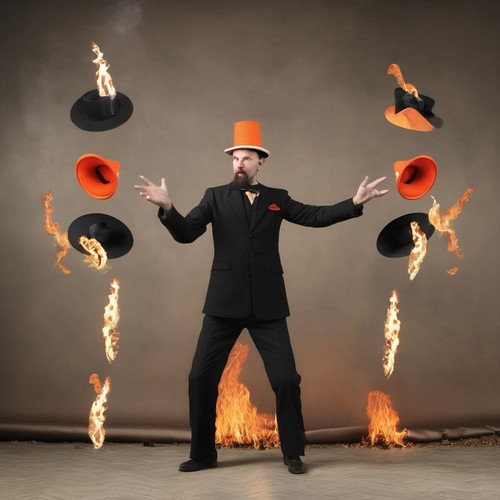} &
        \includegraphics[width=0.157\textwidth]{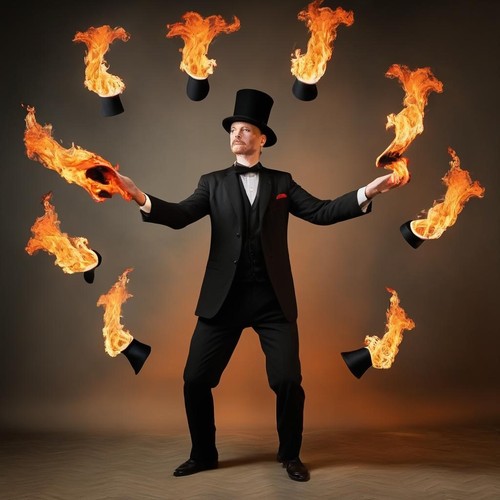} &
        \includegraphics[width=0.157\textwidth]{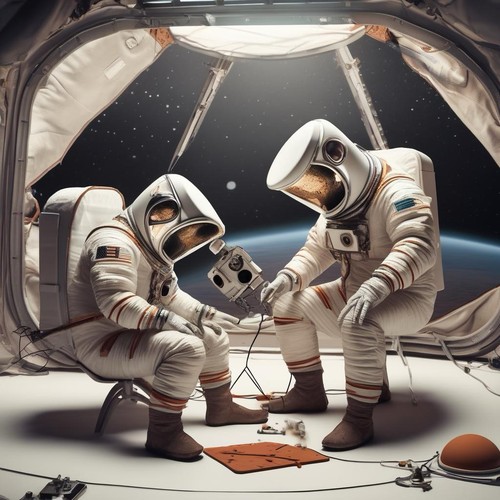} &
        \includegraphics[width=0.157\textwidth]{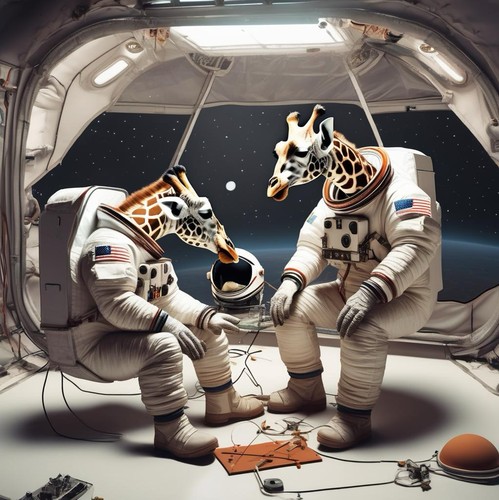} &
        \includegraphics[width=0.157\textwidth]{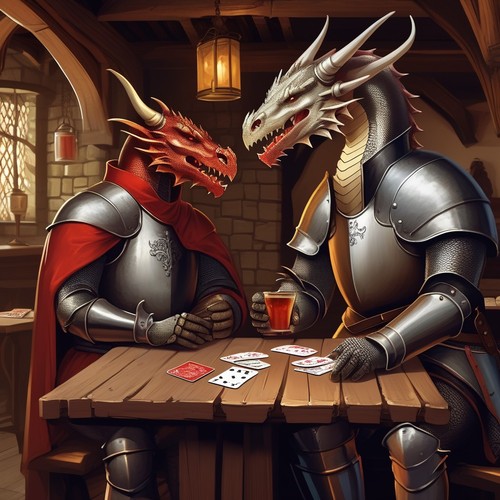} &
        \includegraphics[width=0.157\textwidth]{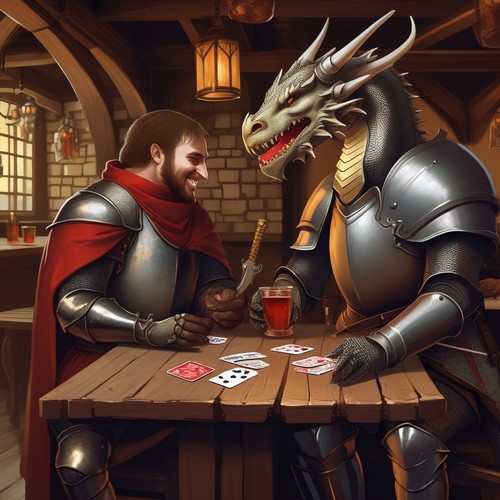} \\[2pt]
        \multicolumn{2}{c}{\shortstack[c]{\tiny "A man juggles flaming hats."}} &
        \multicolumn{2}{c}{\shortstack[c]{\tiny "Two \textcolor{red}{\textbf{giraffes}} in astronaut suits repairing a spacecraft on Mars."}} &
\multicolumn{2}{c}{\shortstack[c]{\tiny "A dragon and a \textcolor{red}{\textbf{knight}} playing cards in a tavern."}}\\[4pt]
        \includegraphics[width=0.157\textwidth]{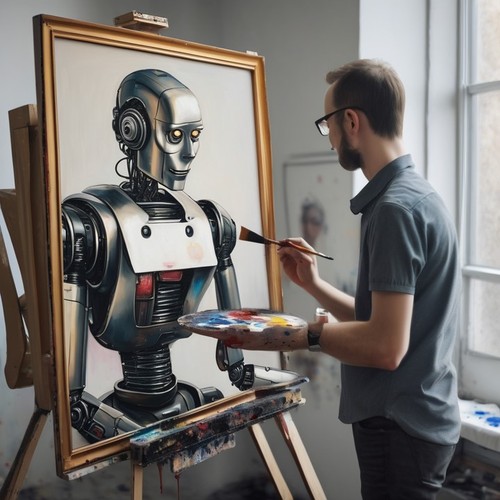} &
        \includegraphics[width=0.157\textwidth]{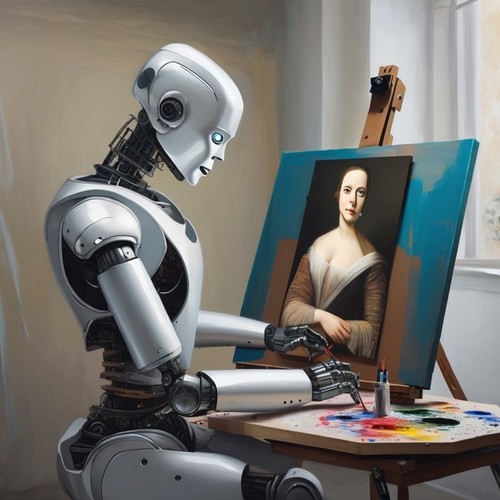} &
        \includegraphics[width=0.157\textwidth]{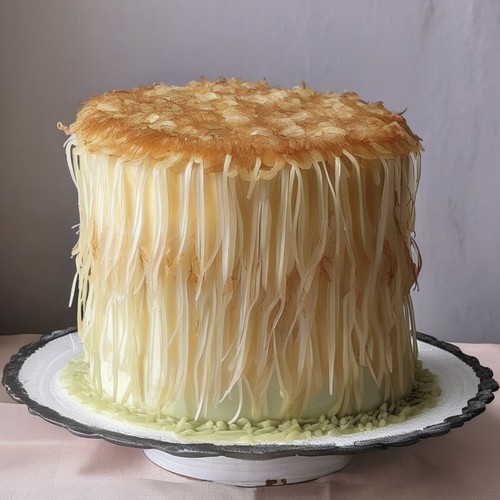} &
        \includegraphics[width=0.157\textwidth]{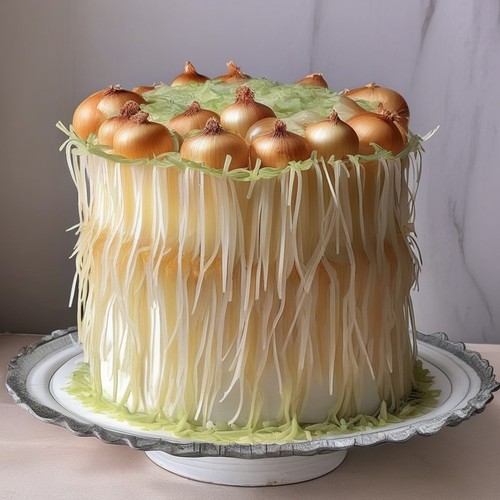} &
        \includegraphics[width=0.157\textwidth]{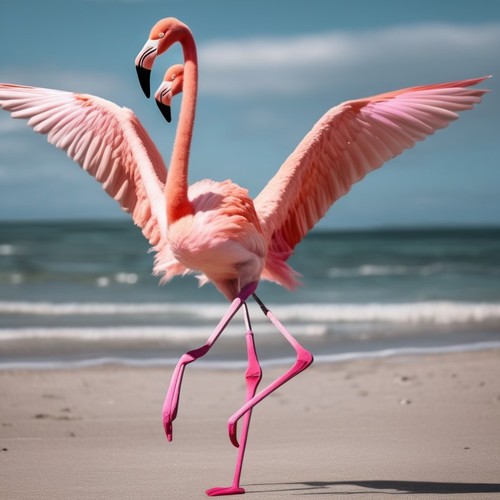} &
        \includegraphics[width=0.157\textwidth]{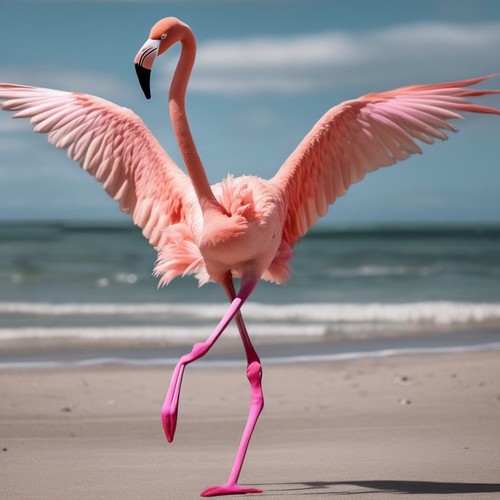} \\[2pt]
        \multicolumn{2}{c}{\shortstack[c]{\tiny "A \textcolor{red}{\textbf{robot}} painting a portrait."}} &
        \multicolumn{2}{c}{\shortstack[c]{\tiny "A cake with \textcolor{red}{\textbf{onions}} on top of it."}} &
        \multicolumn{2}{c}{\shortstack[c]{\tiny "A photo of a ballerina flamingo dancing on the beach."}} \\[4pt]
        \includegraphics[width=0.157\textwidth]{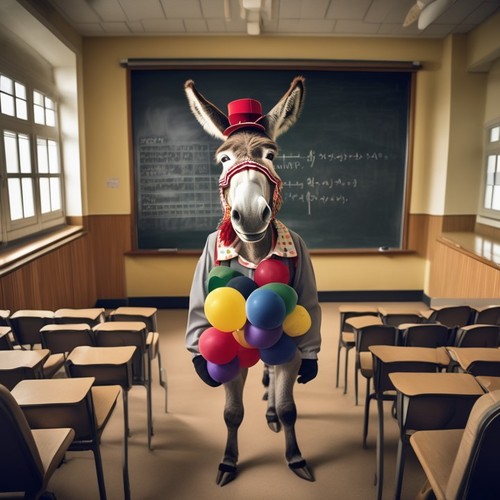} &
        \includegraphics[width=0.157\textwidth]{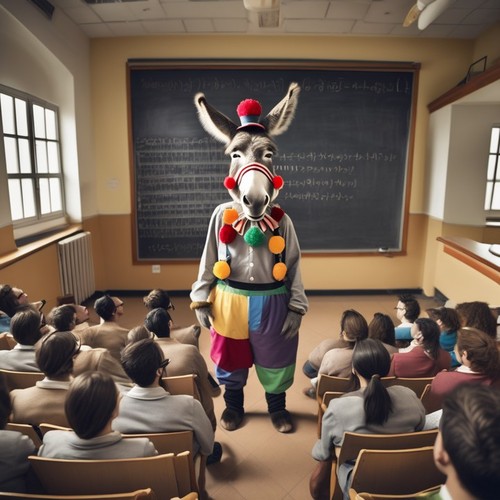} &
        \includegraphics[width=0.157\textwidth]{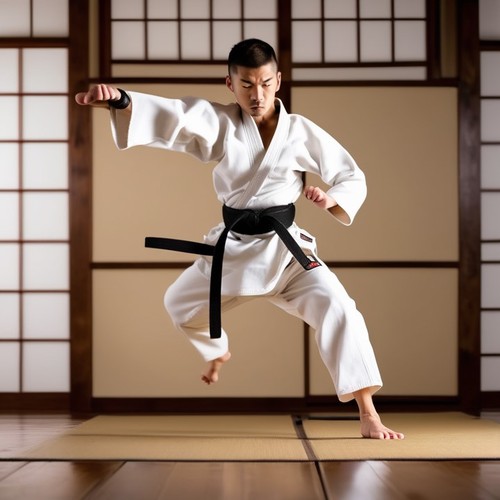} &
        \includegraphics[width=0.157\textwidth]{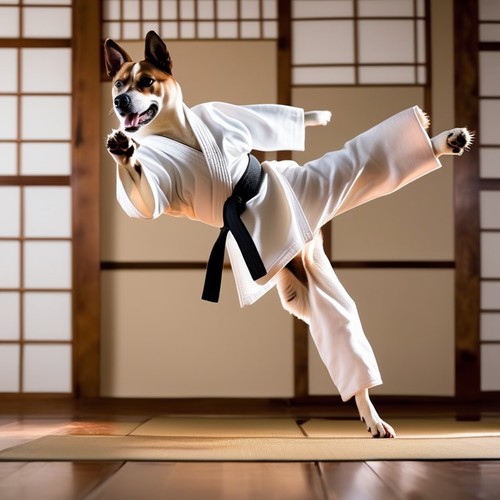} &
        \includegraphics[width=0.157\textwidth]{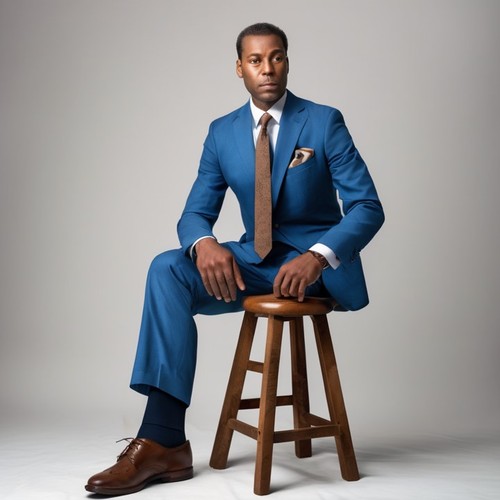} &
        \includegraphics[width=0.157\textwidth]{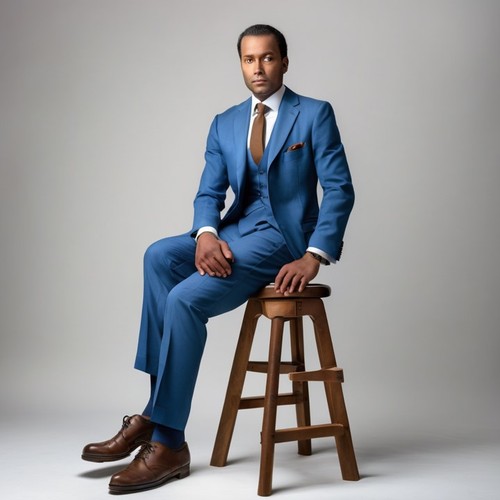} \\[2pt]
        \multicolumn{2}{c}{\shortstack[c]{\tiny "A donkey in a clown costume giving a lecture at the front of a lecture hall... \\  \tiny There are \textcolor{red}{\textbf{many students}} in the lecture hall."}} &
        \multicolumn{2}{c}{
  \shortstack[c]{\tiny
    "A mid-air \textcolor{red}{\textbf{dog}} practicing karate in a Japanese dojo, wearing a white gi\\\tiny with a black belt on wooden floors..."
  }
} &
        \multicolumn{2}{c}{
  \shortstack[c]{\tiny
    "A man is seated on a wooden stool against a white background,\\\tiny dressed in a blue suit with a tie and brown shoes."
  }
}
    \end{tabular}
    \captionof{figure}{Our annealing guidance scheduler significantly enhances image quality and alignment with the text prompt.}
    \label{fig:2x2subfigures}
\end{teaserfigure}

\ccsdesc[500]{Computing methodologies~Image processing}
\maketitle
\makeatletter
\def\thefootnote{*}\footnotetext{Denotes equal contribution.}
\makeatother

\section{Introduction}
\label{sec:intro}

Denoising diffusion models~\cite{ho2020denoising,song2020denoising,nichol2021improved,sohl2015deep,song2019generative,song2020score} have shown outstanding abilities in text-based generation of images~\cite{rombach2022high,ramesh2022hierarchical,saharia2022photorealistic,dhariwal2021diffusion,podell2023sdxl}. %
At training, these models learn to denoise a noisy signal $z_t$, e.g. an image latent, based on its existing lower frequency structure, and a text prompt $c$. However, while these models are tasked with iteratively pushing the signal towards the conditional distribution $p\left( z | c \right)$, in practice, step corrections must be applied to sample high-quality results.

\ifthenelse{\boolean{archive}}
    {}
  {\begin{figure}[ht!]
     \setlength{\belowcaptionskip}{-10pt}
    \includegraphics[width=\columnwidth]{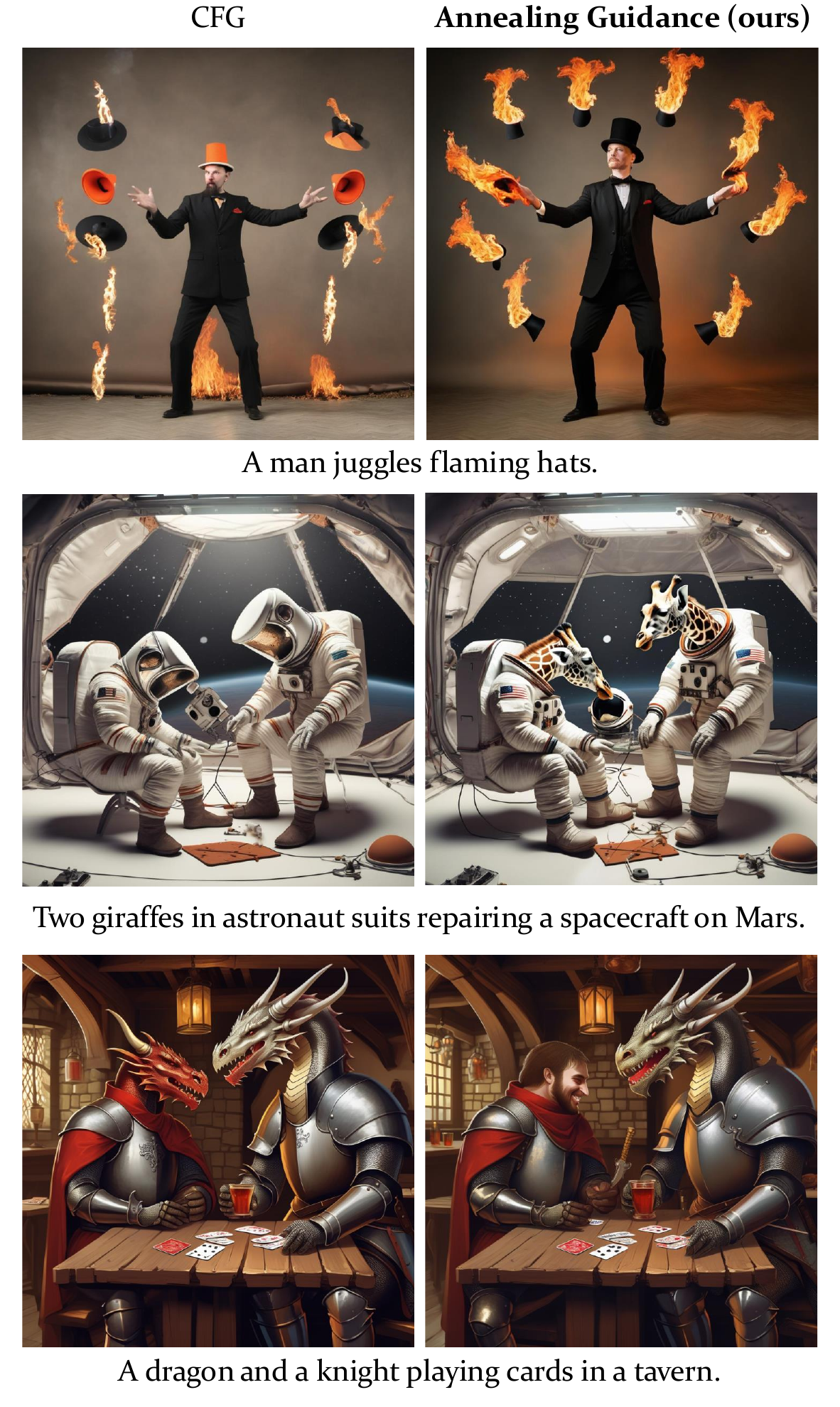}\\[-4pt]
    \caption{Our annealing guidance scheduler significantly enhances image quality and alignment with the text
prompt.}
    \label{fig:2x2subfigures}
\end{figure}
}         %

The widely used approach, classifier-free guidance (CFG)~\cite{ho2022classifier}, suggests corrections
by extrapolating predictions away from the unconditional distribution $p\left( z \right)$.
In practice, this is performed by guiding the latent in the direction of difference between the conditional and unconditional predictions $\delta_{t}\left(z_t\right)\equiv\epsilon_{t}^c\left(z_t\right)-\epsilon_{t}^{\varnothing}\left(z_t\right)$,
using a step size of $w$. 
This mechanism requires special care setting the guidance scale $w$, which directly affects image quality, diversity and prompt alignment of the generated image.
Selecting a proper guidance scale is extremely challenging. The VAE latent space, which we refer to as \textit{diffusion space}, has a complex high-dimensional landscape with non-uniform densities. Properly navigating through this landscape requires skipping low-likelihood regions towards a nearby mode which aligns well with the prompt.

From this perspective, we can think of CFG as a tool that assists navigating in diffusion space. CFG applies correction steps, by which the latent is iteratively refined to agree with both the prompt and the prior distribution of the diffusion model. 
The size $w$ of the correction steps fundamentally affects the success of proper convergence to an image which is both visually appealing and adheres to the prompt.

Recent works have attempted to address the instability of CFG by proposing schedulers for the guidance scale \( w \), typically defined as functions of the timestep \( t \). However, these schedules are often manually designed and based on opposing heuristics. Crucially, such methods do not adapt to the initial noise or the evolving denoising trajectory—factors that are essential for navigating the diffusion space effectively.

To address this limitation, we propose a learning-based scheduler that adapts the guidance scale throughout the generation process. Our approach leverages the signal \( \delta_t = \epsilon_t^c - \epsilon_t^\varnothing \), which captures the discrepancy between the model's conditional and unconditional predictions at each step. We train a lightweight MLP to predict \( w \) as a function of both the timestep \( t \) and \( \|\delta_t\| \), enabling trajectory-aware, sample-specific guidance. %

Our method builds upon CFG++~\cite{chung2024cfg++}, an improved variant of CFG that casts the sampling process as an optimization problem. Specifically, it views guidance as a gradient descent step that minimizes the Score Distillation Sampling (SDS) loss~\cite{poole2022dreamfusion}, which measures the model’s accuracy in predicting the true noise based on the prompt. In this framework, the signal \( \delta_t \) naturally emerges as a proxy for the gradient of the SDS loss, providing a principled way to steer the denoising trajectory toward prompt-consistent samples.

\ifthenelse{\boolean{archive}}
{Fig.~\ref{fig:2x2subfigures} presents examples where our annealing scheduler enhances prompt alignment and corrects generation artifacts, resulting in visually pleasing images that more accurately reflect the user's intent.}
{Fig.~\ref{fig:2x2subfigures} showcases examples where our annealing scheduler improves alignment on challenging prompts, producing visually pleasing images that more accurately reflect the user’s intent.}

\begin{figure}[t]
    \centering
    \includegraphics[width=0.9\linewidth]{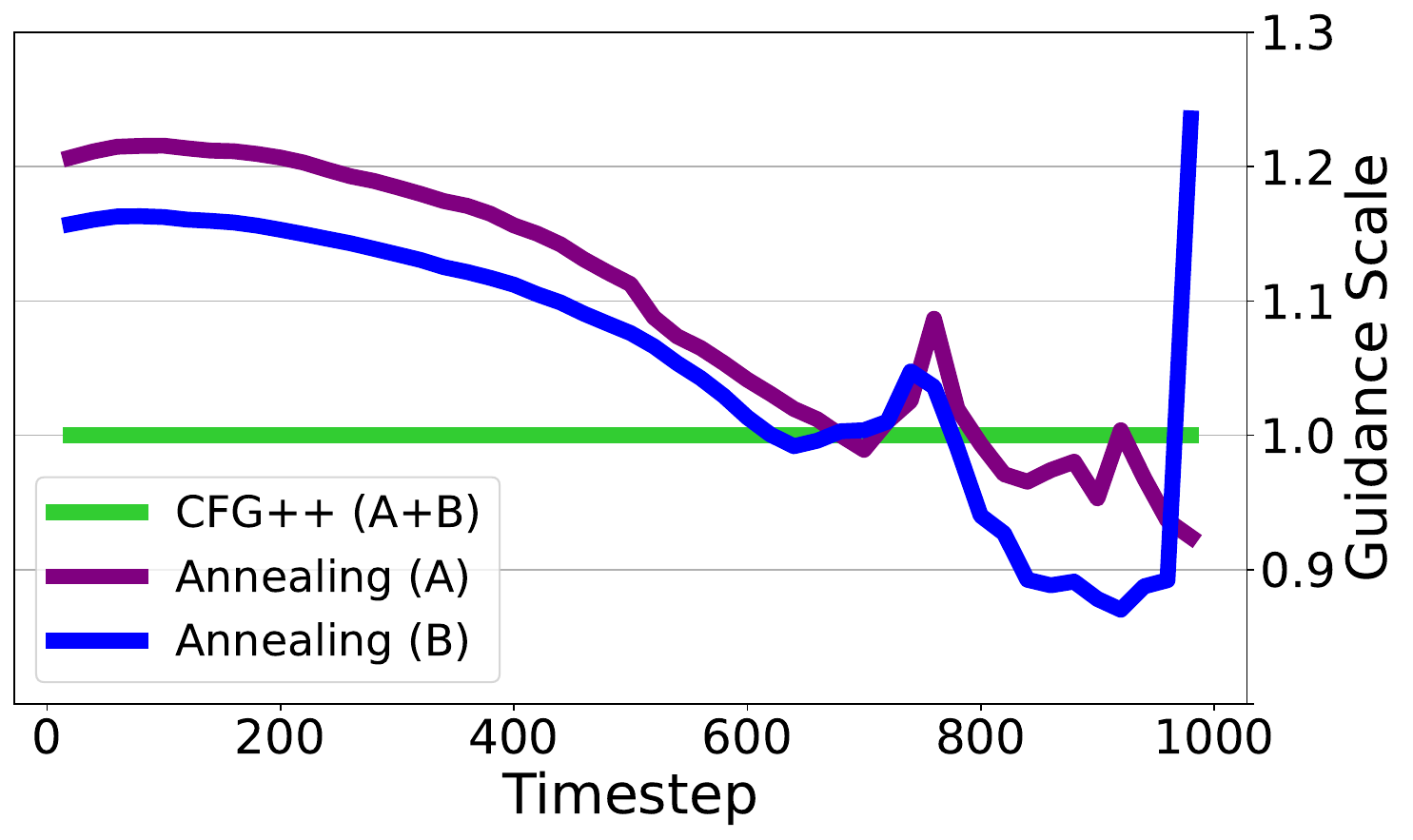}
    \vspace{-4pt}
    \begin{tabular}{c@{\hskip 2pt}c@{\hskip 2pt}c}
    \makecell{CFG} & 
            \makecell{CFG++} & 
        \makecell{  Annealing (Ours)} \\

        \includegraphics[width=0.32\linewidth]{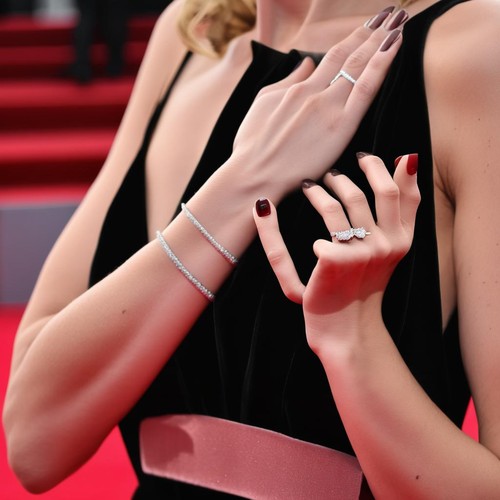} &
        \includegraphics[width=0.32\linewidth]{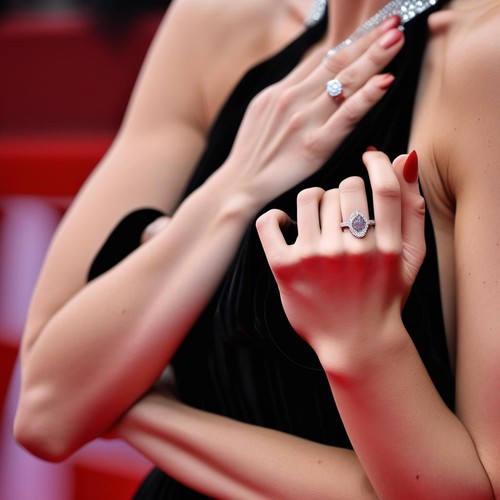} &
        \includegraphics[width=0.32\linewidth]{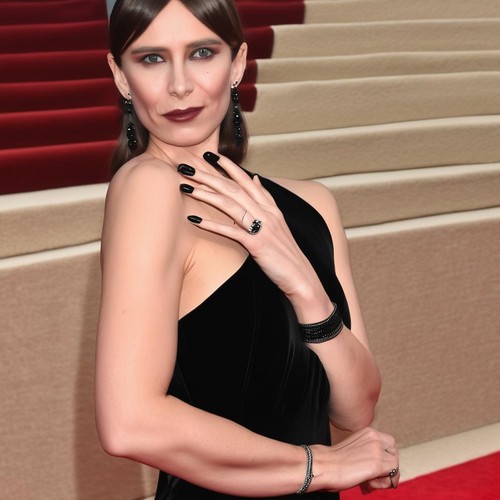} \\[-2pt]
        \multicolumn{3}{c}{\small  \textbf{\textcolor{violet}{A:}} "Woman in black dress on the red carpet wearing a ring on the finger."} \\[4pt]

        \includegraphics[width=0.32\linewidth]{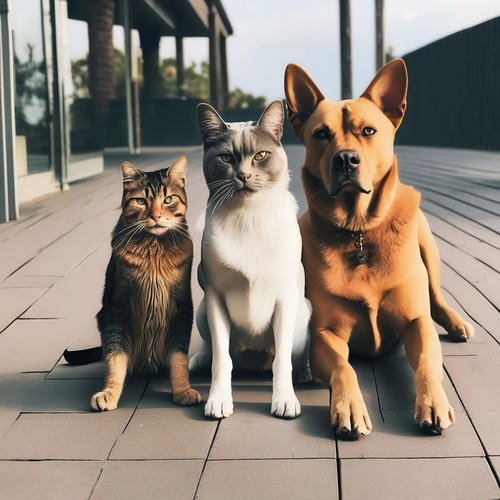} &
        \includegraphics[width=0.32\linewidth]{images/trajectories_new/two_dogs_one_cat_1/anneal_figure_cfgpp.jpg} &
        \includegraphics[width=0.32\linewidth]{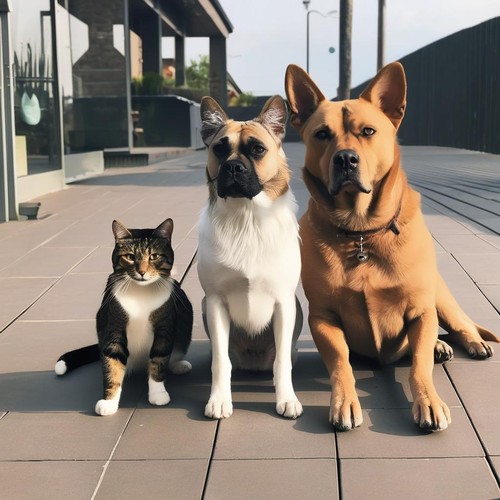} \\[-2pt]
        \multicolumn{3}{c}{\small \textbf{\textcolor{blue}{B:}} "Two dogs, one cat"} \\[-1pt]
    \end{tabular}

    \vspace{-6pt}
    \caption{\textbf{Guidance Scale Over Time.}
    Top: Guidance scale trajectories for two prompts: \textbf{\textcolor{violet}{A}} and \textbf{\textcolor{blue}{B}}. CFG++ uses a constant scale for both prompts, while our annealing scheduler dynamically adapts the scale per prompt. CFG is omitted from the plot for clarity but uses a fixed scale of \( w = 10 \). Bottom: Comparison of generations from CFG (left), CFG++ (center) and our method (right). Our scheduler improves both quality and alignment: resolving visual artifacts (distorted hands, scene \textbf{\textcolor{violet}{A}}) and correcting object counts (scene \textbf{\textcolor{blue}{B}}).
    }
    \vspace{-17pt}

    \label{fig:sds_scale_over_time}
\end{figure}

Fig.~\ref{fig:sds_scale_over_time} illustrates the behavior of our annealing scheduler. As shown in the plots, the predicted scale \( w \) evolves differently across two generations (\textbf{\textcolor{violet}{A}} and \textbf{\textcolor{blue}{B}}), exhibiting non-monotonic fluctuations that adapts to each denoising trajectory.
This adaptive behavior contrasts with the fixed guidance scales used in CFG and CFG++, which cannot account for such variations. For scene \textbf{\textcolor{violet}{A}}, our scheduler corrects artifacts present in the baselines, most notably the distorted anatomy of the woman's hands, resulting in a higher-quality image. For scene \textbf{\textcolor{blue}{B}}, our method produces an image that is more faithfully aligned with the prompt, accurately capturing the specified number of objects, unlike the generations produced by the baselines.

We further explore the behavior of the annealing scheduler over a toy example, and demonstrate quantitatively and qualitatively that our navigation scheme improves the quality and prompt alignment of generated images. Notably, our scheduler achieves state-of-the-art performance on FID/CLIP and FD-DINOv2/CLIP when evaluated on MSCOCO17~\cite{lin2014microsoft}, outperforming prior methods by a considerable margin.

\section{Related works}

\subsection{Guidance in Diffusion Models}

Diffusion-based models have emerged as the driving force behind advanced generative modeling, defining the state-of-the-art in the synthesis of high-quality, diverse, and coherent data across various domains. A significant aspect of diffusion-based generative models is their ability to perform sampling guided by specific conditions, with text-based conditioning being the most commonly employed.

The conditioning mechanism in diffusion-based generative models can be implemented in various ways, with classifier-free guidance (CFG) \cite{ho2022classifier} emerging as a foundational and widely adopted technique. CFG replaces the use of external gradients \cite{dhariwal2021diffusion} by combining conditional and unconditional model outputs in a linear manner, offering a powerful and flexible method for controlling generation. 

This approach has become a standard in most modern sampling algorithms, significantly enhancing both the quality and controllability of generated outputs. Additionally, other approaches extend conditioning through internal feature corrections \cite{voynov2023sketch}, domain-specific architectural adaptations \cite{zhang2023adding, ye2023ip-adapter}, and alternative strategies \cite{liu2023more, Tumanyan_2023_CVPR}, further enriching the capabilities of diffusion-based models.

\subsection{Advanced Sampling}

Classifier-Free Guidance (CFG) sampling with a simple solver produces plausible results; however, models often struggle to generate complex scenes, such as those with intricate compositions or multiple elements \cite{chefer2023attend,dahary2025yourself}. Despite its widespread use, CFG introduces an inherent tradeoff between faithfulness to the desired prompt and diversity, where increasing the guidance scale enhances alignment with the conditioning but reduces output variability. Moreover, simply increasing the guidance scale is not always effective, as it can result in unnatural artifacts or over-saturated images that compromise realism. Additionally, certain seeds have been shown to consistently produce low-quality images~\cite{xu2024good}.

Several works have proposed improved sampling techniques to address these challenges. One approach considers various non-learnable hyperparameter configurations for the noise scheduler and guidance scales \cite{karras2022elucidating}. Another method introduces guidance distillation, enabling the use of a single model to streamline the sampling process \cite{meng2023distillation}. To mitigate issues at higher guidance scales, some techniques suggest clipping the guidance step size to prevent over-saturation \cite{lin2024common,sadat2024eliminating}, while others propose controlling 
the step size using empirically designed schedulers \cite{sadat2023cads,kynkaanniemi2024applying,wang2024analysis}.

Other studies have proposed modifications to CFG to address its limitations. Some approaches restrict the guidance to the image manifold, ensuring more coherent outputs \cite{chung2024cfg++}, while others redefine the guidance process by introducing a new basis that better separates the denoising and prompt-guidance components \cite{sadat2024eliminating}. More relevant to our work are techniques that employ non-constant guidance, such as adjusting the steps at which guidance is applied \cite{dinh2024compress,kynkaanniemi2024applying}, modifying guidance based on segmentation of generated objects \cite{shen2024rethinking}, or altering the unconditional component in the CFG formulation \cite{karras2024guiding}.

\begin{figure}
    \centering
    \includegraphics[width=1.\linewidth]{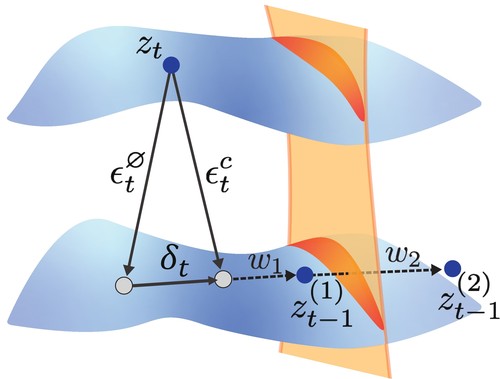}
    \vspace{-14pt}
    \caption{\textbf{Classifier-Free Guidance step.} The denoising step of a sample $z_{t}$ is illustrated as a linear combination of the conditional noise prediction $\epsilon_{t}^c$ and the unconditional noise prediction $\epsilon_{t}^\varnothing$. We denote the difference between predictions as $\delta_t = \epsilon_t^c - \epsilon_t^\varnothing$. The dashed line represents possible $z_{t-1}$ predictions using CFG. For simplicity, we do not depict the rescaling of $z_t$ which is performed at each denoising step. Here, $z_{t-1}^{(1)}$ and $z_{t-1}^{(2)}$ denote predictions corresponding to two different guidance scales, $w_1$ and $w_2$, respectively. The blue manifold represents the density $p_t(z)$, while the orange manifold illustrates the conditional density $p_t(z | c)$.}
    \vspace{-8pt}
    \label{fig:cfg}
\end{figure}

\section{Overview}

The Classifier-Free Guidance (CFG) sampling equation in the simplest case is given by:

\begin{equation}
    \mathbf{\hat{\epsilon}}_t = \mathbf{\epsilon}_t^{\varnothing} + w \cdot \left(\mathbf{\epsilon}_t^{c} - \mathbf{\epsilon}_t^{\varnothing}\right),
    \label{eq:cfg_sampling}
\end{equation}
where \( \mathbf{\hat{\epsilon}}_t \) is the guided noise prediction  at time step \( t \), \( \mathbf{\epsilon}_t^{\varnothing} \) is the unconditional model output, \( \mathbf{\epsilon}_t^{c} \) is the conditional model output, and \( w \) is the guidance scale that controls the extent to which we extrapolate from the unconditional to the conditional outputs (see Algo\ifthenelse{\boolean{acm}}
{
2
}
{
~\ref{alg:cfg_inference}
}for detailed algorithm).

The guidance scale \( w \) determines the strength of alignment with the conditioning input, with higher values improving alignment but potentially reducing diversity or introducing artifacts. 
This is illustrated in Fig.~\ref{fig:cfg}, which depicts a denoising step of $z_{t}$ over the density manifold $p_t(z)$ toward the density manifold $p_{t-1}(z)$.
 We show the unconditional noise direction \( \mathbf{\epsilon}_t^{\varnothing} \) and the conditional noise direction \( \mathbf{\epsilon}_t^{c} \).

\begin{figure}
    \centering
    \includegraphics[width=1.\linewidth]{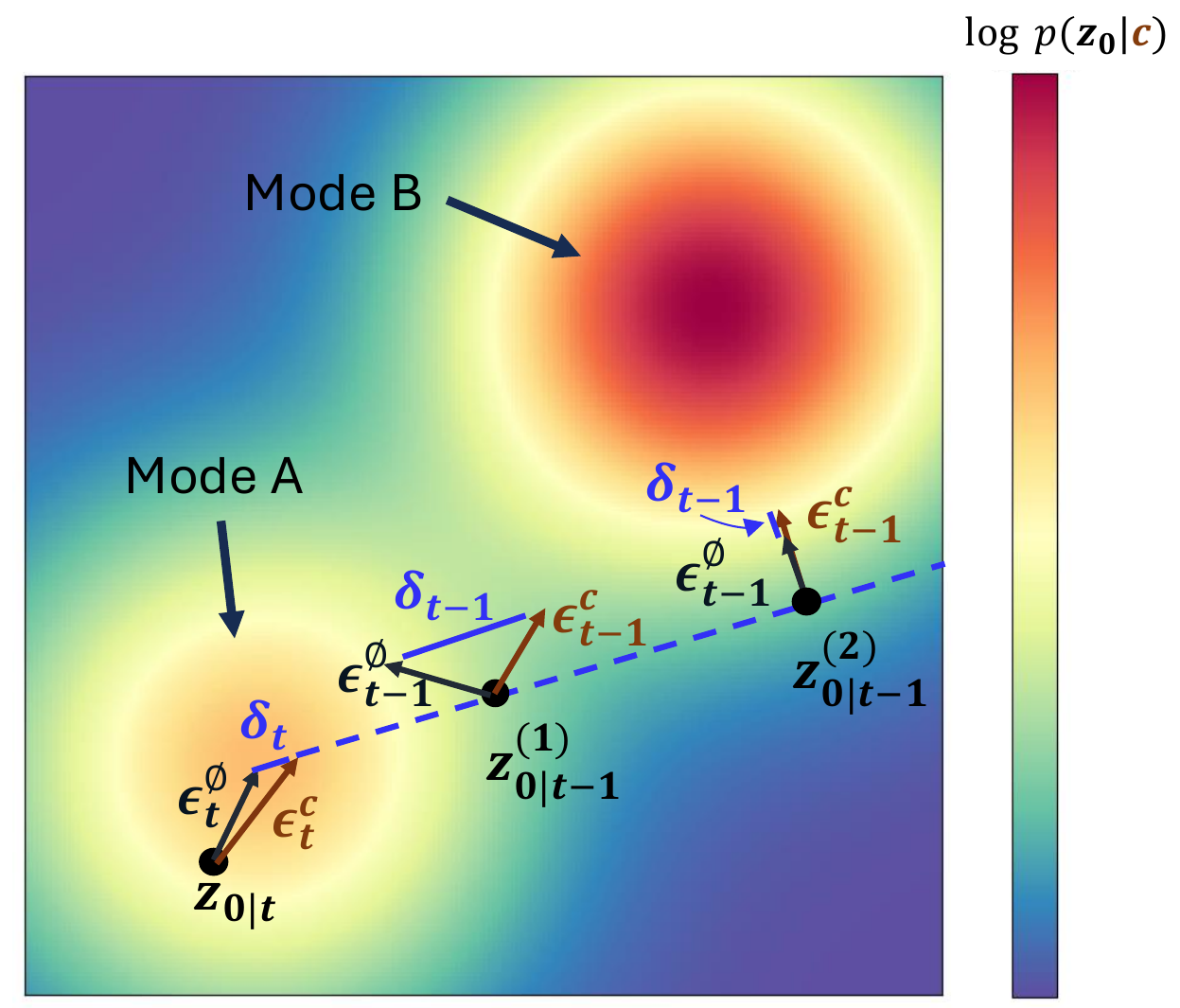}
    \caption{\textbf{Geometric intuition of \(\delta_t\).} A 2D illustration showing how the magnitude of \( \delta_t = \epsilon_t^c - \epsilon_t^{\varnothing} \) reflects alignment with the prompt. At time \( t \), the sample \( z_{0|t} \) lies near mode A, which partially aligns with the prompt, resulting in a small \( \|\delta_t\| \). As the denoising progresses, following the direction of \(\delta_t\) leads toward mode B that better aligns with the prompt. The candidate points along the line correspond to different guidance scales \(w\). Among these, \( z^{(2)}_{0|t-1} \) lies closest to mode B, where the conditional and unconditional predictions are best aligned, yielding a minimal \( \|\delta_{t-1}\| \).}
    
    \label{fig:intuition}
\end{figure}

The CFG operation aims to increase the probability \( p_t(c \mid z) \) while staying on the manifold defined by natural images by extrapolating between \( \epsilon_t^c \) and the unconditional prediction \( \epsilon_t^\varnothing \), weighted by a factor \( w \). The figure illustrates extrapolations with two scales: one with \( w_1 \), which undershoots the target distribution, and another with \( w_2 \), which overshoots.

Determining the optimal size of the guidance scale \( w \) is a non-trivial task, as it depends on the distribution's local geometry, the target prompt, the initial noise, and the model itself.

The commonly used approach is to keep \( w \) constant throughout the generation process. 
While other works have explored relations between \( w \) and timesteps, we argue that \( w \) should also depend on the difference defined as
\begin{equation}
 \delta_t = \mathbf{\epsilon}_t^{c} - \mathbf{\epsilon}_t^{\varnothing}, 
 \label{eq:define_delta}
\end{equation}
which is shown in Fig.~\ref{fig:cfg}. Specifically, $\delta_t$ is affected by the model's predictions on the current noisy latent in relation to the prompt, and thus encapsulates information specific to the denoising trajectory.
This dependency suggests that a fixed or simplistic scheduling of \( w \) may not be sufficient 
for achieving optimal results, stressing the need for more adaptive approaches.

\begin{figure*}[t]
    \centering

    \begin{subfigure}[b]{0.315\textwidth}
        \centering
        \caption{\( \lambda = 0.6 \)}
        \includegraphics[width=\textwidth]{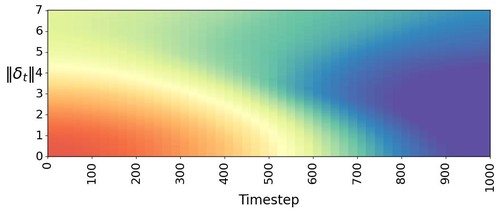}
        
        \label{fig:w_heatmap_a}
    \end{subfigure}
    \begin{subfigure}[b]{0.315\textwidth}
        \centering
        \caption{\( \lambda = 0.7 \)}
        \includegraphics[width=\textwidth]{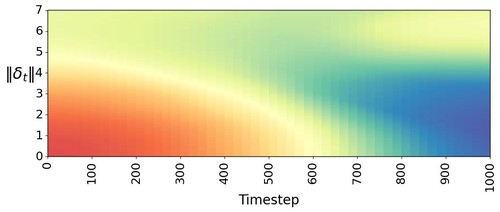}
        
        \label{fig:w_heatmap_b}
    \end{subfigure}
    \begin{subfigure}[b]{0.315\textwidth}
        \centering
        \caption{\( \lambda =0.8 \)}
        \includegraphics[width=\textwidth]{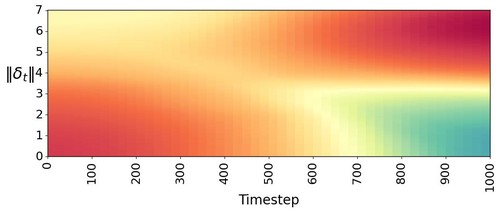}
        
        \label{fig:w_heatmap_c}
    \end{subfigure}
    \begin{subfigure}[b]{0.031\textwidth}
        \centering
        \includegraphics[width=\textwidth]{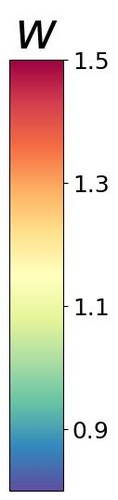}
        \label{fig:w_heatmap_color_bar}
    \end{subfigure}
    \vspace{-8pt}
    \caption{%
    Heatmaps showing the predicted guidance scale \( w_\theta \) as a function of timestep \( t \) and \( \|\delta_t\| \), for three values of \( \lambda \). The color represents the value of \( w_\theta \left( t, \|\delta_t\|, \lambda \right) \), with the colormap shown on the right. Larger \( t \) corresponds to earlier diffusion steps, with \( t = 0 \) marking the end of denoising. %
    At each step, \( \|\delta_t\| \) is recomputed and used to dynamically predict the guidance scale, forming a trajectory over time as demonstrated in Fig.~\ref{fig:sds_scale_over_time}. %
    \vspace{-4pt}
    }
    \label{fig:w_heatmap}
\end{figure*}

Given the temperamental behavior of \( w \), we propose a learning-based approach to determine its optimal value. 
Specifically, we learn \( w \) as a function of the 
timestep \( t \) and $\|\delta_t\|$,
enabling a more adaptive and context-aware guidance scale.

\ifthenelse{\boolean{archive}}
    {}
  {\begin{figure*}[t]
    \centering

    \begin{subfigure}[b]{0.315\textwidth}
        \centering
        \caption{\( \lambda = 0.6 \)}
        \includegraphics[width=\textwidth]{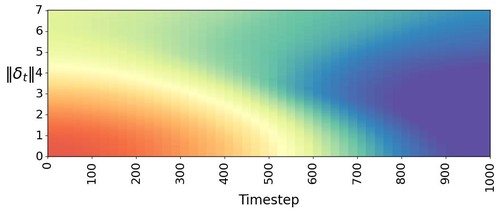}
        
        \label{fig:w_heatmap_a}
    \end{subfigure}
    \begin{subfigure}[b]{0.315\textwidth}
        \centering
        \caption{\( \lambda = 0.7 \)}
        \includegraphics[width=\textwidth]{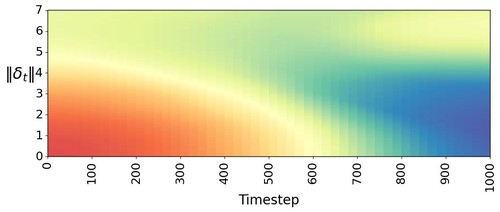}
        
        \label{fig:w_heatmap_b}
    \end{subfigure}
    \begin{subfigure}[b]{0.315\textwidth}
        \centering
        \caption{\( \lambda =0.8 \)}
        \includegraphics[width=\textwidth]{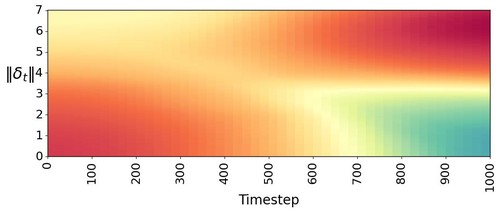}
        
        \label{fig:w_heatmap_c}
    \end{subfigure}
    \begin{subfigure}[b]{0.031\textwidth}
        \centering
        \includegraphics[width=\textwidth]{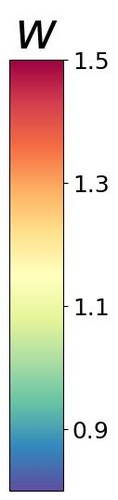}
        \label{fig:w_heatmap_color_bar}
    \end{subfigure}
    \vspace{-8pt}
    \caption{%
    Heatmaps showing the predicted guidance scale \( w_\theta \) as a function of timestep \( t \) and \( \|\delta_t\| \), for three values of \( \lambda \). The color represents the value of \( w_\theta \left( t, \|\delta_t\|, \lambda \right) \), with the colormap shown on the right. Larger \( t \) corresponds to earlier diffusion steps, with \( t = 0 \) marking the end of denoising. %
    At each step, \( \|\delta_t\| \) is recomputed and used to dynamically predict the guidance scale, forming a trajectory over time as demonstrated in Fig.~\ref{fig:sds_scale_over_time}. %
    \vspace{-4pt}
    }
    \label{fig:w_heatmap}
\end{figure*}
}         %

\subsection{SDS and CFG++} \label{sec:sds}

Score Distillation Sampling (SDS)~\cite{poole2022dreamfusion} is a technique for aligning input data with a target distribution defined by a pre-trained diffusion model, by leveraging gradients extracted from the model. It operates using the explicit SDS loss~\cite{zhu2024hifahighfidelitytextto3dgeneration}:
\begin{equation}
L^{\text{SDS}}\left(z_0\right) = \mathbb{E}_{t,\epsilon} \left\|{\epsilon}_t^c\left(z_t\right) - \epsilon\right\|_2^2,
\end{equation}
which encourages the optimized input \( z_0 \) to both align with the conditional signal \( c \) and remain consistent with the model distribution. Here, \( z_t \) is the noisy latent corresponding to \( z_0 \), \( \epsilon \) is the sampled true noise, and \( \epsilon_t^c \) is the conditional prediction of the model.

Recent work~\cite{chung2024cfg++} adopts this loss formulation to reinterpret CFG within the framework of diffusion-based inverse problems~\cite{chung2022diffusion}. From this perspective, CFG can be seen as implicitly seeking a solution $z_0$ that minimizes the SDS loss, under the constraint that it lies on the clean data manifold $p_0(z)$.

This reformulation, termed CFG++, introduces inference-time modifications to ensure \( z_0 \) is on the image manifold:  
(1) it restricts the guidance scale \( w \)~{\footnote{For simplicity, we use $w$ to interchangeably denote the guidance scale of both CFG and CFG++.} to the interval \( [0, 1] \)
; and  
(2) in contrast to CFG, which uses the guided noise prediction \( \hat{\epsilon}_t \) for both denoising and renoising when computing \( z_{t-1} \) from \( z_t \), CFG++ uses \( \hat{\epsilon}_t \) for denoising but reintroduces noise using the unconditional prediction \( \epsilon_t^\varnothing \). For more details, we refer to Algorithm\ifthenelse{\boolean{acm}}
{
3
}
{
~\ref{alg:cfgpp_inference}
}in the supplement.

With these adjustments, Eq.~\eqref{eq:cfg_sampling} can be interpreted as a manifold-constrained gradient descent (MCG) step~\cite{chung2022improving,chung2023decomposed} toward minimizing the SDS loss, thereby enhancing prompt alignment.
Notably, the MCG is approximated at each step by
\begin{equation}
\nabla_{z_{0|t}} L^{\text{SDS}} = 2 \gamma_t (\epsilon_t^c - \epsilon_t^\varnothing),
\label{eq:sds_approx}
\end{equation}
where \(z_{0|t}\) is the current estimate of denoised latent, and \( \gamma_t = \sqrt{\bar{\alpha}_t}/\sqrt{1-\bar{\alpha}_t} \) is a time-dependent coefficient that scales the current noise level to the latent space.

Substituting Eq.~\eqref{eq:define_delta} into Eq.~\eqref{eq:sds_approx} reveals that \( \delta_t \) can serve as a time-normalized proxy for the SDS gradients. Consequently, smaller values of \( \|\delta_t\| \) indicate proximity to stationary points of the SDS loss. Intuitively, if \( z_t \) is within the model's distribution, stronger alignment between the conditional and unconditional predictions corresponds to better adherence to the prompt.

This observation is consistent with the original inspiration behind CFG~\cite{ho2022classifier}, 
which frames \( \delta_t \) as a direction proportional to the gradients of an implicit classifier 
\( p^i(c \mid z_t) \) predicting the condition from the noisy latent. 
Intuitively, smaller values of \( \|\delta_t\| \) correspond to weaker classifier gradients, 
suggesting that the sample is already likely to satisfy the desired condition.

In Sec.~\ref{sec:anneal} we build upon this insight to design our scheduler. While constraining the guidance scale \( w \) to the interval \( [0,1] \) is theoretically well-motivated within CFG++, we argue that this restriction can hinder the guidance mechanism’s ability to explore diverse modes of the conditional distribution \( p(z \mid c) \), ultimately limiting prompt adherence. To overcome this limitation, we lift the constraint on \( w \) and instead train our scheduler to robustly balance between mode exploration and fidelity to the data manifold.

\subsection{Geometric Intuition}
\label{sec:delta_intuition}

To provide further intuition into \( \| \delta_t \| \) as a navigational tool, we present a 2D illustration depicting two modes of the conditional distribution \( p(z_0 \mid c) \) in Figure~\ref{fig:intuition}.

The point \( z_{0|t} \) represents the estimated clean image at time \( t \), and the vectors \( \epsilon_t^{c} \) and \( \epsilon_t^{\varnothing} \) denote the conditional and unconditional noise predictions, respectively (scaling factors omitted for clarity). Their difference \( \delta_t \), shown in blue, reflects the guidance direction.

At time \( t \), the sample \( z_{0|t} \) is close to mode A, which corresponds to a high-quality image that partially matches the prompt \( c \). As a result, \( \epsilon_t^c \) is only slightly biased toward mode B relative to \( \epsilon_t^{\varnothing} \), leading to a small \( \|\delta_t\| \).

From a navigation perspective, we aim to reach mode B, which even better aligns with the prompt. We consider candidate estimates along the blue dashed line.

The point \( z^{(1)}_{0|t} \) is the clean image estimate at the next step when a small guidance scale \( w \) is used. At this location, there is a larger gap between \( \epsilon_{t-1}^{c} \) and \( \epsilon_{t-1}^{\varnothing} \), indicating misalignment.

In contrast, the point \( z^{(2)}_{0|t} \), which lies near mode B, represents a better solution. Here, both \( \epsilon_{t-1}^{c} \) and \( \epsilon_{t-1}^{\varnothing} \) are already aligned toward mode B, resulting in a minimal \( \|\delta_{t-1}\| \).

This geometric view suggests that $\|\delta_{t-1}\|$ reflects the degree of alignment with the condition and can serve as a signal for adaptively selecting the guidance scale $w$. We build on this property in the following section.

\makeatletter
\newcommand{\removelatexerror}{\let\@latex@error\@gobble}
\makeatother

\begin{figure}[t]
\removelatexerror
\begin{algorithm}[H]
\caption{Annealing Scheduler - Training}
\label{alg:annealing_scheduler}
\SetKwInOut{Input}{Require}
\Input{ \BlankLine
  $w_\theta$: \textbf{trainable} guidance scale model\;
  $\epsilon^{(\cdot)}_t$: \textbf{frozen} noise predictor, accepts $\varnothing$ or a condition, at timestep $t$\;
  $T$: total number of denoising steps
}

\Repeat{converged}{

  \tcp{--- Sample data and noise ---}
  Sample $(\mathbf{z}_0, \mathbf{c}) \sim p(\mathbf{z}_0, \mathbf{c})$, $t \sim U[1, T]$, $\boldsymbol{\epsilon} \sim \mathcal{N}(0, \mathbf{I})$, $\lambda\sim[0,1]$\;
  $\mathbf{z}_t \gets \text{AddNoise}(\mathbf{z}_0, t, \boldsymbol{\epsilon})$\;
  $\tilde{\mathbf{c}} \gets \text{Perturb}(\mathbf{c})$\;
  \BlankLine \BlankLine \BlankLine

  \tcp{--- Step at time $t$ ---}
    $\delta_{t} \gets \epsilon_t^{\tilde{c}}(\mathbf{z}_{t}) - \epsilon_t^{\varnothing}(\mathbf{z}_{t})$\;
    
  $\hat{\boldsymbol{\epsilon}}_t \gets \epsilon^{\varnothing}_t(\mathbf{z}_t) + w_\theta(t,\delta_{t},\lambda)\cdot \left( \epsilon_t^{\tilde{c}}(\mathbf{z}_t) - \epsilon_t^{\varnothing}(\mathbf{z}_t) \right)$\                             \tcp*{CFG}
  $z_{0|t}=(z_t-\sqrt{1-\bar{\alpha}_t}\hat{\epsilon}_t)/\sqrt{\bar{\alpha}_t}$\ \tcp*{Denoise}
  $z_{t-1}=\sqrt{\bar{\alpha}_{t-1}}z_{0|t}+\sqrt{1-\bar{\alpha}_{t-1}}\epsilon_t^{\varnothing}(z_t)$\ \tcp*{Renoise}
  \BlankLine \BlankLine \BlankLine

  \tcp{--- Step at time $t-1$ ---}
  $\delta_{t-1} \gets \epsilon_t^{\tilde{c}}(\mathbf{z}_{t-1}) - \epsilon_t^{\varnothing}(\mathbf{z}_{t-1})$\;
  \BlankLine \BlankLine \BlankLine

  \tcp{--- Compute loss and update ---}
  $\mathcal{L} \gets \lambda  \|\delta_{t-1}\|^2+(1 - \lambda)  \|\boldsymbol{\epsilon} - \hat{\boldsymbol{\epsilon}}_t\|^2 $\;
  Take gradient step on $\nabla_\theta \mathcal{L}$                                                           \tcp*{Update scheduler}
}
\end{algorithm}
\vspace{-10pt}
\end{figure}

\begin{figure*}[ht!] %
    \centering
    \begin{subfigure}[b]{0.24\textwidth} %
        \centering
        \setlength{\unitlength}{1\textwidth} %
        \begin{picture}(1,1) %
            \put(0,0){\includegraphics[width=\textwidth]{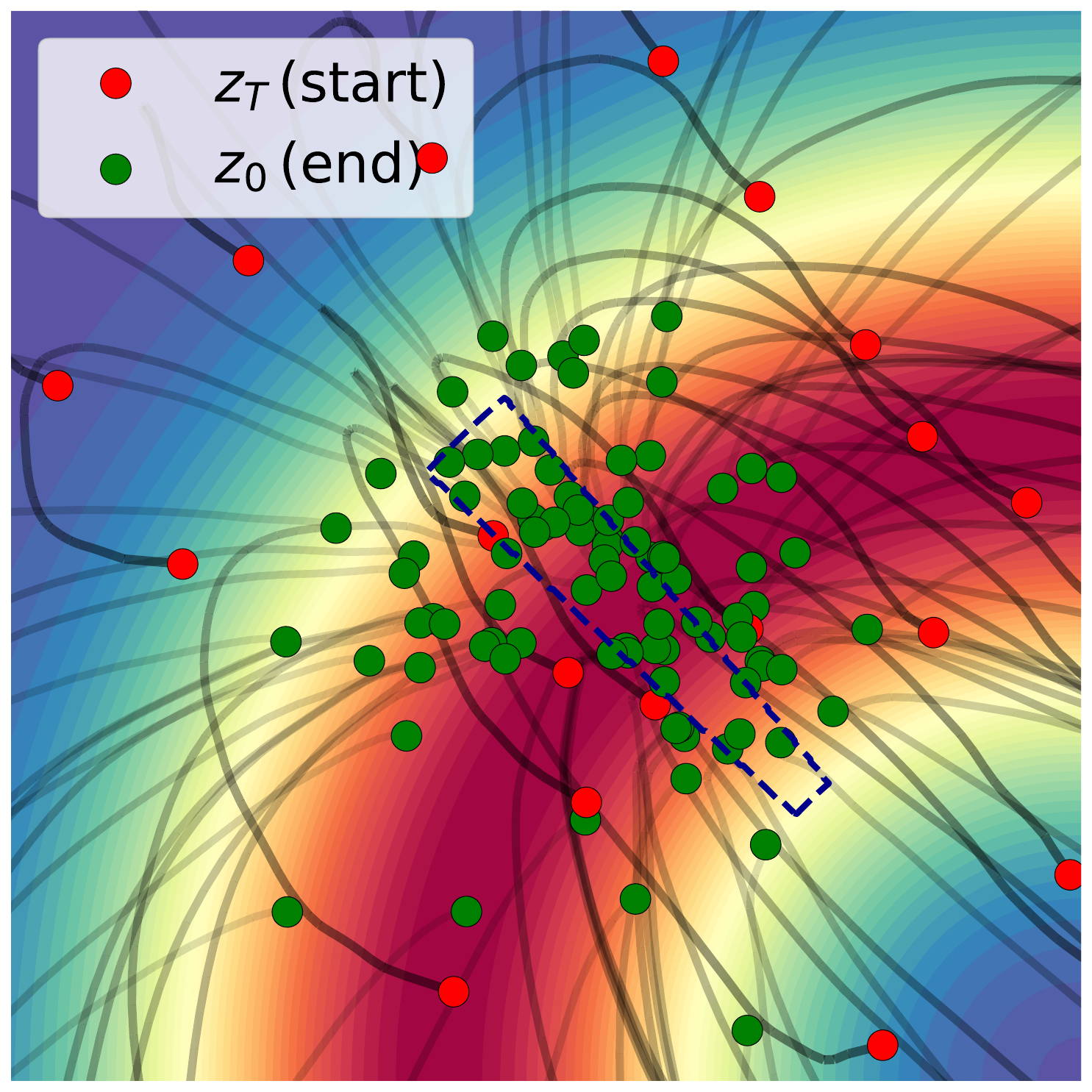}}
            \put(0.5,0.5){\includegraphics[width=0.5\textwidth]{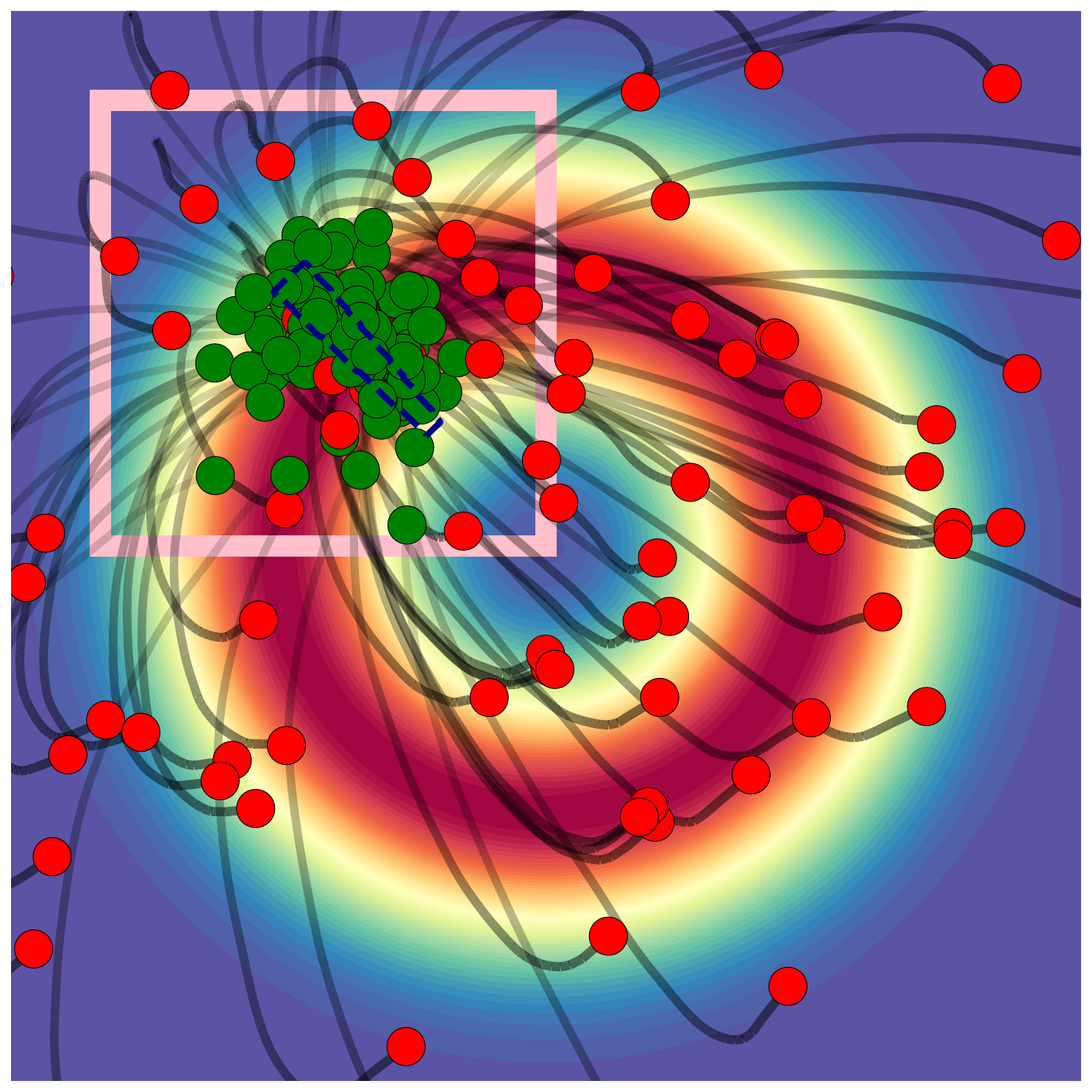}}
        \end{picture}
        \vspace{-15pt}
        \caption{\(w=0.1\)}
    \end{subfigure}
    \begin{subfigure}[b]{0.24\textwidth}
        \centering
        \includegraphics[width=\textwidth]{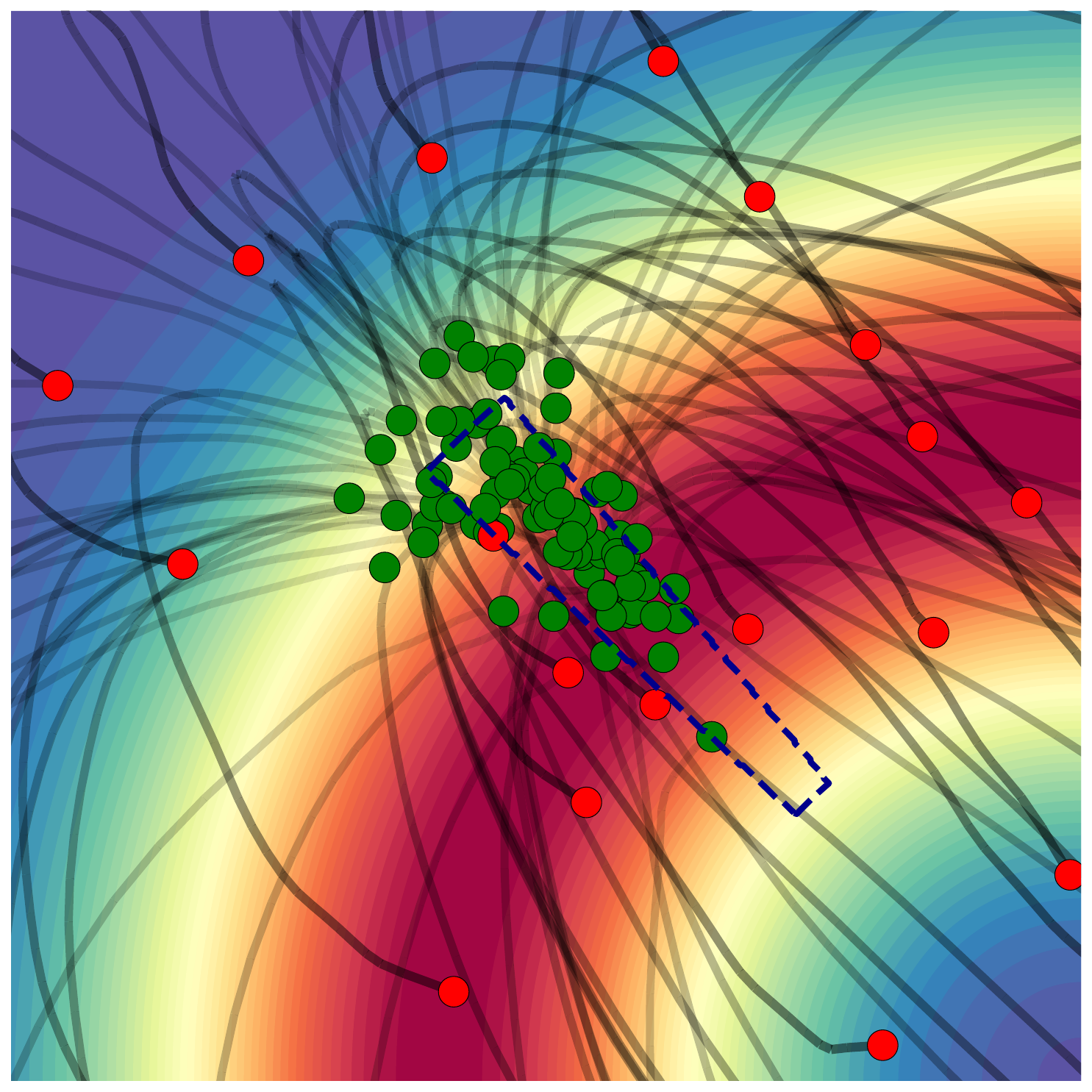}
        \vspace{-15pt}
        \caption{\(w=0.15\)}
    \end{subfigure}
    \begin{subfigure}[b]{0.24\textwidth}
        \centering
        \includegraphics[width=\textwidth]{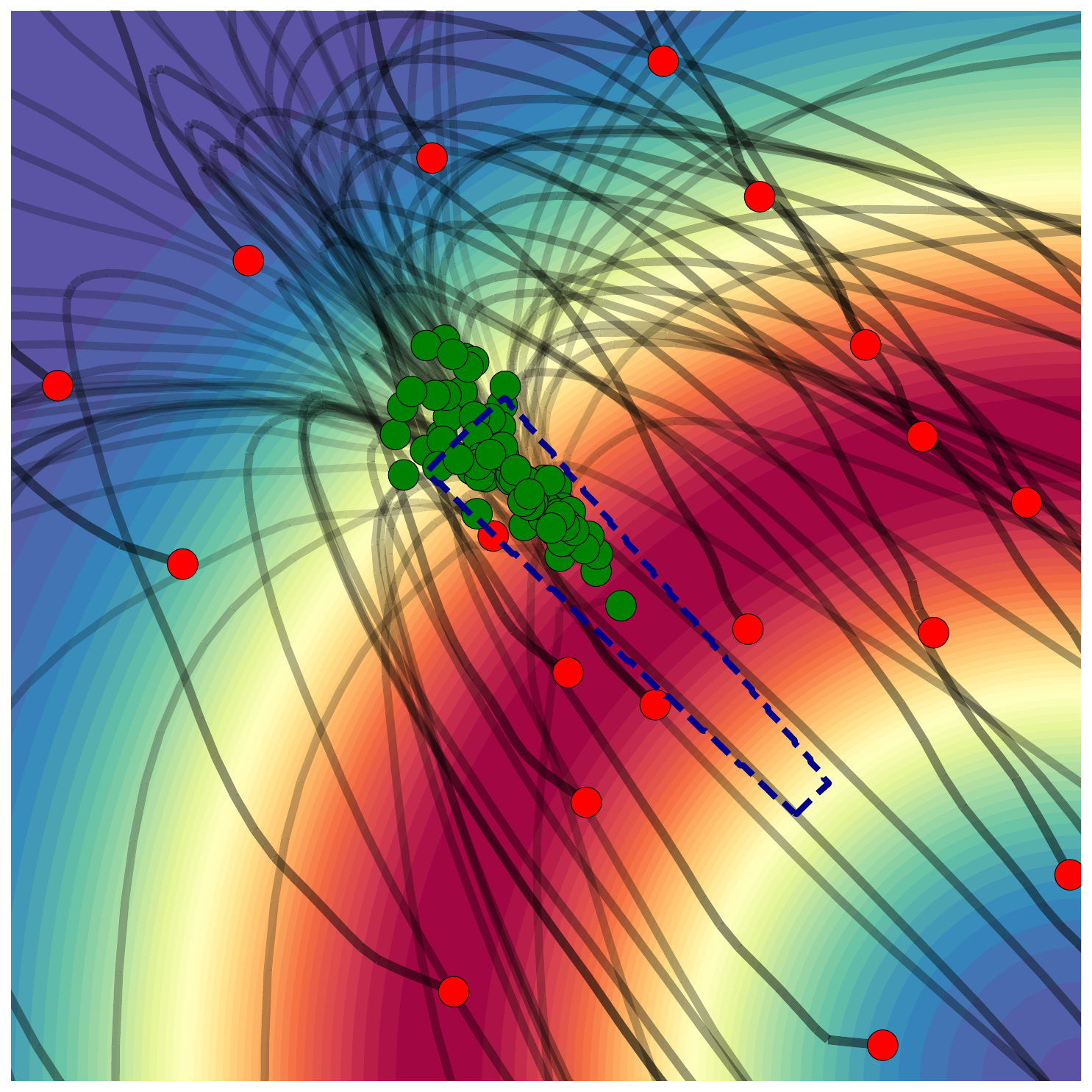}
        \vspace{-15pt}
        \caption{\(w=0.2\)}
    \end{subfigure}
    \begin{subfigure}[b]{0.24\textwidth}
        \centering
        \includegraphics[width=\textwidth]{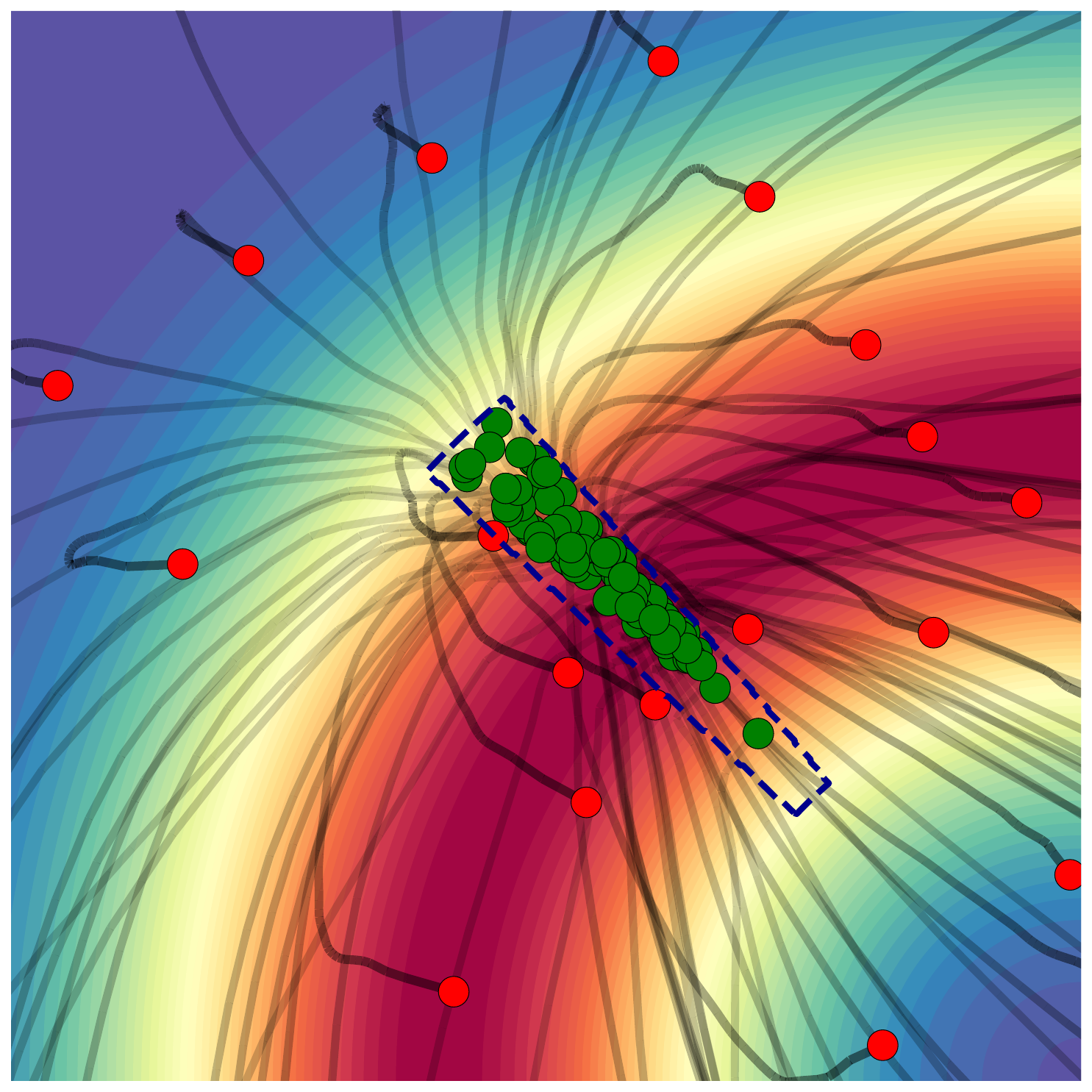}
        \vspace{-15pt}
        \caption{\textbf{Annealing}}
        \label{fig:toy_adaptive}
    \end{subfigure}
    \vspace{-10pt}
    \caption{A 2D diffusion toy example with a distribution density shaped as a wide ring. Random seeds conditioned on \( c = 3\pi/4 \) are plotted, with their denoising trajectories shown in gray. The dashed section highlights a region within a tolerance of \( \pm \pi/64 \) from \( c = \frac{3\pi}{4} \) where the manifold density is high. \textbf{(a)} \(w = 0.1\): Sampling with low guidance scale shows sub-optimal condition adherence. \textbf{(b)} \(w = 0.15\): Moderate guidance improves alignment, though some samples remain out of distribution. \textbf{(c)} \(w = 0.2\): Stronger guidance overfits the condition, at the expense of the sample quality. \textbf{(d)} Our Annealing scheduler achieves better condition alignment while remaining on the sample manifold.
    }
    \vspace{-15pt}    
    \label{fig:toy_diff}
\end{figure*}

\begin{figure} %
    \centering
    \begin{tabular}{@{}c@{}c@{}c@{}}
        \includegraphics[width=0.379\columnwidth]{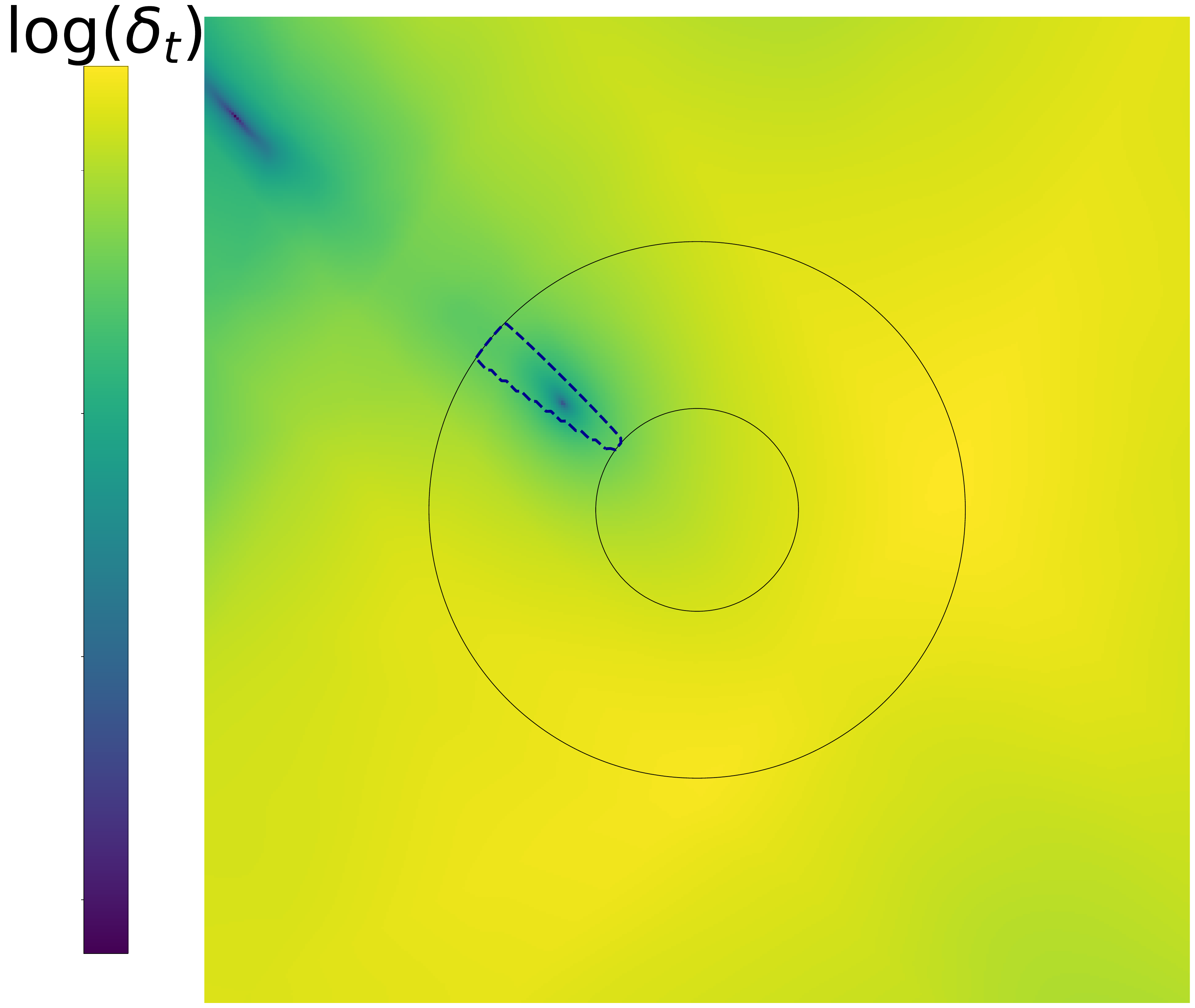} &
        \includegraphics[width=0.3160\columnwidth]{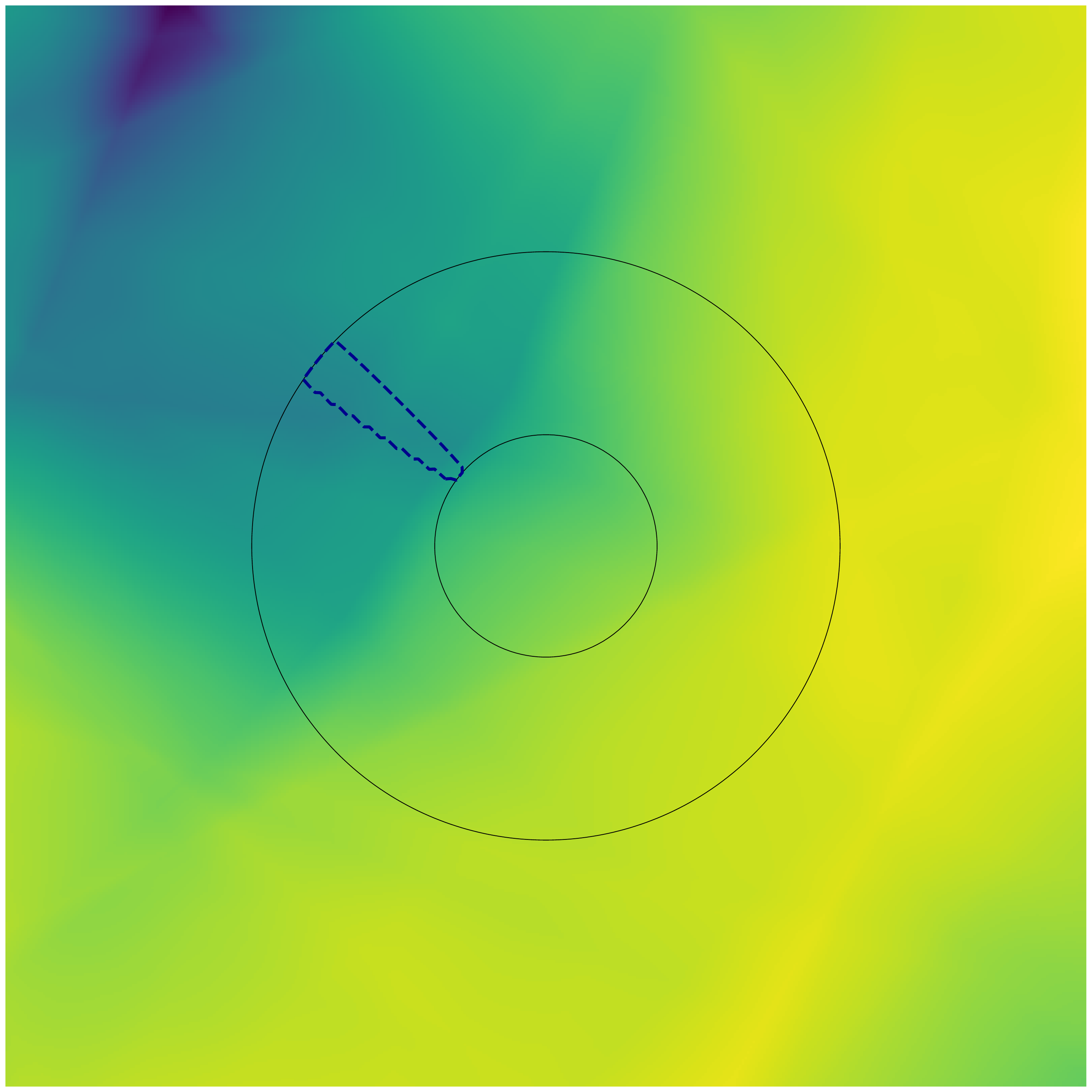} &
        \includegraphics[width=0.3160\columnwidth]{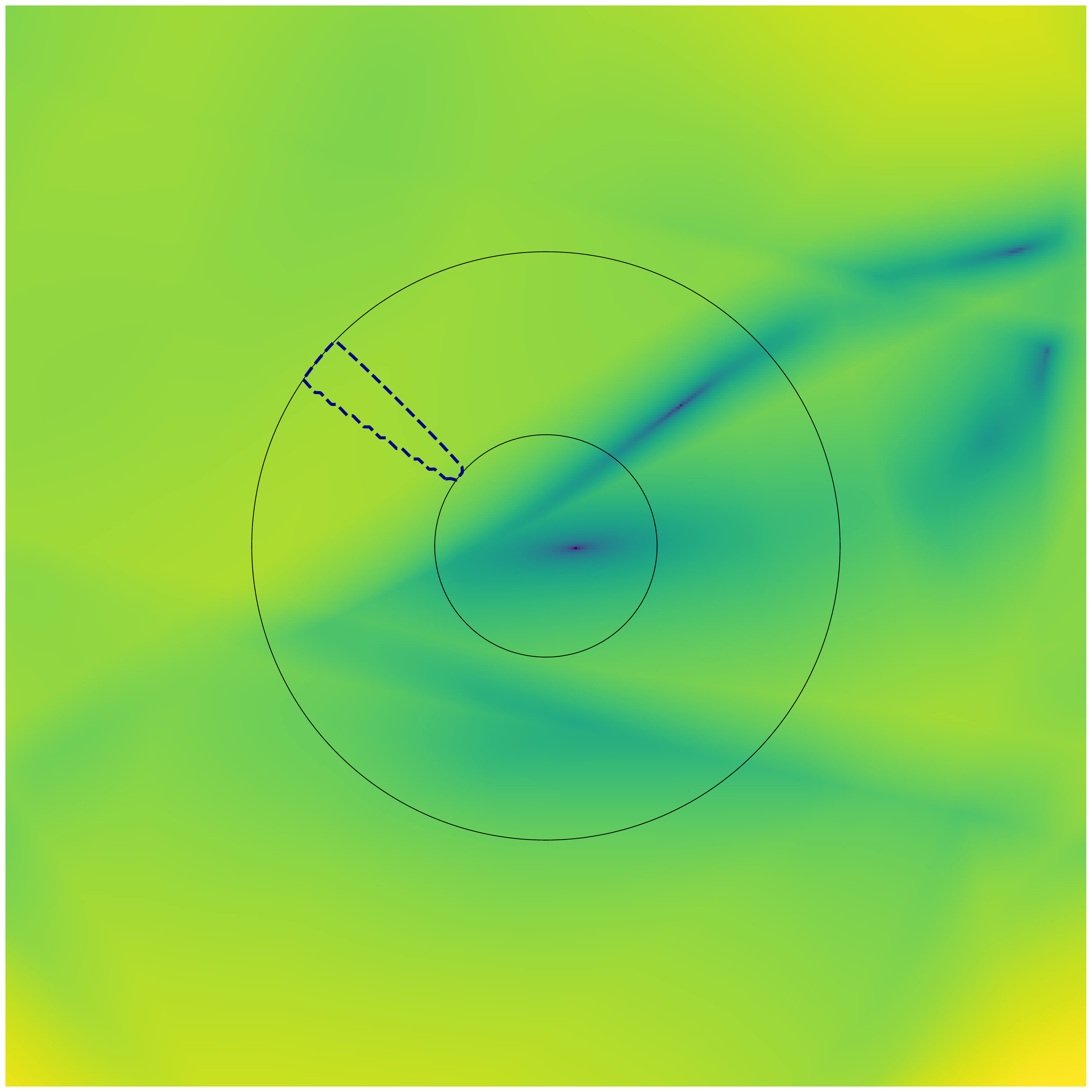} \\[-5pt]
        \(t=1\) & \(t=24\) & \(t=49\) %
    \end{tabular}
    \vspace{-10pt}
    \caption{\textbf{\(\log||\delta^t||\) heatmap for \(c = 3\pi/4\).} This measures the alignment between conditional and unconditional predictions across timesteps. The region between black circles indicates high sample density; the blue dashed line marks the target condition. \textbf{\(t=49\)}: Noise dominates, and predictions cluster near the source distribution center. \textbf{\(t=24\)}: Alignment improves near the target, though lower values persist off-distribution. \textbf{\(t=1\)}: A local minimum emerges at the target location on the ring.
} 
    \vspace{-12pt}
    \label{fig:diff_heatmap}
\end{figure}

\section{Annealing Scheduler}
\label{sec:anneal}

Building upon our insight that $\delta_t$ captures trajectory-specific information and that its norm is representative of the SDS convergence, we propose a learnable model \(w_\theta\left(t, \|\delta_t\|,\lambda\right)\) that maps the timestep $t$ and the magnitude \(\|\delta_t\|\) to a guidance scale.
The scalar \( \lambda \in [0,1] \) serves as a user-defined input that controls the trade-off between image quality and prompt alignment, offering an interpretable alternative to manually selecting a fixed guidance scale \( w \). Instead of directly tuning \( w \) as in vanilla CFG and CFG++, the user specifies a high-level preference via \( \lambda \), and the scheduler adaptively determines the optimal \( w \) throughout the generation process. Through experimentation (Sec.~\ref{sec:experiments}), we show that this formulation yields more consistent and controllable outcomes.

During inference, we incorporate our scheduler to the CFG++ sampling mecahnism by replacing the constant guidance scale $w$ in Eq.~\eqref{eq:cfg_sampling} to achieve:
\begin{equation}
    \mathbf{\hat{\epsilon}}_t = \mathbf{\epsilon}_t^{\varnothing} + {w_\theta\left(t, \delta_t, \lambda\right)} \cdot \left(\mathbf{\epsilon}_t^{c} - \mathbf{\epsilon}_t^{\varnothing}\right).
    \label{eq:scheduler}
\end{equation}

We implement \( w_\theta\) as a lightweight MLP and train it with a subset of the LAION-POP dataset~\cite{schuhmann2022laion}, which was curated for high resolution and high prompt-aligned images. We provide implementation details in Sec.\ifthenelse{\boolean{acm}}
{
A.1
}
{
~\ref{sec:imp_details}
}of the supplement.

During training, the pre-trained diffusion model is kept frozen. At each iteration, we sample an image with its corresponding caption $c$, together with a random timestep $t$ and noise \(\epsilon\) to compute $z_t$. The guided noise prediction \( \hat{\epsilon}_t \) is obtained from Eq.~\eqref{eq:scheduler}. The parameter \( \lambda \in \left[0, 1\right]\) is sampled uniformly.

Our training loss balances between two objectives, as governed by $\lambda$:
\begin{equation}
\mathcal{L} = \lambda   L_{t}^{\delta} + \left(1-\lambda\right)L_{t}^{\epsilon}.
\end{equation}
Here, \(L_t^\delta\) and \(L_t^\epsilon\) are loss terms that promote prompt alignment and image quality, respectively. We now turn to formally define these losses,
with the complete training procedure detailed in Algorithm~\ref{alg:annealing_scheduler}.

\paragraph{$\delta$-loss}

Following our observation in sections~\ref{sec:sds} and \ref{sec:delta_intuition}, we introduce a novel loss, leveraging
\(\|\delta_t\|\) as a proxy value that aims to reflect prompt alignment. This loss is designed to encourage the scheduler to select guidance scales that move the denoising trajectory toward regions where the model's conditional and unconditional predictions begin to agree.

In practice, for a given $z_{t}$, we perform denoising with $\hat{\epsilon}_t$ and renoising with $\epsilon_t^\varnothing$ to obtain \(z_{t-1}\). By evaluating \(\|\delta_{t-1}\|\) at this point, we introduce our $\delta$-loss:
\begin{equation}
L_{t}^{\delta} = \|\delta_{t-1}\|_2^2.
\label{eq:delta_loss}
\end{equation}
This loss leverages the diffusion model's prior of the alignment with the target prompt, as illustrated in Fig.~\ref{fig:intuition}. However, solely optimizing on \(L_{t}^{\delta}\) results in very high guidance scales, leading to out-of-distribution samples, similar to \(z_{t-1}^{(2)}\) in Fig.~\ref{fig:cfg} (see Sec. \ref{sec:toy_example},\ref{sec:experiments} for further analysis). Therefore, we opt to maintain fidelity to the data manifold using the second loss term \(L_t^\epsilon\).

\paragraph{$\epsilon$-loss}

To ensure that the predicted guided noise \( \hat{\epsilon}_t \) from Eq.~\eqref{eq:scheduler} matches the sampled noise \( \epsilon \), we introduce a denoising objective, namely, the reconstruction loss:
\begin{equation}
L_{t}^\epsilon = \|\hat{\epsilon}_{t} - \epsilon\|_2^2.
\label{eq:eps_loss}
\end{equation}
This loss resembles the standard denoising diffusion objective, but instead of applying to the conditional model prediction, it operates on the guided prediction \( \hat{\epsilon}_t \), which combines both conditional and unconditional signals. Its primary role is to regularize the $\delta$-loss by preventing the guidance scale from pushing the generation toward implausible regions. By encouraging \( \hat{\epsilon}_t \) to remain close to the true noise \( \epsilon \), this loss helps preserve visual quality and ensures that the denoising trajectory remains within realistic bounds.

\vspace{-2pt}
\paragraph{Prompt Perturbation}

During training, each latent \( z_t \) is paired with a prompt \( c \) that closely matches the corresponding image. Even after applying noise to obtain \( z_t \), semantic information about the prompt remains encoded in the latent~\cite{lin2024common}, preserving alignment throughout the denoising trajectory. In contrast, inference begins from pure noise, and the prompt is injected through the denoising process. As shown by prior work~\cite{samuel2024generating,ma2025inference,singhal2025general}, the alignment of complex prompts remains highly sensitive to the initial seed, often leading to greater variability at inference time. To simulate this mismatch, we inject Gaussian noise into the prompt embeddings during training (see Sec.\ifthenelse{\boolean{acm}}
{
A.6
}
{
~\ref{sec:perturb}
}for details). This exposes the scheduler to imperfect prompt-image alignment, improving its robustness.

Our approach was motivated by CADS~\cite{sadat2023cads}, where noise is injected into the prompt embeddings \textit{during inference} to encourage mode diversity. Their analysis showed that this perturbation smooths the conditional score \( \nabla_{z_t} \log p(c \mid z_t) \), acting as a regularizer that prevents the model from collapsing onto dominant modes. In contrast, we apply this principle \textit{during training} to enhance robustness, enabling the scheduler to generalize across a range of prompt-image alignment scenarios.

This technique improves the scheduler's behavior across different guidance regimes. When \( \lambda \) is low and \( L_t^\epsilon \) dominates, it promotes the generation of high-quality images even under imprecise alignment. When \( \lambda \) is high and \( L_t^\delta \) dominates, it helps the scheduler adaptively shift toward nearby modes that better satisfy the prompt.

\paragraph{Predicted guidance scales.} We present the learned guidance scales predicted by the trained scheduler in Figure~\ref{fig:w_heatmap}. As shown, the scheduler adapts its annealing strategy based on different values of the user-specified parameter \( \lambda \).

\vspace{-4pt}
\section{2D Toy Example}
\label{sec:toy_example}
\vspace{-1pt}

We now turn to investigating the behavior of our annealing scheduler in a controlled and interpretable setting using a 2D toy example.

In Fig.~\ref{fig:toy_diff}, we illustrate the behavior of a diffusion model trained to approximate a target distribution shaped as a wide ring. The conditional distribution is defined over the angular variable \( c \sim U(0, 2\pi) \), while the initial noise samples \( z_T \) are drawn from a standard normal distribution. We condition generation on \( c = \frac{3\pi}{4} \) and visualize the denoising trajectories under different constant guidance scales with CFG++, as well as our adaptive scheduler.

The formed trajectories demonstrate both the strengths and limitations of classifier-free guidance. Increasing the guidance scale enforces stronger adherence to the conditioning signal but also pushes samples away from the data manifold. In contrast, our annealing scheduler (Fig.~\ref{fig:toy_adaptive}) adaptively modulates the guidance strength during denoising, resulting in improved condition alignment while preserving fidelity to the data manifold. As can be seen, our scheduler also achieves better coverage of the conditional distribution, reflecting more diverse and representative generations.

To support our insight into the \(\delta_t\)-loss (Eq.~\ref{eq:delta_loss}), we display the norm \( \|\delta_t\| \) across different denoising steps in Fig.~\ref{fig:diff_heatmap}, for the same conditioning value \( c = \frac{3\pi}{4} \). As \( t \) decreases, we observe that \( \|\delta_t\| \) becomes small near the correct region of the ring, indicating that the conditional and unconditional predictions are well aligned and the sample is approaching the target mode. This implies that promoting low \(\|\delta_t\|\) throughout the denoising process can lead to better alignment with the conditioning signal.

However, $\| \delta_t \|$ also tends to have low values away from the ring (e.g., Fig.~\ref{fig:diff_heatmap}, \( t = 1 \)), suggesting that minimizing $\| \delta_t \|$ alone may guide samples off the data manifold. This highlights the need for an additional regularization term, such as the \(\epsilon\)-loss (Eq.~\eqref{eq:eps_loss}), to ensure that generations remain faithful to the data manifold.
\vspace{-2pt}

\ifthenelse{\boolean{archive}}
    {}
  {\begin{figure} %
    \centering
     \setlength{\belowcaptionskip}{-10pt}

    \begin{tabular}{@{}c@{}c@{}c@{}c@{}}
        \makecell{   CFG} & 
        \makecell{  APG} & 
        \makecell{  CFG++} & 
        \makecell{  \textbf{Annealing}} \\[0pt]

        \includegraphics[width=0.25\columnwidth]{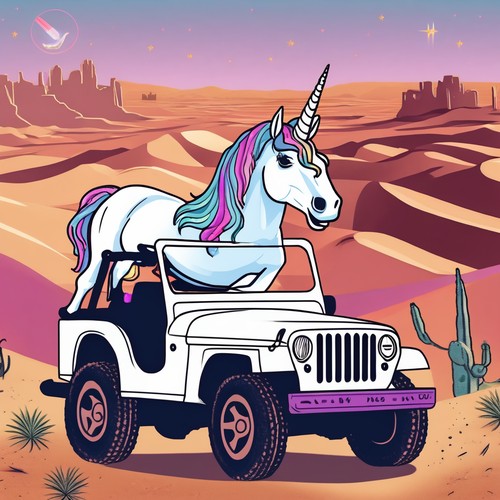} &
        \includegraphics[width=0.25\columnwidth]{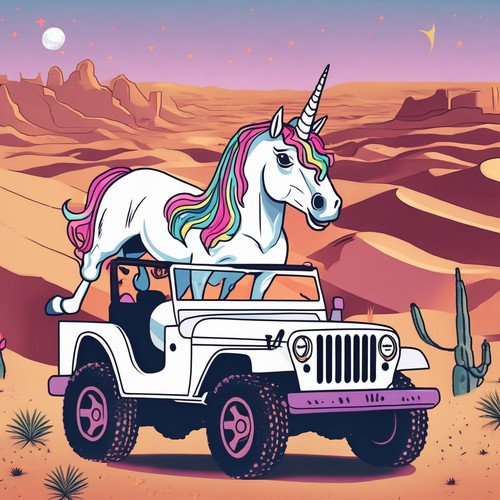} &
        \includegraphics[width=0.25\columnwidth]{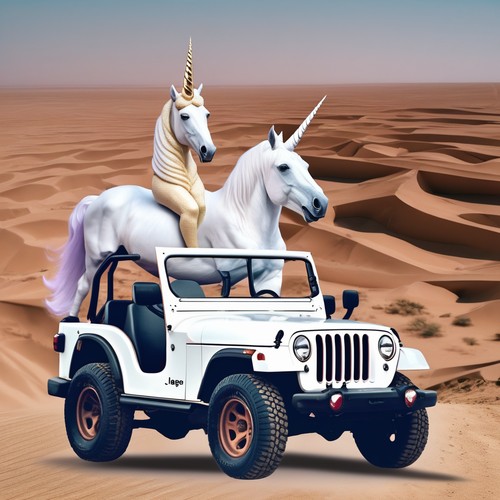}\hspace{0.3em} &
        \includegraphics[width=0.25\columnwidth]{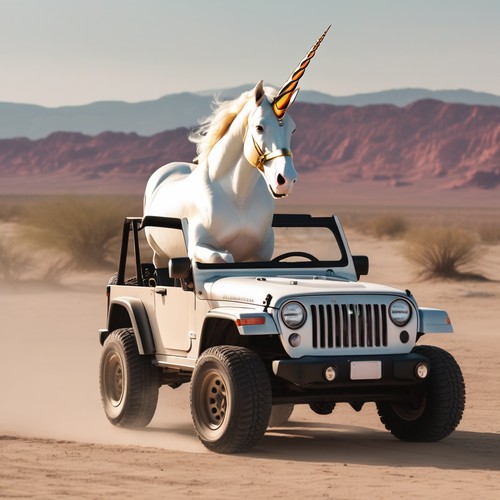} \\[-5pt]
        \multicolumn{4}{c}{\scriptsize "A \textcolor{red}{\textbf{photo}} of unicorn driving a jeep in the desert"} \\[-1pt]

        \includegraphics[width=0.25\columnwidth]{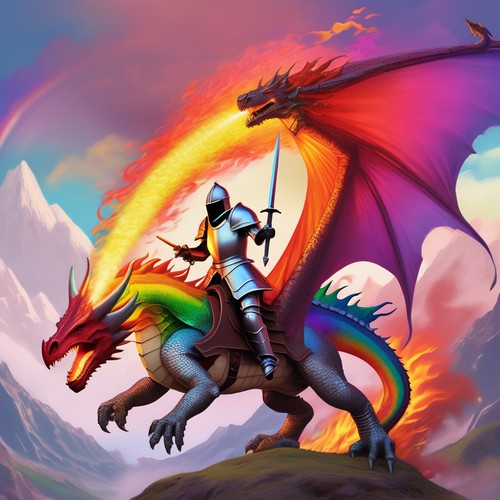} &
        \includegraphics[width=0.25\columnwidth]{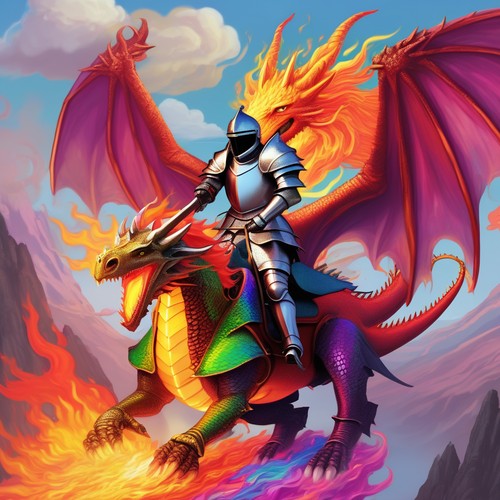} &
        \includegraphics[width=0.25\columnwidth]{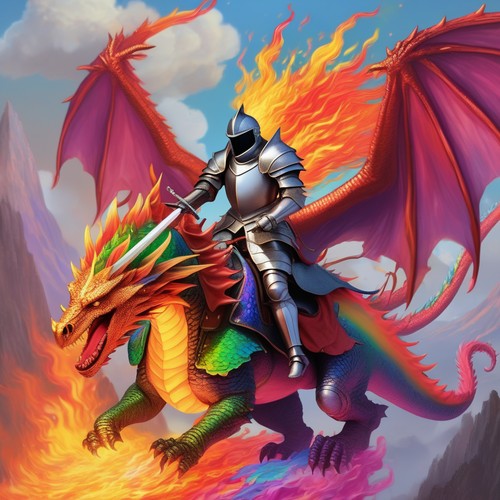}\hspace{0.3em} &
        \includegraphics[width=0.25\columnwidth]{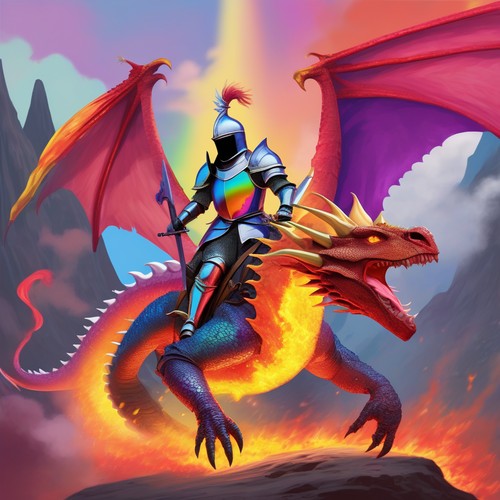} \\[-5pt]
        \multicolumn{4}{c}{\scriptsize \shortstack{%
            "A \textcolor{red}{\textbf{knight in rainbow armor}} riding a \textcolor{red}{\textbf{dragon made of fire}}"
        }} \\[-1pt]

        \includegraphics[width=0.25\columnwidth]{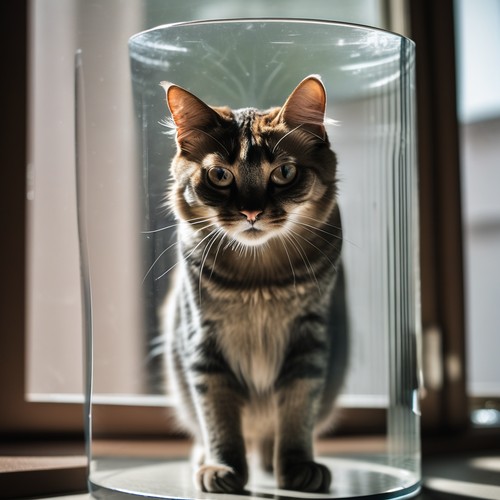} &
        \includegraphics[width=0.25\columnwidth]{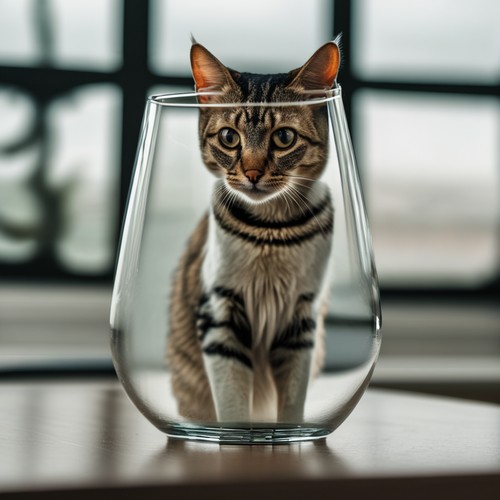} &
        \includegraphics[width=0.25\columnwidth]{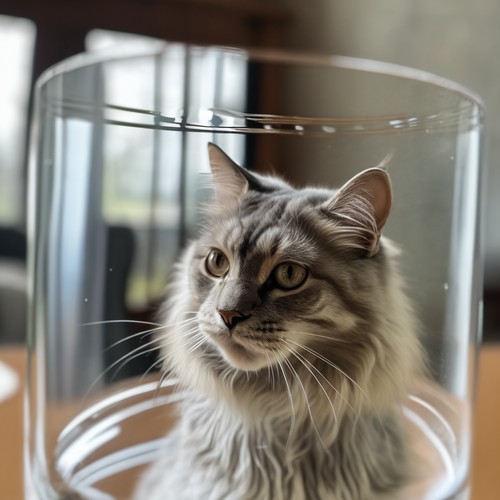}\hspace{0.3em} &
        \includegraphics[width=0.25\columnwidth]{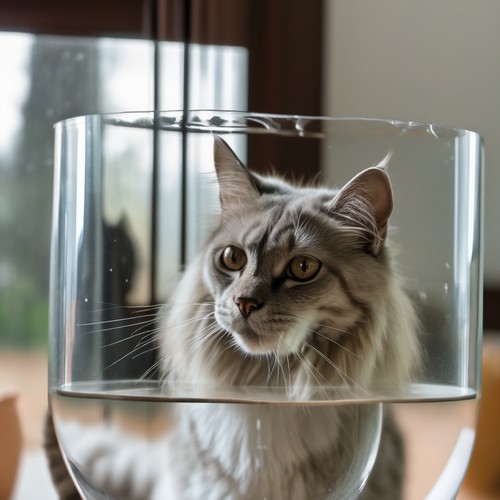} \\[-5pt]
        \multicolumn{4}{c}{\scriptsize "A cat looking through a  \textcolor{red}{\textbf{glass of water}}"} \\[-1pt]

        \includegraphics[width=0.25\columnwidth]{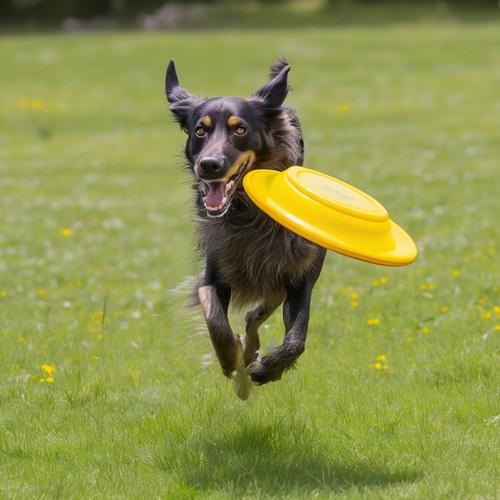} &
        \includegraphics[width=0.25\columnwidth]{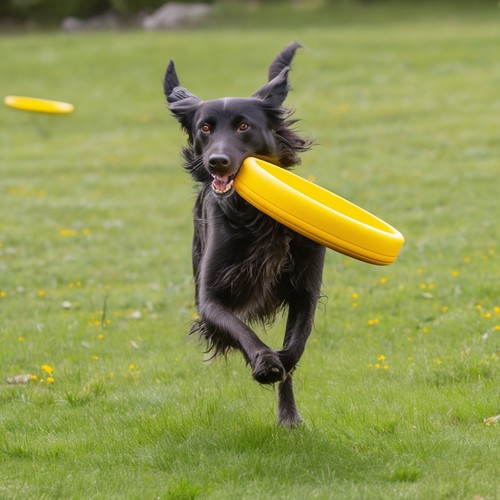} &
        \includegraphics[width=0.25\columnwidth]{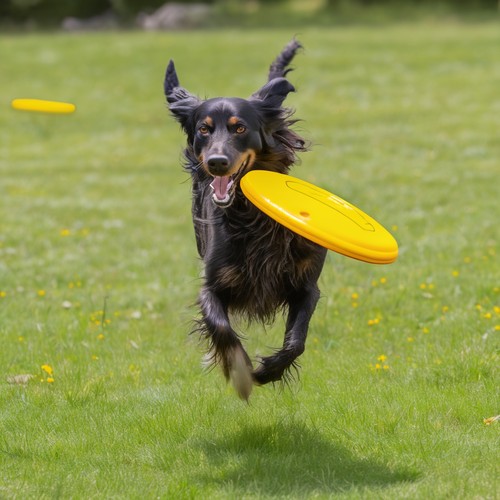}\hspace{0.3em} &
        \includegraphics[width=0.25\columnwidth]{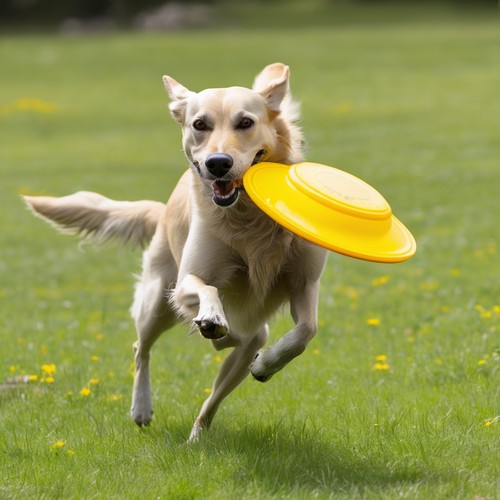} \\[-5pt]
        \multicolumn{4}{c}{\scriptsize \shortstack{%
            "A \textcolor{red}{\textbf{yellow dog}} runs to grab a yellow frisbee in the grass."
        }} \\[-1pt]
        
        \includegraphics[width=0.25\columnwidth]{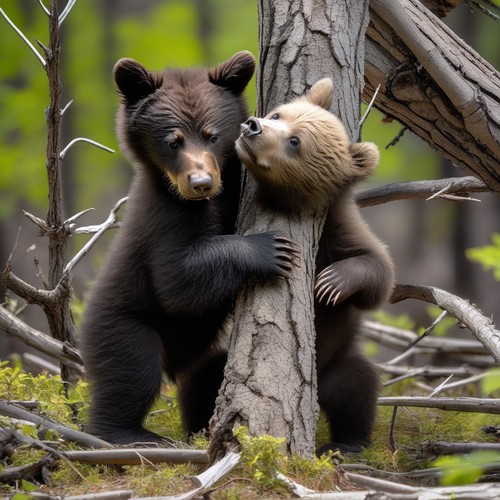} &
        \includegraphics[width=0.25\columnwidth]{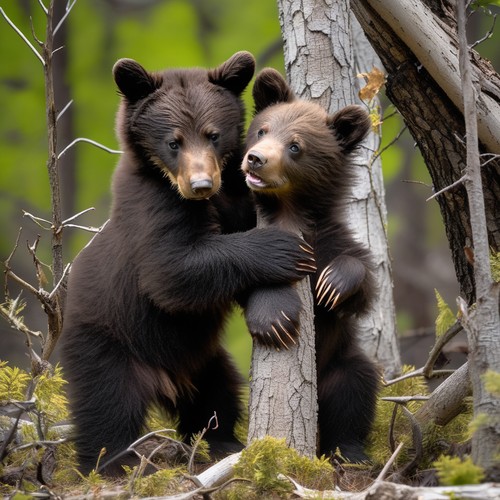} &
        \includegraphics[width=0.25\columnwidth]{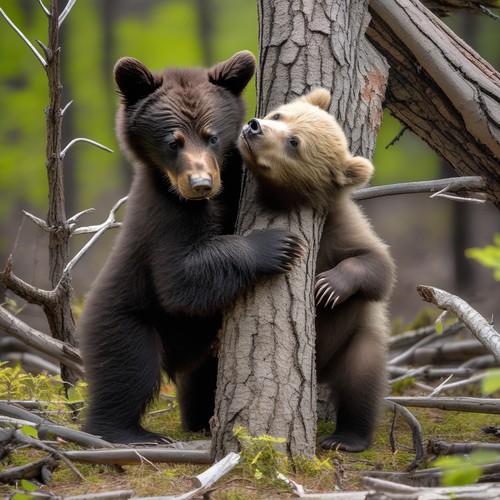}\hspace{0.3em} &
        \includegraphics[width=0.25\columnwidth]{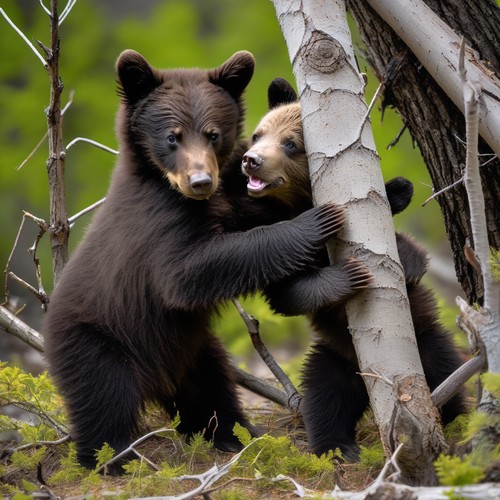} \\[-5pt]
        \multicolumn{4}{c}{\scriptsize "Bear cubs play among the fallen tree limbs."} \\[-1pt]

        \includegraphics[width=0.25\columnwidth]{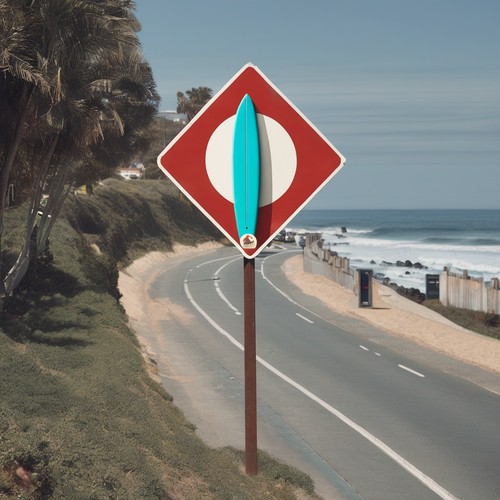} &
        \includegraphics[width=0.25\columnwidth]{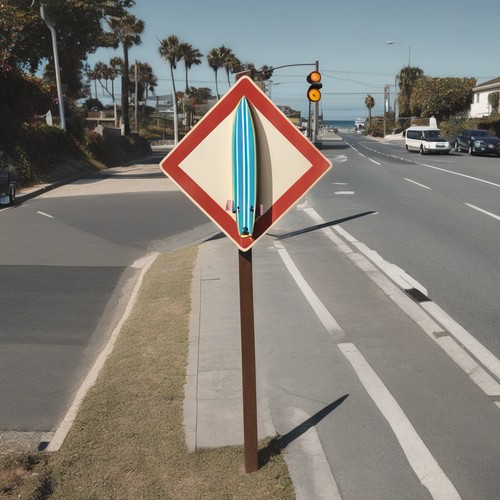} &
        \includegraphics[width=0.25\columnwidth]{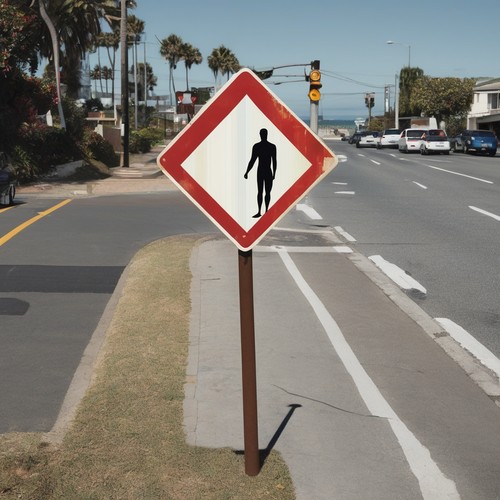}\hspace{0.3em} &
        \includegraphics[width=0.25\columnwidth]{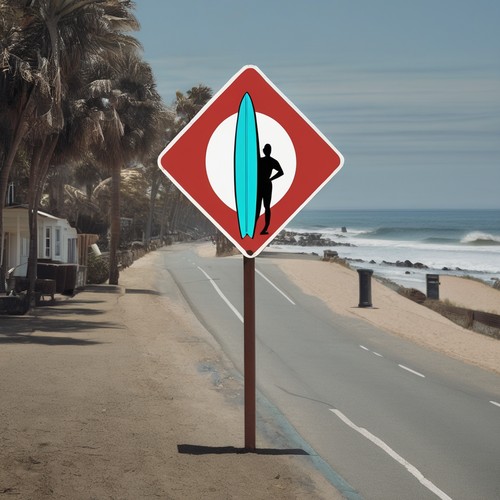} \\[-5pt]
        \multicolumn{4}{c}{\scriptsize \shortstack{%
            "A traffic sign that has a picture of a  \textcolor{red}{\textbf{man holding a surfboard}} on it."
        }} \\[-1pt]

    \end{tabular}
    \vspace{-10pt}
    \caption{Qualitative comparison of our Annealing method $\lambda=0.8$ (right column) vs. three guidance methods: CFG ($w=15$), APG ($w=20$) and CFG++ ($w=1.2$).}
    \vspace{-6pt}
    \label{fig:method_comparison}
\end{figure}
}         %

\section{Experiments and Results}
\label{sec:experiments}
\begin{figure}[t]
    \centering
    \setlength{\tabcolsep}{0.0em} %
    \begin{tabular}{ccc}
        \includegraphics[width=0.48\textwidth]{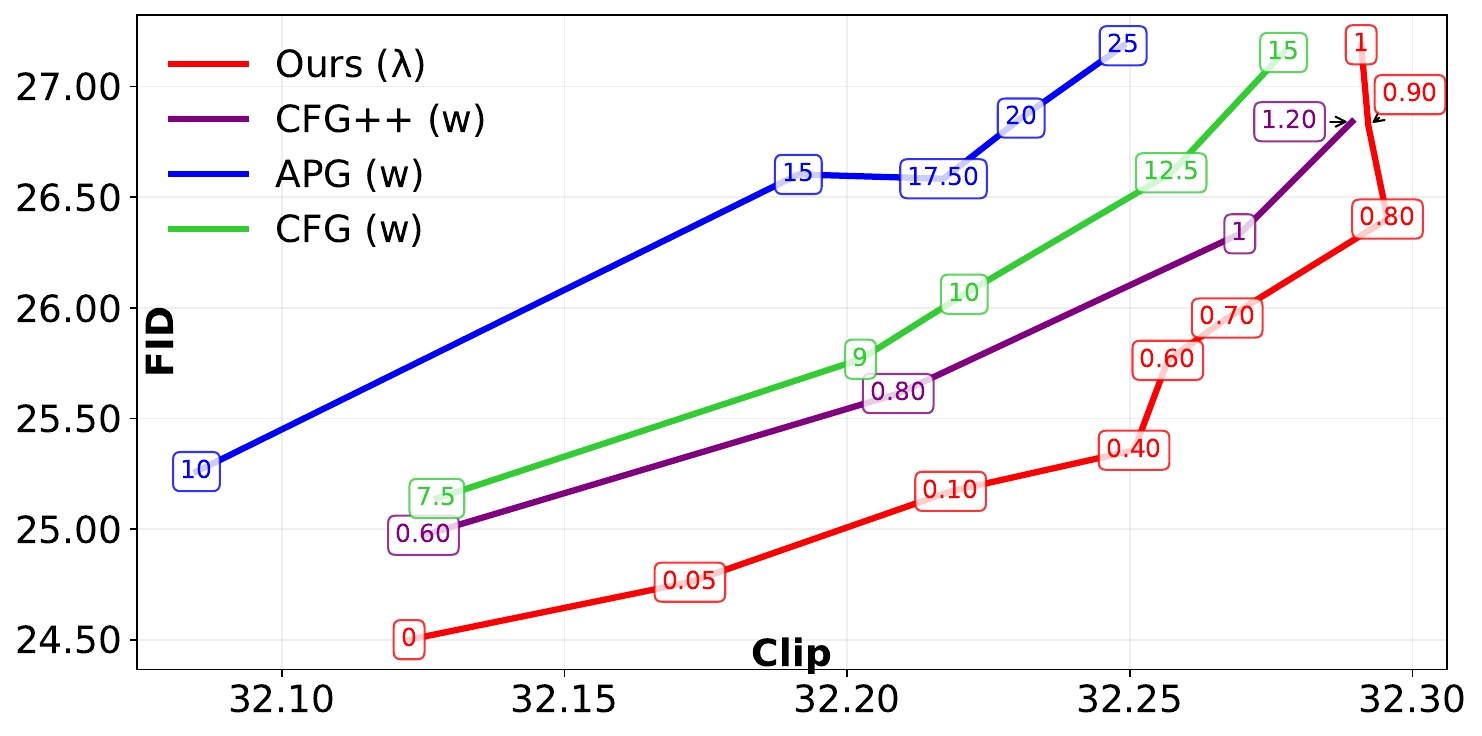}&

    \end{tabular}
    \vspace{-15pt}
    \caption{\textbf{Quantitative Metrics.} FID versus CLIP.}
    \vspace{-15pt}
    \label{fig:metrics_curves}
\end{figure}

\begin{table}[t]
    \centering
    \caption{Comparison of CFG, APG, CFG++, and our method across FID, CLIP similarity, Image Reward (IR), Precision (P), and Recall (R).}
    \vspace{-5pt}
    \begin{tabular}{l l ccccc}
        \toprule
        \textbf{Method} & \textbf{Scale} & \textbf{FID ↓} & \textbf{CLIP ↑} & \textbf{IR ↑} & \textbf{P ↑} & \textbf{R ↑} \\
        \midrule
        CFG&$w=7.5$ & 25.13 & 32.12 & 0.817 & \textbf{0.863} & 0.630 \\
        APG&$w=10$ & 25.25 & 32.08 & \textbf{0.818} & 0.862 & \textbf{0.631} \\
        CFG++& $w=0.6$ & 24.97 & 32.12 & 0.808 & 0.859 & 0.629 \\
        Ours&$\lambda=0.05$ & \textbf{24.76} &\textbf{32.16} & 0.809 &  0.860 &  0.620\\
        \midrule
        CFG&$w=10$ & 26.06 & 32.22 & 0.859 & 0.859 & 0.594 \\
        APG&$w=15$ & 26.60 & 32.19 & \textbf{0.865} & \textbf{0.864} & 0.592 \\
        CFG++ &$w=0.8$ & 25.61 & 32.20 & 0.857 & 0.855 & 0.601 \\
        Ours & $\lambda=0.4$ & \textbf{25.35} & \textbf{32.25} & \textbf{0.865} & 0.859 & \textbf{0.606} \\
        \midrule
        CFG&$w=12.5$ & 26.61 & 32.25 & 0.881 & 0.850 & 0.570 \\
        APG&$w=17.5$ & 26.58 & 32.21 & \textbf{0.887} &\textbf{0.861} & 0.586 \\
        CFG++& $w=1$ & 26.33 & \textbf{32.26} & 0.882 & 0.848 & 0.570 \\
        Ours& $\lambda=0.7$ & \textbf{25.95} & \textbf{32.26} & 0.884 & 0.852 & \textbf{0.594} \\
        \midrule
        CFG&$w=15$ & 27.15 & 32.27 & 0.883 & 0.844 & 0.570 \\
        APG&$w=20$ & 26.85 & 32.23 & 0.893 &\textbf{ 0.855} & 0.577 \\
        CFG++& $w=1.2$ & 26.84 & 32.28 & 0.894 & 0.847 & 0.551 \\
        Ours& $\lambda=0.8$ & \textbf{26.40} & \textbf{32.29} & \textbf{0.898} & 0.846 & \textbf{0.586} \\
        \bottomrule
    \end{tabular}
    \vspace{-15pt}
    \label{tab:subset_metric_comparison}
\end{table}

To evaluate our annealing guidance scheduler, we conduct a comprehensive set of experiments, including qualitative comparisons and quantitative evaluations. We compare our method against existing guidance scheduling approaches, including APG~\cite{sadat2024eliminating}, CFG++~\cite{chung2024cfg++}, and the commonly used CFG~\cite{ho2022classifier} baseline. All experiments are performed using SDXL~\cite{podell2023sdxl}. In the supplementary materials, we provide 
results on different prompt categories (Sec.\ifthenelse{\boolean{acm}}
{
B.1
}
{
~\ref{sec:parti_prompts}
}), ablation studies (Sec.\ifthenelse{\boolean{acm}}
{
B.2
}
{
~\ref{sec:ablations}
}), experiments on different solvers and noise schedules (Sec.\ifthenelse{\boolean{acm}}
{
B.3
}
{
~\ref{sec:other_solvers_schedulers}
}) and a flow-matching extension (Sec.\ifthenelse{\boolean{acm}}
{
B.5
}
{
~\ref{sec:flow_matching}
}).

\subsection{Qualitative comparisons}

We compare our annealing guidance scheduler qualitatively in Fig.~\ref{fig:method_comparison}. As shown, our method consistently delivers superior results both in image quality and prompt alignment.

In the first row, the left prompt asks for a unicorn driving a jeep, where baselines produce cartoonish results or artifacts and fail to place the unicorn correctly. On the right, our approach is the only one that renders the knight in rainbow armor, while others leak the rainbow onto the dragon and even hallucinate an extra head.

In the second row, only our scheduler generates the water in the glass (left), and produces a dog with yellowish fur (right), and in the third row, our method separates the bear from the tree (left) and faithfully renders the surfboard sign (right), whereas baselines blur or distort the content. Additional qualitative comparisons are shown in Figs.~\ref{fig:extra_cfgpp_1} and~\ref{fig:extra_1}, against CFG++ and CFG, respectively. We provide further results, including comparisons across additional guidance scales, in Sec.\ifthenelse{\boolean{acm}}
{
C
}
{
~\ref{sec:add_results}
}of the supplement.

\vspace{-1pt}
\subsection{Quantitative comparisons}

We conduct a quantitative evaluation on the MSCOCO 2017 validation set by generating 5,000 images per model using identical seeds. Image quality is assessed using FID~\cite{heusel2018ganstrainedtimescaleupdate} and the recently proposed FD-DINOv2~\cite{oquab2024dinov2learningrobustvisual}, while prompt alignment is measured via CLIP similarity~\cite{radford2021learningtransferablevisualmodels}.

To visualize the trade-off between image quality and prompt alignment at commonly used CFG guidance scales (\( w \geq 7.5 \)), we plot FID vs. CLIP in Fig.~\ref{fig:metrics_curves}, and DINOv2 vs. CLIP in Fig.\ifthenelse{\boolean{acm}}
{
12
}
{
~\ref{fig:fd_dino_curves}
}of the supplement.
As shown, APG doesn't improve over CFG across these metrics, while CFG++ provides gains only in the FID/CLIP space. In contrast, our method consistently enhances both alignment and image quality, outperforming all baselines across evaluation criteria.

For direct comparison, we select multiple operating points of our scheduler by varying $\lambda$, and match each to the closest configuration of CFG, CFG++, or APG in terms of FD-DINOv2. Table~\ref{tab:subset_metric_comparison} reports the corresponding FID, CLIP similarity, and additionally ImageReward~\cite{xu2023imagerewardlearningevaluatinghuman} for human-preference, and precision and recall~\cite{kynkaanniemi2019improved} for quality and diversity respectively. Across all settings, our scheduler achieves the lowest FID, the highest CLIP similarity, and consistently outperforms baselines in recall at higher guidance strengths. Notably, it also attains the highest ImageReward in two out of four matched configurations. A full table including FD-DINOv2 scores, and implementation details for the evaluations are provided in Sec.\ifthenelse{\boolean{acm}}
{
A.4
}
{
~\ref{sec:metrics_calc}
}of the supplement.

\section{Conclusions}

We have presented an annealing guidance scheduler that adaptively adjusts the guidance scale throughout the denoising process. Unlike the widely used CFG, which relies on a fixed guidance scale, and its improved variant CFG++, our method dynamically determines step sizes based on the evolving structure of the latent space. This approach is grounded in viewing guidance as an optimization problem aimed at minimizing the SDS loss, steering latents to better match the prompt while remaining faithful to the model’s prior distribution.

We find that this adaptive strategy is particularly beneficial for complex prompts, where balancing prompt fidelity and sample quality is most challenging. Nonetheless, our results highlight a fundamental trade-off between strict adherence to the prompt and staying within the data manifold.

Navigating the high-dimensional diffusion space remains inherently difficult due to its intricate and multimodal structure. Nevertheless, our work opens the door to future exploration of more principled, context-aware guidance mechanisms that better adapt to the geometry of the denoising trajectory.

\makeatletter
\begin{acks}
We would like to thank Oren Katzir, Daniel Garibi, Or Patashnik, and Shelly Golan for their early feedback and insightful discussions. We also thank the anonymous reviewers for their thorough and constructive comments, which helped improve this work.
\end{acks}

\bibliographystyle{ACM-Reference-Format}
\bibliography{main}

\newpage

\begin{figure*}[t]
    \centering
    \begin{tabular}{@{}c@{\hspace{0.005\textwidth}}c@{\hspace{0.03\textwidth}}c@{\hspace{0.005\textwidth}}c@{}}

        CFG++ ($w=0.8$)& \textbf{Annealing ($\lambda=0.4$)} & CFG++ ($w=0.8$) & \textbf{Annealing ($\lambda=0.4$)} \\
        \includegraphics[width=0.44\columnwidth]{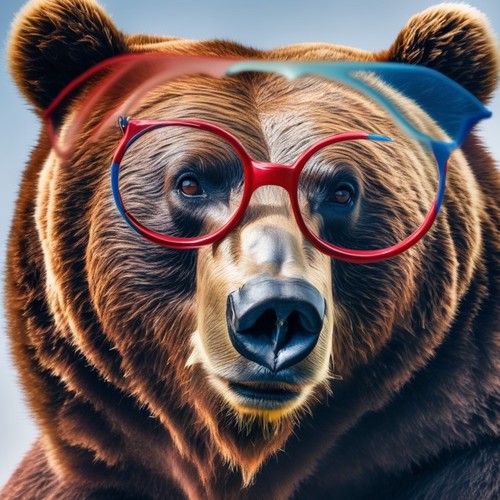} &
        \includegraphics[width=0.44\columnwidth]{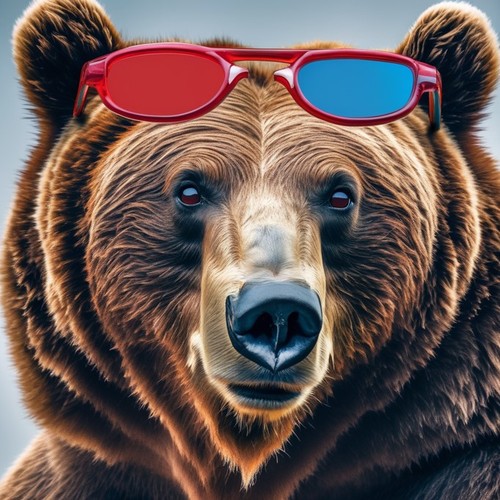} &
        \includegraphics[width=0.44\columnwidth]{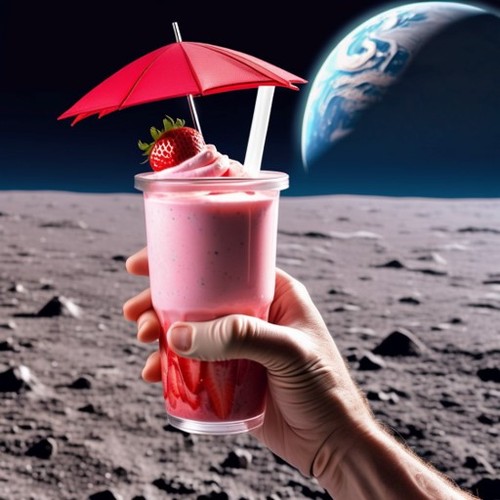} &
        \includegraphics[width=0.44\columnwidth]{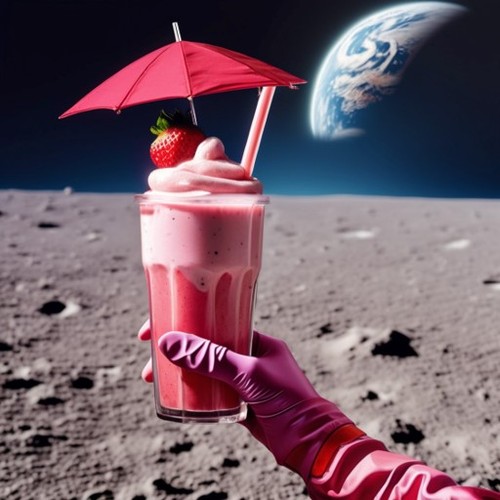} \\
        \multicolumn{2}{c}{\small \shortstack{
            \textit{``Photo of a bear wearing colorful glasses:} \\
            \textit{``\textcolor{red}{\textbf{left glass is red, right is blue}}.`` }
        }} &
        \multicolumn{2}{c}{\small \shortstack{
            \textit{``A \textcolor{red}{\textbf{gloved hand}} holding a strawberry milkshake... } \\
            \textit{Earth visible in the distance on the moon horizon.''}
        }} \\

        \includegraphics[width=0.44\columnwidth]{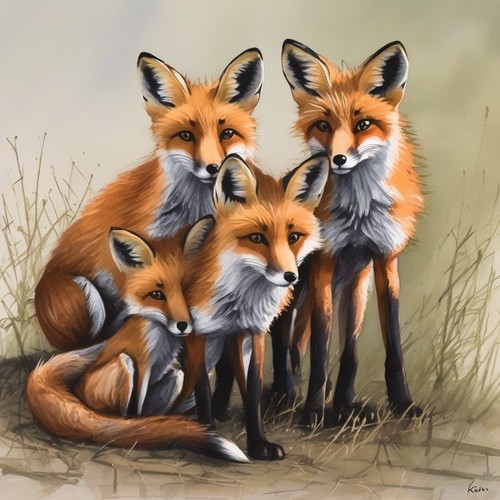} &
        \includegraphics[width=0.44\columnwidth]{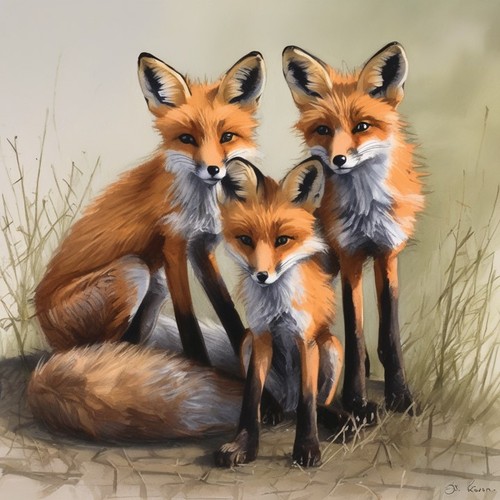} &
        \includegraphics[width=0.44\columnwidth]{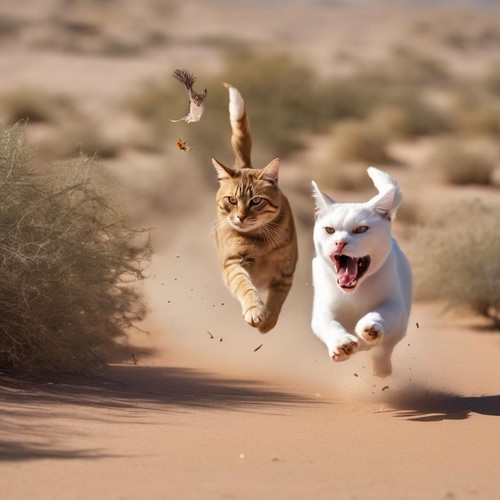} &
        \includegraphics[width=0.44\columnwidth]{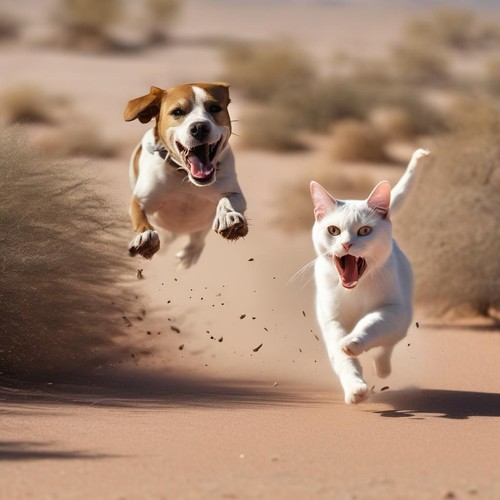} \\[-3pt]
        
        \multicolumn{2}{c}{\small \shortstack{
            \textit{``\textcolor{red}{\textbf{Three}} Young Foxes by Kain Shannon.``} 
        }} &
        \multicolumn{2}{c}{\small \shortstack{
            \textit{``A \textcolor{red}{\textbf{dog}} chasing a cat in a desert, at high speed.''}
        }} \\

        \includegraphics[width=0.44\columnwidth]{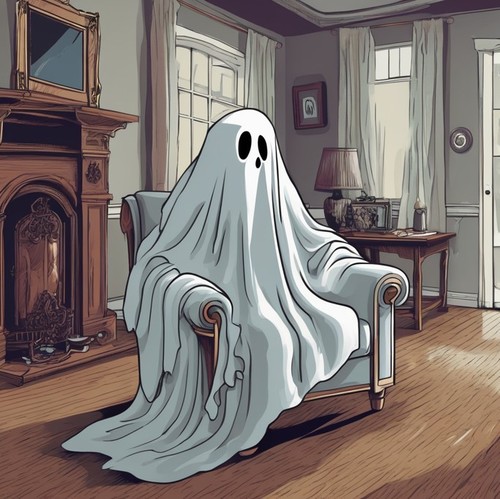} &
        \includegraphics[width=0.44\columnwidth]{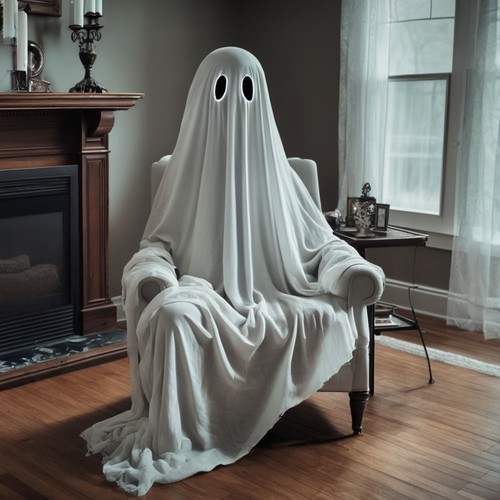} &
        \includegraphics[width=0.44\columnwidth]{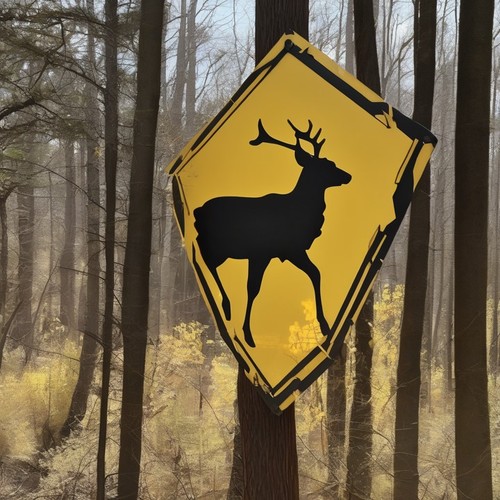} &
        \includegraphics[width=0.44\columnwidth]{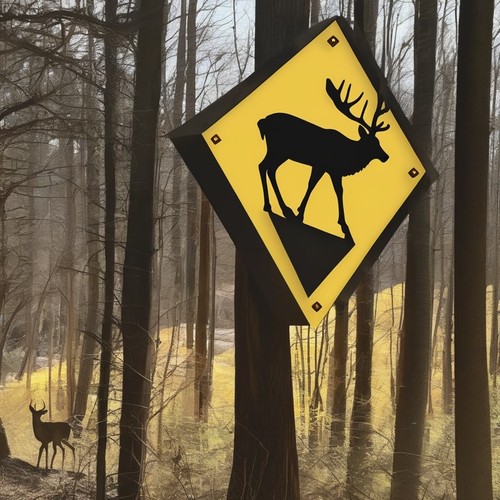} \\[-3pt]
        
        \multicolumn{2}{c}{\small \shortstack{
            \textit{``A ghost sitting on a living room chair.''}
        }} &
        \multicolumn{2}{c}{\small \shortstack{
            \textit{``A yellow \textcolor{red}{\textbf{diamond-shaped}} sign with a deer silhouette.''}
        }}

        \ifthenelse{\boolean{submission}}
        {

        }
        {
        \includegraphics[width=0.44\columnwidth]{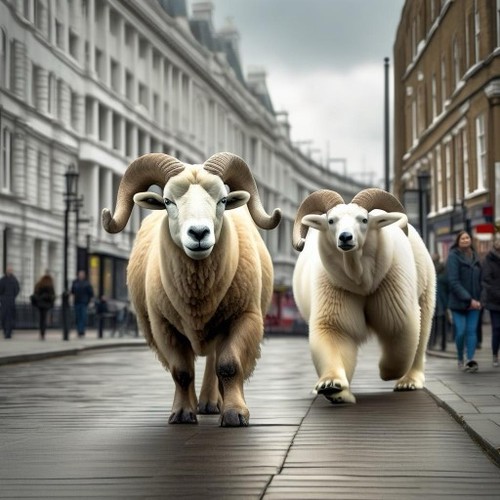} &
        \includegraphics[width=0.44\columnwidth]{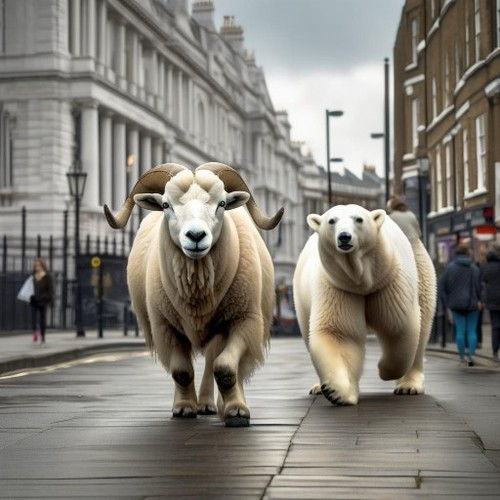} &
        \includegraphics[width=0.44\columnwidth]{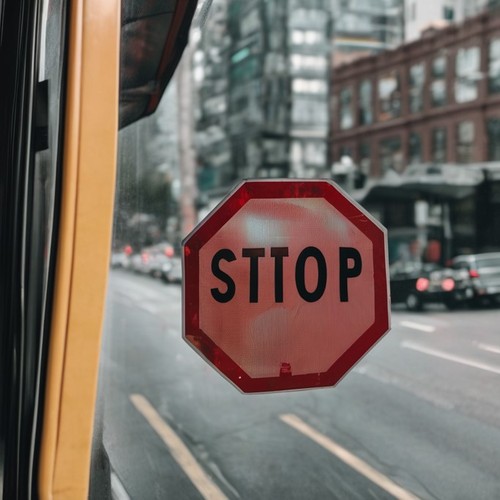} &
        \includegraphics[width=0.44\columnwidth]{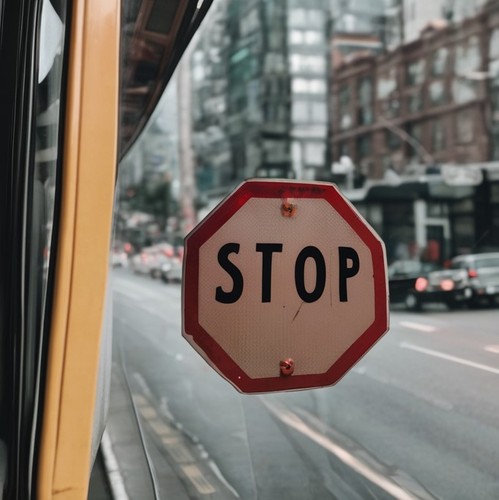} \\[-3pt]
        
        \multicolumn{2}{c}{\small \shortstack{
            \textit{``A photo of a ram and a \textcolor{red}{\textbf{polar bear}} walking in London.''}
        }} &
        \multicolumn{2}{c}{\small \shortstack{
            \textit{``There is a \textcolor{red}{\textbf{stop}} sign outside of a window.''}
        }} \\

        \includegraphics[width=0.44\columnwidth]{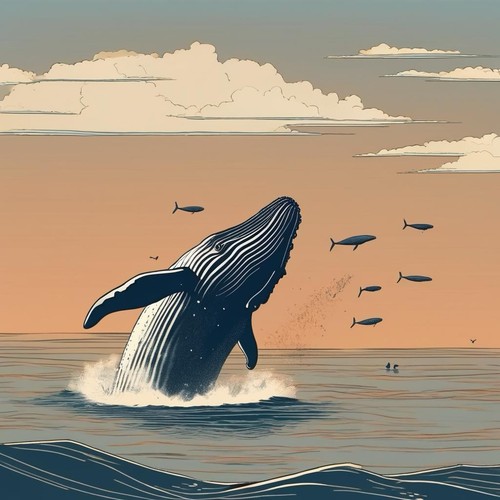} &
        \includegraphics[width=0.44\columnwidth]{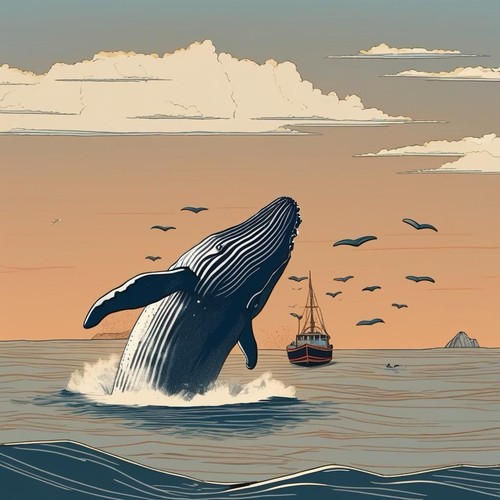} &
        \includegraphics[width=0.44\columnwidth]{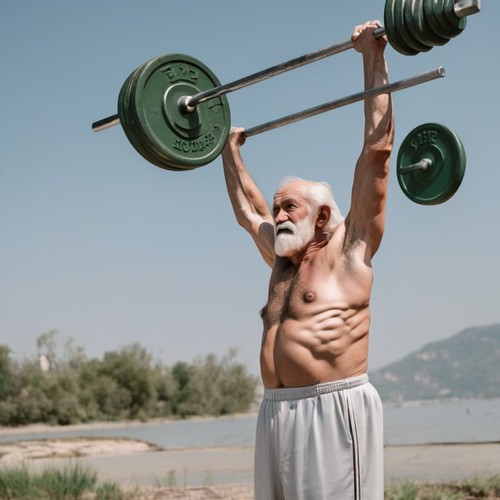} &
        \includegraphics[width=0.44\columnwidth]{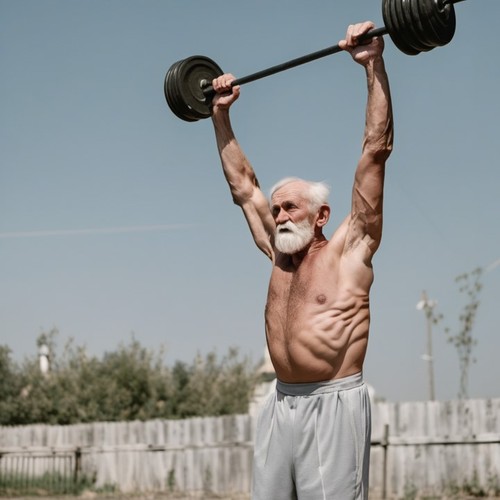} \\[-3pt]
        
        \multicolumn{2}{c}{\small \shortstack{
            \textit{``A whale in an open ocean jumps over a \textcolor{red}{\textbf{small boat}}.''}
        }} &
        \multicolumn{2}{c}{\small \shortstack{
            \textit{``An old man lifts a barbell above his head.''}
        }}
        \\
        }
    \end{tabular}
    \vspace{-10pt}
        \caption{Qualitative comparison of our Annealing method $\lambda=0.4$ (right) vs.\ CFG++ $w=0.8$ (left).}
    \label{fig:extra_cfgpp_1}
\end{figure*}

\ifthenelse{\boolean{submission}}
{
\begin{figure*}
    \centering
    \setlength{\belowcaptionskip}{-10pt}

    \begin{tabular}{@{}c@{}c@{}c@{}c@{\hspace{1em}}c@{}c@{}c@{}c@{}}
        \makecell{CFG} & \makecell{APG} & \makecell{CFG++} & \makecell{\textbf{Annealing}} &
        \makecell{CFG} & \makecell{APG} & \makecell{CFG++} & \makecell{\textbf{Annealing}} \\[0pt]

        \includegraphics[width=0.1175\textwidth]{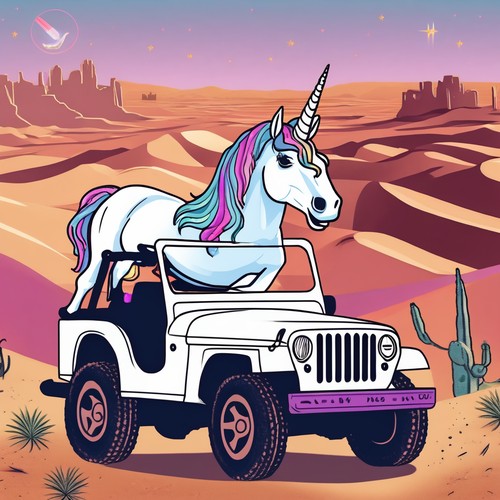} &
        \includegraphics[width=0.1175\textwidth]{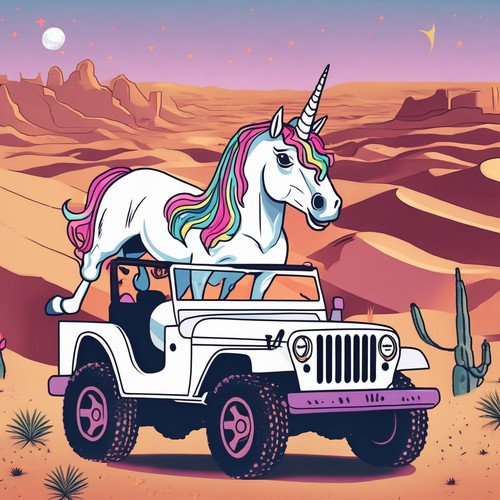} &
        \includegraphics[width=0.1175\textwidth]{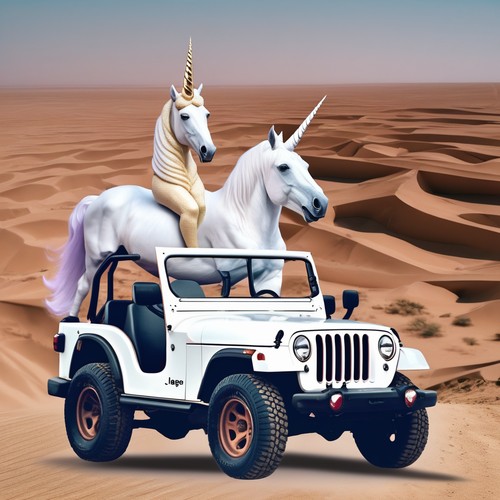} &
        \includegraphics[width=0.1175\textwidth]{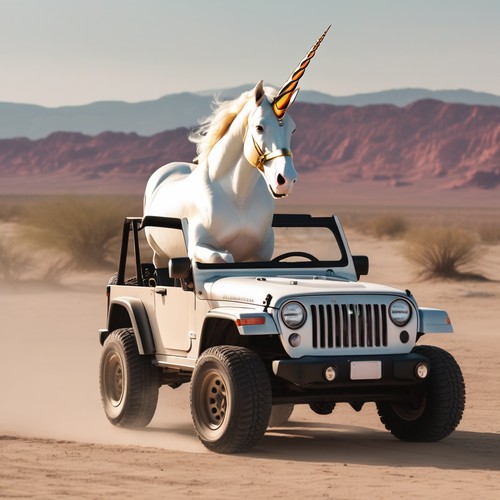} &
        \includegraphics[width=0.1175\textwidth]{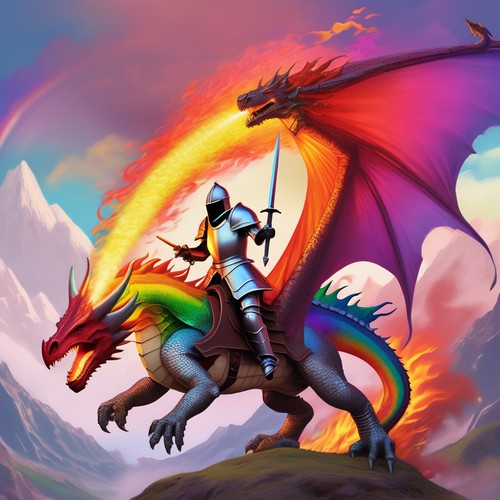} &
        \includegraphics[width=0.1175\textwidth]{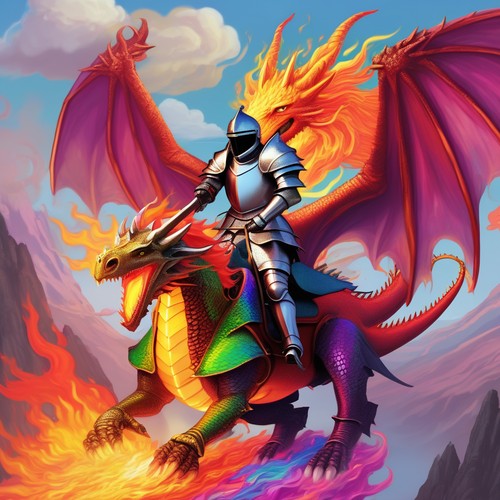} &
        \includegraphics[width=0.1175\textwidth]{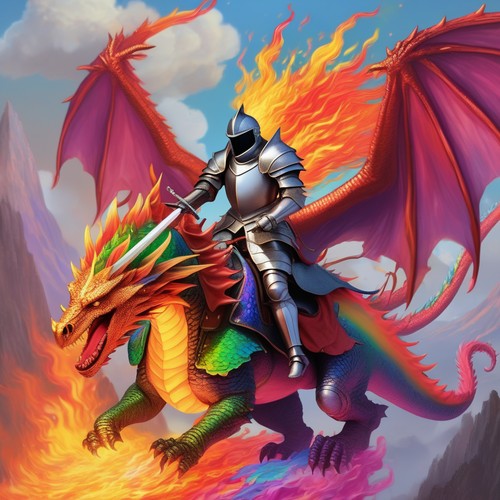} &
        \includegraphics[width=0.1175\textwidth]{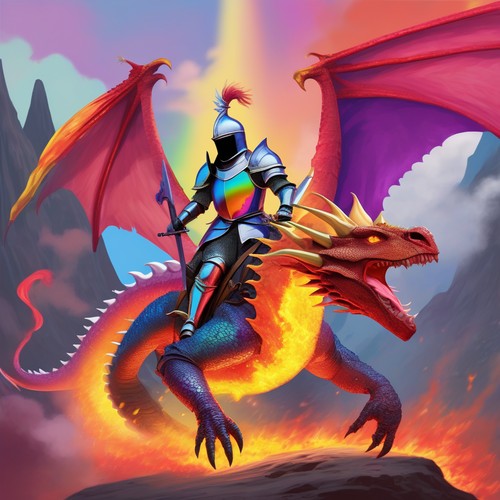} \\[-5pt]
        \multicolumn{4}{c}{\scriptsize "A \textcolor{red}{\textbf{photo}} of unicorn driving a jeep in the desert"} &
        \multicolumn{4}{c}{\scriptsize "A \textcolor{red}{\textbf{knight in rainbow armor}} riding a \textcolor{red}{\textbf{dragon made of fire}}"} \\[-1pt]

        \includegraphics[width=0.1175\textwidth]{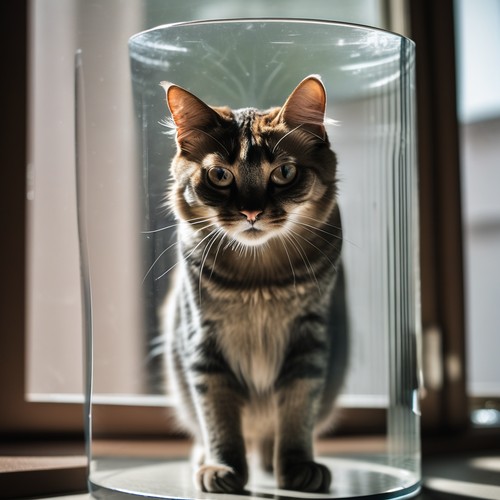} &
        \includegraphics[width=0.1175\textwidth]{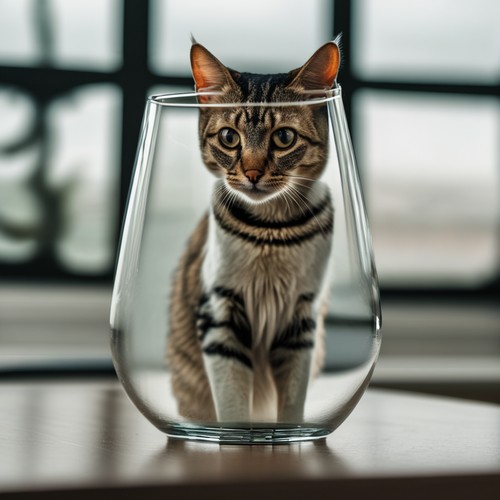} &
        \includegraphics[width=0.1175\textwidth]{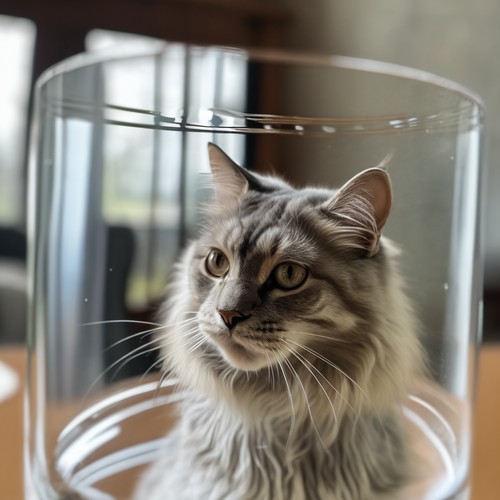} &
        \includegraphics[width=0.1175\textwidth]{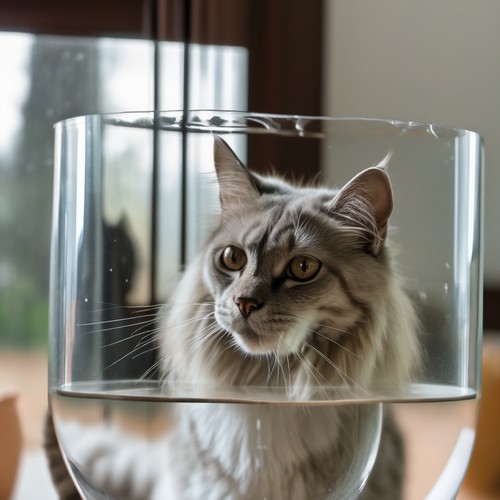} &
        \includegraphics[width=0.1175\textwidth]{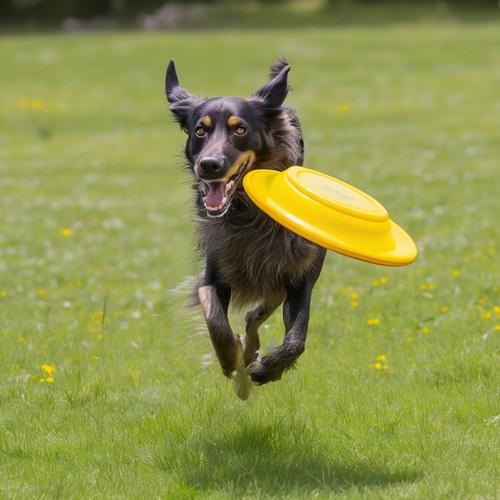} &
        \includegraphics[width=0.1175\textwidth]{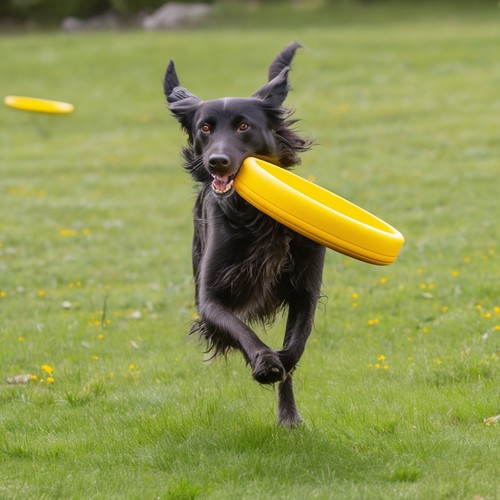} &
        \includegraphics[width=0.1175\textwidth]{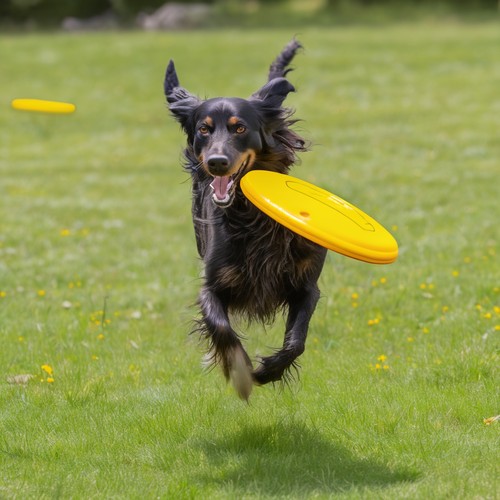} &
        \includegraphics[width=0.1175\textwidth]{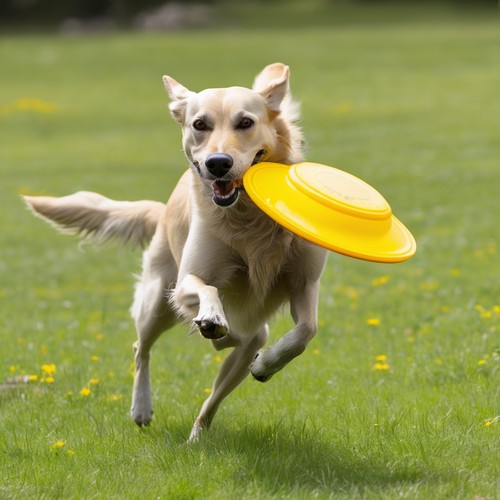} \\[-5pt]
        \multicolumn{4}{c}{\scriptsize "A cat looking through a \textcolor{red}{\textbf{glass of water}}"} &
        \multicolumn{4}{c}{\scriptsize "A \textcolor{red}{\textbf{yellow dog}} runs to grab a yellow frisbee in the grass."} \\[-1pt]

        \includegraphics[width=0.1175\textwidth]{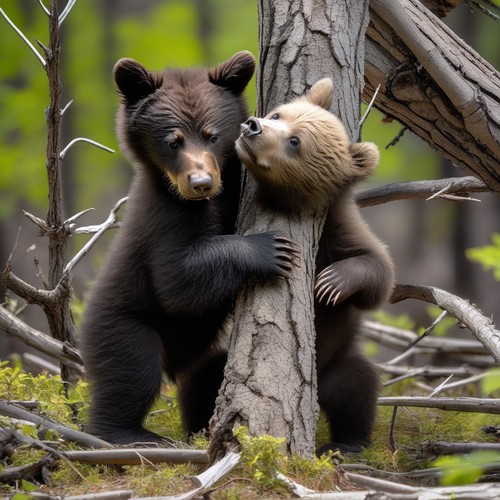} &
        \includegraphics[width=0.1175\textwidth]{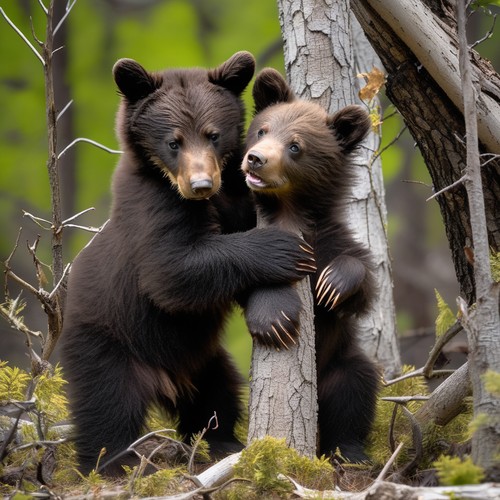} &
        \includegraphics[width=0.1175\textwidth]{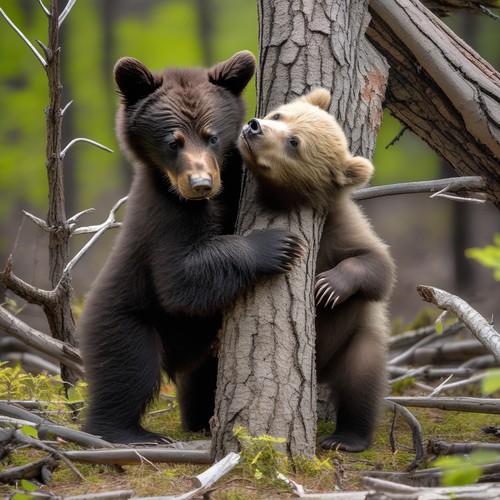} &
        \includegraphics[width=0.1175\textwidth]{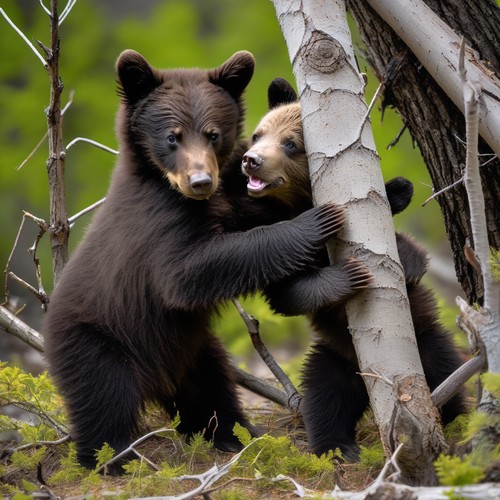} &
        \includegraphics[width=0.1175\textwidth]{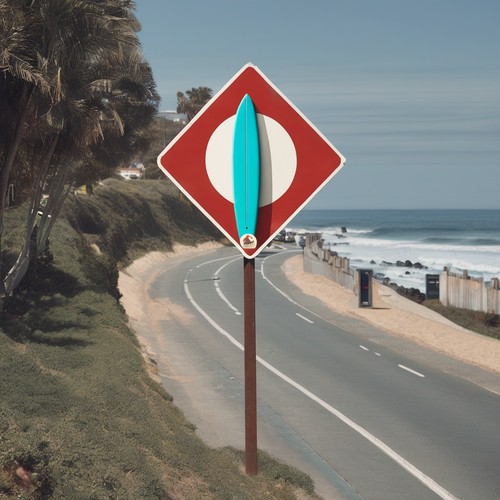} &
        \includegraphics[width=0.1175\textwidth]{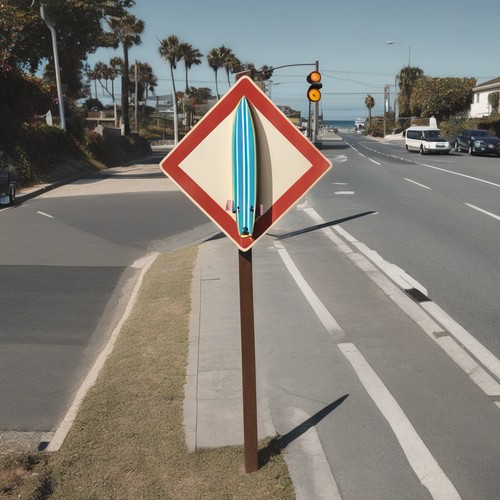} &
        \includegraphics[width=0.1175\textwidth]{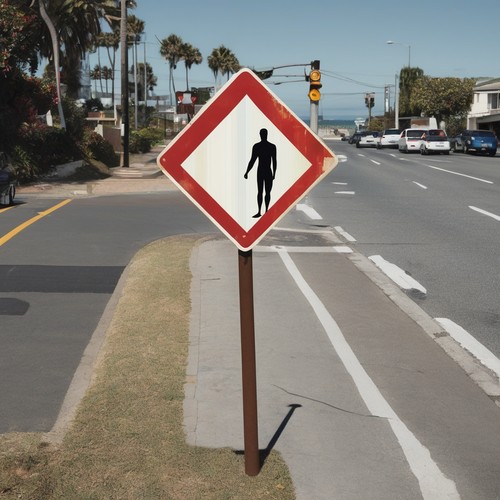} &
        \includegraphics[width=0.1175\textwidth]{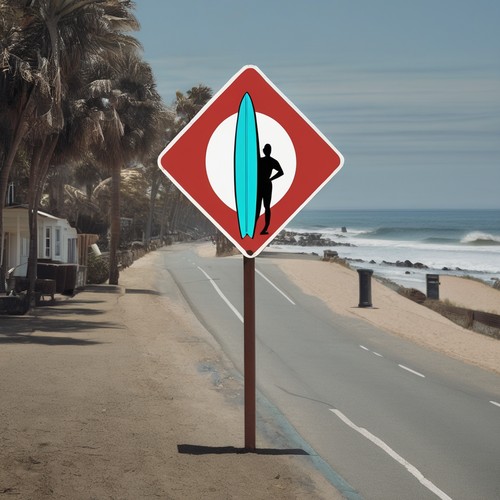} \\[-5pt]
        \multicolumn{4}{c}{\scriptsize "Bear cubs play among the fallen tree limbs."} &
        \multicolumn{4}{c}{\scriptsize "A traffic sign that has a picture of a \textcolor{red}{\textbf{man holding a surfboard}} on it."} \\[-1pt]

    \end{tabular}

    \vspace{-10pt}
    \caption{Qualitative comparison of our Annealing method $\lambda=0.8$ vs. three guidance methods: CFG ($w=15$), APG ($w=20$) and CFG++ ($w=1.2$).}
    \vspace{-6pt}
    \label{fig:method_comparison}
\end{figure*}

}
{
}

\DisableACMJournalFooterInAppendix
\pagestyle{appendixpagestyle}
\thispagestyle{appendixpagestyle}
\begin{figure*}[t]
    \centering
    \begin{tabular}{@{}c@{\hspace{0.005\textwidth}}c@{\hspace{0.03\textwidth}}c@{\hspace{0.005\textwidth}}c@{}}

        CFG ($w=10$) & \textbf{Annealing ($\lambda=0.4$)} & CFG ($w=10$) & \textbf{Annealing ($\lambda=0.4$)} \\
        \includegraphics[width=0.49\columnwidth]{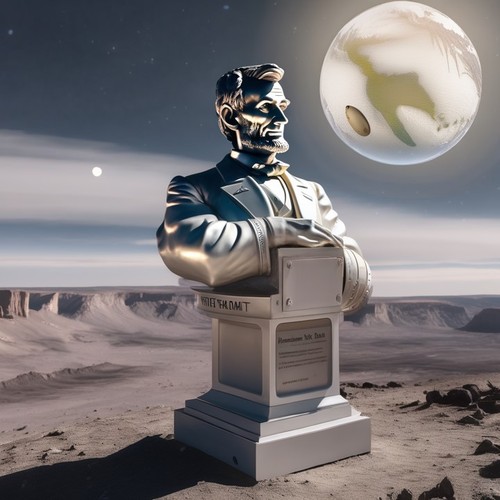} &
        \includegraphics[width=0.49\columnwidth]{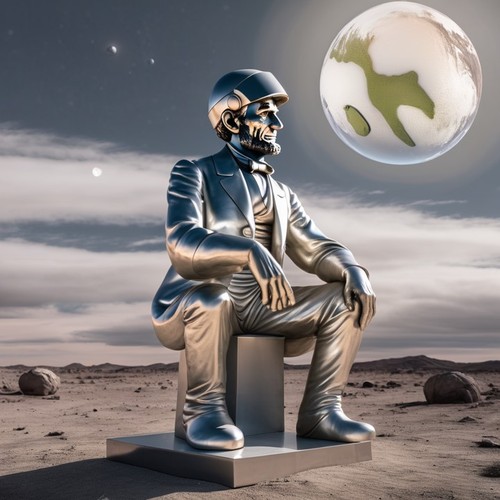} &
        \includegraphics[width=0.49\columnwidth]{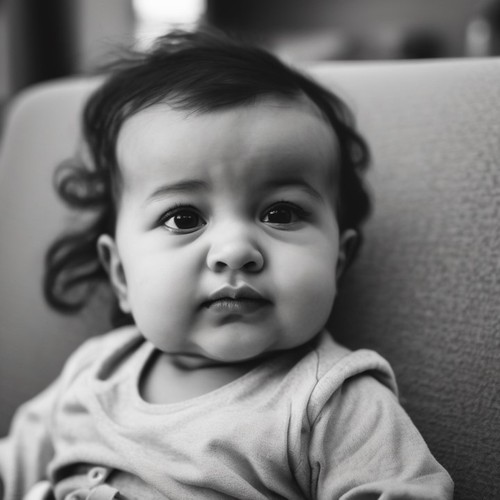} &
        \includegraphics[width=0.49\columnwidth]{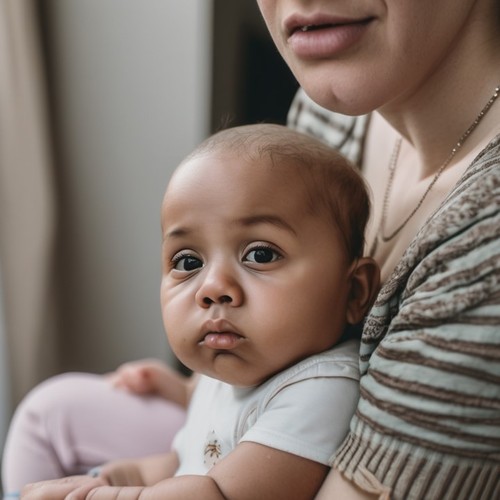} \\

        \multicolumn{2}{c}{\small \shortstack{
            \textit{``A statue of Abraham Lincoln wearing an opaque and } \\
            \textit{\textcolor{red}{\textbf{shiny astronaut's helmet}}. The statue sits on the moon.''}
        }} &
        \multicolumn{2}{c}{\small \shortstack{
            \textit{``A baby \textcolor{red}{\textbf{sitting on a female}}'s lap } \\
            \textit{``staring into the camera.''}
        }} \\

        \includegraphics[width=0.49\columnwidth]{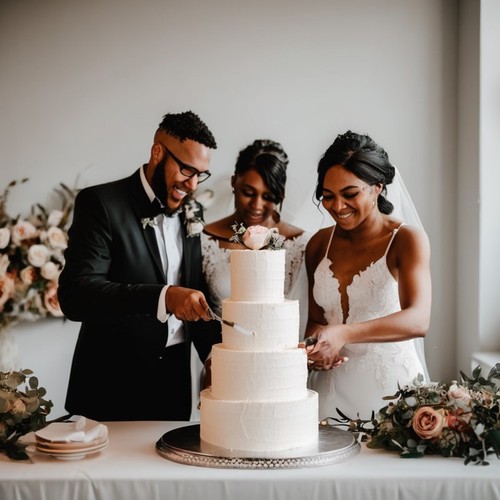} &
        \includegraphics[width=0.49\columnwidth]{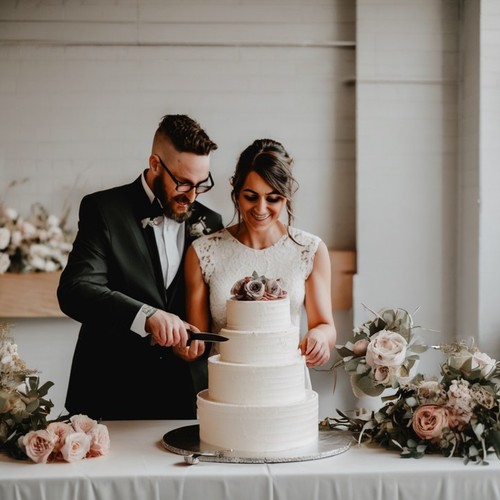} &
        \includegraphics[width=0.49\columnwidth]{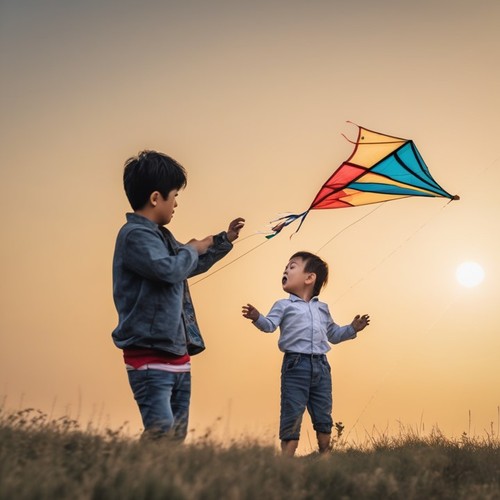} &
        \includegraphics[width=0.49\columnwidth]{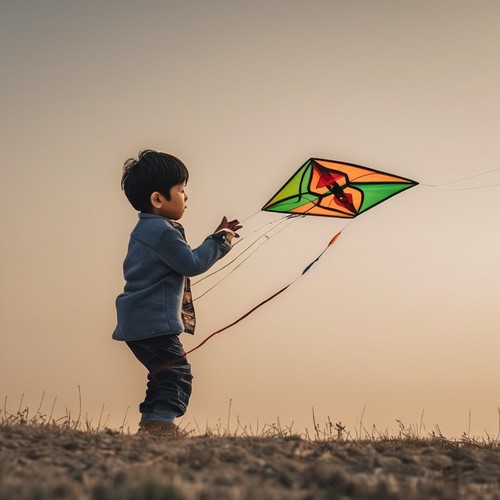} \\[-3pt]
        
        \multicolumn{2}{c}{\small \shortstack{
            \textit{``\textcolor{red}{\textbf{A bride}} and groom cutting their wedding cake.''}
        }} &
         \multicolumn{2}{c}{\small \shortstack{
            \textit{``\textcolor{red}{\textbf{A small boy}} trying to fly a small kite.''}
        }}\\
        
        \includegraphics[width=0.49\columnwidth]{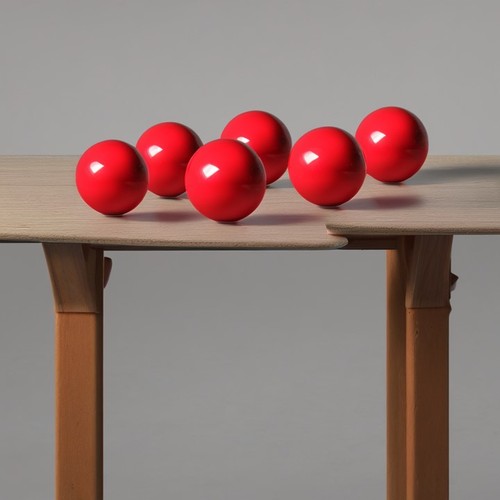} &
        \includegraphics[width=0.49\columnwidth]{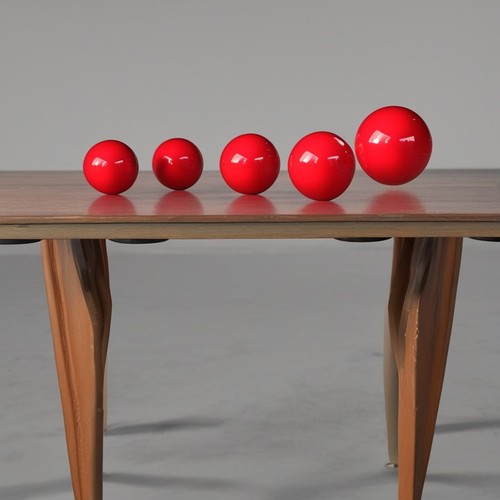} &
        \includegraphics[width=0.49\columnwidth]{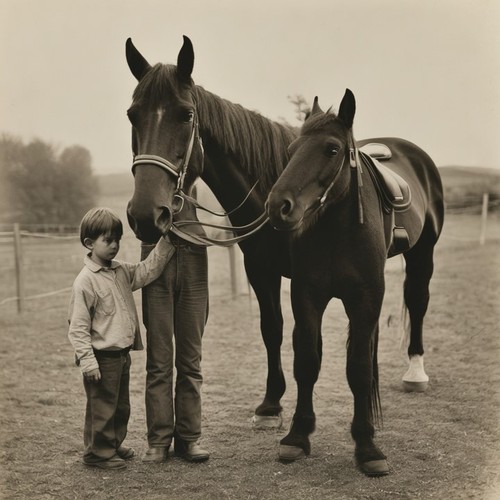} &
        \includegraphics[width=0.49\columnwidth]{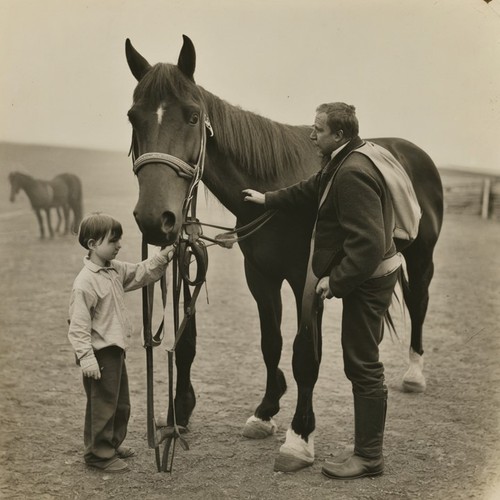} \\[-3pt]

        \multicolumn{2}{c}{\small \shortstack{
            \textit{``\textcolor{red}{\textbf{five}} red balls on a table.''}
        }} &
        \multicolumn{2}{c}{\small \shortstack{
            \textit{``\textcolor{red}{\textbf{A man}} and child next to a horse.''}
        }} \\

        \includegraphics[width=0.49\columnwidth]{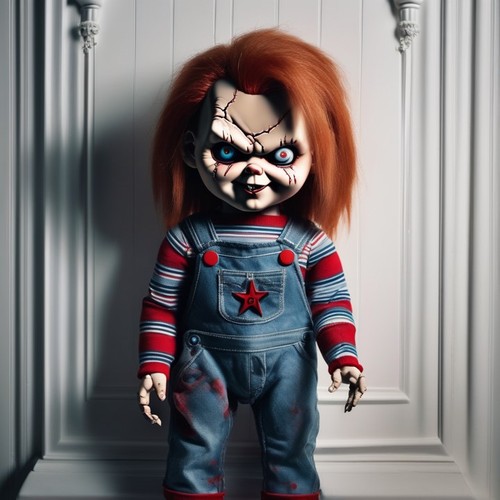} &
        \includegraphics[width=0.49\columnwidth]{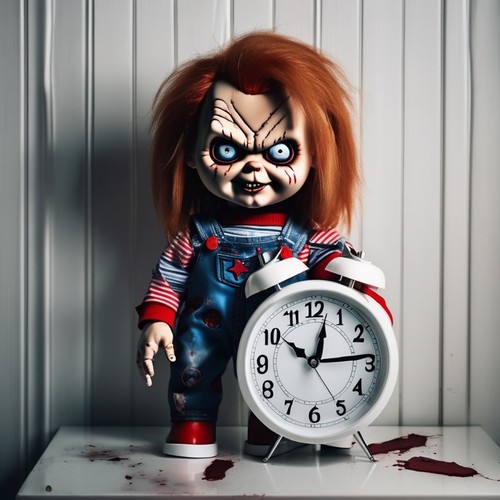} &
        \includegraphics[width=0.49\columnwidth]{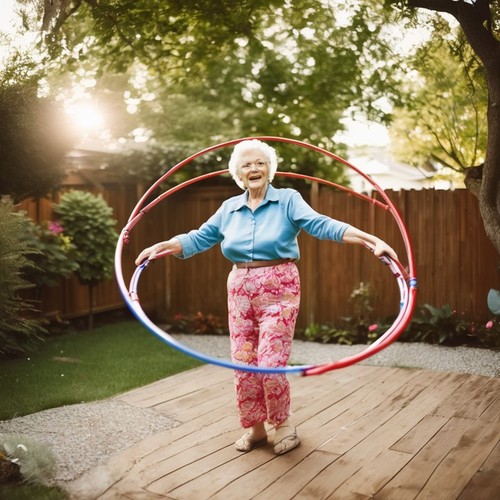} &
        \includegraphics[width=0.49\columnwidth]{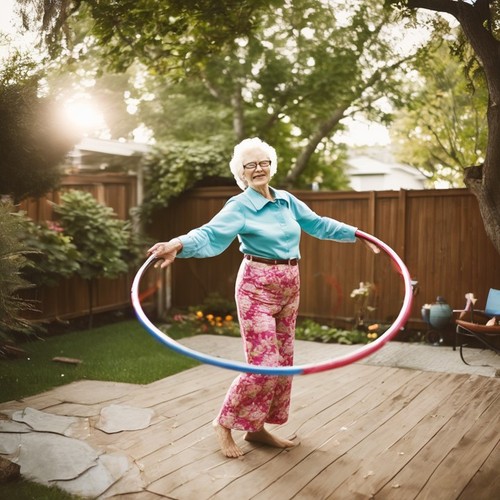} \\[-3pt]
        
        \multicolumn{2}{c}{\small \shortstack{
            \textit{``A demonic looking chucky like doll standing next to a \textcolor{red}{\textbf{white clock}}.''}
        }} &
        \multicolumn{2}{c}{\small \shortstack{
            \textit{``Older woman hula hooping in backyard.''}
        }} \\

        \includegraphics[width=0.49\columnwidth]{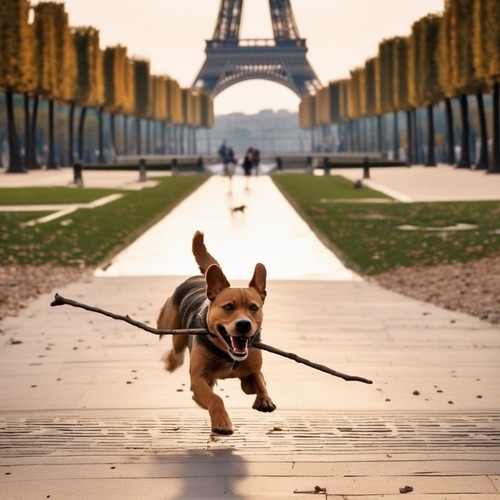} &
        \includegraphics[width=0.49\columnwidth]{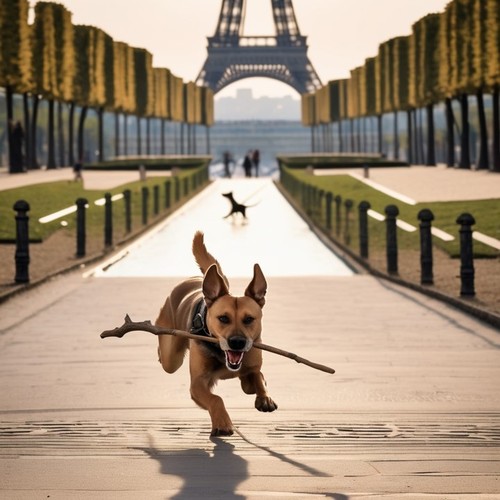} &
        \includegraphics[width=0.49\columnwidth]{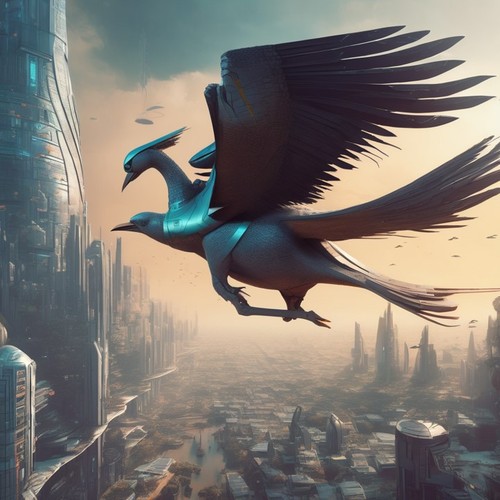} &
        \includegraphics[width=0.49\columnwidth]{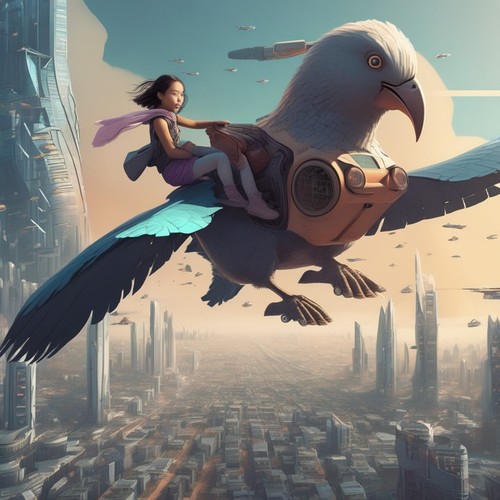} \\[-3pt]
        
        \multicolumn{2}{c}{\small \shortstack{
            \textit{``A dog running with a stick \textcolor{red}{\textbf{in its mouth}}, Eiffel tower in the background.``}}} &
        \multicolumn{2}{c}{\small \shortstack{
            \textit{``A \textcolor{red}{\textbf{girl}} riding a giant bird over a futuristic city.''}
        }} 
    \end{tabular}

        \caption{Qualitative comparison of our Annealing method $\lambda=0.4$ (right) vs.\ CFG $w=10$ (left).}
    \label{fig:extra_1}
\end{figure*}

\clearpage

\clearpage
\begin{appendices}

\section*{\LARGE Appendix}
\global\csname @topnum\endcsname 0   %

\section{Implementation Details}

\subsection{Architecture and Training}
\label{sec:imp_details}

We provide our annealing scheduler training algorithm in Alg.\ref{alg:annealing_scheduler}, and its inference algorithm in Alg.\ref{alg:annealing_inference}.
Our annealing scheduler $w_\theta(t,z_t,\lambda)$ is implemented as a lightweight MLP with three hidden layers of dimension 128, resulting in a total of 52K trainable parameters. The model takes as input sinusoidal embeddings of three features: (1) the normalized timestep \( t/T \), (2) the normalized guidance magnitude \( \|\delta_t\| / \|\delta\|_{\text{max}} \), where \( \|\delta\|_{\text{max}} \) corresponds to the typical maximum norm of \( \delta_t \) observed empirically across the training set and set to 5.0 in SDXL, and (3) the prompt-alignment parameter $\lambda$. Each embedding is 4-dimensional, and the three embeddings are concatenated before being passed through the first layer. ReLU activations are applied after each layer. The network outputs a single scalar corresponding to the predicted guidance scale. When we constrain the guidance scale $w$ in the ablation to $[0,1]$ we add a sigmoid layer at the output.  Training is performed for a maximum of 20,000 steps using the AdamW optimizer with a learning rate of \(1\text{e}{-3}\) and weight decay of 0.01. We train with a per-device batch size of 2 and accumulate gradients for 8 steps before performing an optimizer update. We use the default Kaiming-uniform initialization. All training runs complete within approximately 4.5 hours on a single NVIDIA A6000 GPU (48GB).

\begin{algorithm}[t]
\caption{CFG - Inference (DDIM)}
\label{alg:cfg_inference}
\SetKwInOut{Input}{Require}
\Input{ \BlankLine
  $T$: total number of denoising steps\;
  $w$: guidance scale\;
  $\epsilon^{(\cdot)}_t$: \textbf{frozen} noise predictor, accepts $\varnothing$ or a condition, at timestep $t$\;
  $\mathbf{c}$: condition\;
}
\BlankLine
$\mathbf{z}_T \sim \mathcal{N}(0, \mathbf{I})$\;

\For{$t = T$ \KwTo $1$}{
    $\hat{\epsilon}_t \gets \epsilon_t^{\varnothing}(\mathbf{z}_t) + w\cdot (\epsilon_t^{\mathbf{c}}(\mathbf{z}_t) - \epsilon_t^{\varnothing}(\mathbf{z}_t))$\tcp*{CFG}
    
    $z_{0|t} \gets (\mathbf{z}_t - \sqrt{1 - \bar{\alpha}_t} \cdot \hat{\epsilon}_t) / \sqrt{\bar{\alpha}_t}$ \tcp*{Denoise}

    $\mathbf{z}_{t-1} \gets \sqrt{\bar{\alpha}_{t-1}} \cdot z_{0|t} + \sqrt{1 - \bar{\alpha}_{t-1}} \cdot \hat{\epsilon}_t$ \tcp*{Renoise}
}
\Return{$\mathbf{z}_0$}
\end{algorithm}

\begin{algorithm}[t]
\caption{CFG++ - Inference (DDIM)}
\label{alg:cfgpp_inference}
\SetKwInOut{Input}{Require}
\Input{ \BlankLine
  $T$: total number of denoising steps\;
  $w\in[0,1]$: guidance scale\;
  $\epsilon^{(\cdot)}_t$: \textbf{frozen} noise predictor, accepts $\varnothing$ or a condition, at timestep $t$\;
  $\mathbf{c}$: condition\;
}
\BlankLine
$\mathbf{z}_T \sim \mathcal{N}(0, \mathbf{I})$\;

\For{$t = T$ \KwTo $1$}{
    $\hat{\epsilon}_t \gets \epsilon_t^{\varnothing}(\mathbf{z}_t) + w\cdot (\epsilon_t^{\mathbf{c}}(\mathbf{z}_t) - \epsilon_t^{\varnothing}(\mathbf{z}_t))$\tcp*{CFG}
    
    $z_{0|t} \gets (\mathbf{z}_t - \sqrt{1 - \bar{\alpha}_t} \cdot \hat{\epsilon}_t) / \sqrt{\bar{\alpha}_t}$ \tcp*{Denoise}

    $\mathbf{z}_{t-1} \gets \sqrt{\bar{\alpha}_{t-1}} \cdot z_{0|t} + \sqrt{1 - \bar{\alpha}_{t-1}} \cdot \boldsymbol{\epsilon_t^{\varnothing}(\mathbf{z}_t)}$ \tcp*{Renoise}
}

\Return{$\mathbf{z}_0$}
\end{algorithm}

\begin{algorithm}[t]
\caption{Annealing Scheduler - Inference (DDIM)}
\label{alg:annealing_inference}
\SetKwInOut{Input}{Require}
\Input{ \BlankLine
  $\lambda\in[0,1]$: prompt alignment weighting parameter\;
  $T$: total number of denoising steps\;
  $w_\theta$: \textbf{trained} guidance scale model\;
  $\epsilon^{(\cdot)}_t$: \textbf{frozen} noise predictor, accepts $\varnothing$ or a condition, at timestep $t$\;
  $\mathbf{c}$: condition\;
}
\BlankLine
$\mathbf{z}_T \sim \mathcal{N}(0, \mathbf{I})$\;

\For{$t = T$ \KwTo $1$}{
  $\delta_{t} \gets \epsilon_t^{\tilde{c}}(\mathbf{z}_{t}) - \epsilon_t^{\varnothing}(\mathbf{z}_{t})$\;
    $\hat{\epsilon}_t \gets \epsilon_t^{\varnothing}(\mathbf{z}_t) + \boldsymbol{w_\theta(t,\|\delta_{t}\|,\lambda)}\cdot (\epsilon_t^{\mathbf{c}}(\mathbf{z}_t) - \epsilon_t^{\varnothing}(\mathbf{z}_t))$\tcp*{CFG}

    $z_{0|t} \gets (\mathbf{z}_t - \sqrt{1 - \bar{\alpha}_t} \cdot \hat{\epsilon}_t) / \sqrt{\bar{\alpha}_t}$ \tcp*{Denoise}

    $\mathbf{z}_{t-1} \gets \sqrt{\bar{\alpha}_{t-1}} \cdot z_{0|t} + \sqrt{1 - \bar{\alpha}_{t-1}} \cdot \boldsymbol{\epsilon_t^{\varnothing}(\mathbf{z}_t)}$ \tcp*{Renoise}
}

\Return{$\mathbf{z}_0$}
\end{algorithm}

\subsection{Training Data}
\label{sec:training_data}

We use the LAION-POP subset of LAION-5B dataset~\cite{schuhmann2022laion} with high-resolution images with detailed descriptions, and selected 20,000 images based on the highest similarity scores.    

\subsection{Memory and Time Consumption}
\label{sec:time_con}

We evaluated the inference time of our lightweight model, which has a footprint of only 700KB. Running the model 10{,}000 times on a NVIDIA RTX A5000 yielded a mean inference time of 0.001434 seconds with a standard deviation of 0.000123 seconds. Given that the model is activated for 50 timesteps during a typical diffusion process, this results in an additional computational cost of approximately \( 0.0717 \) seconds per sample.

\begin{table*}[t]
    \centering
    \caption{Comparison of CFG, APG, CFG++, and our method across FID, FD-DINOv2, CLIP score, Image Reward (IR), Precision, and Recall, with the corresponding guidance scale ($w$ or $\lambda$) explicitly shown in a dedicated \textbf{Scale} column. Arrows indicate whether higher (↑) or lower (↓) values are better.}
    \begin{tabular}{l l ccccccc}
        \toprule
        \textbf{Method} & \textbf{Scale} & \textbf{FID ↓} & \makecell{\textbf{FD-DINOv2 ↓}} & \textbf{CLIP ↑} & \textbf{Image Reward ↑} & \textbf{Precision ↑} & \textbf{Recall ↑} \\
        \midrule
        CFG & $w=7.5$ & 25.13 & 269.44 & 32.12 & 0.817 & \textbf{0.863} & 0.630 \\
        APG & $w=10$ & 25.25 & 268.00 & 32.08 & \textbf{0.818} & 0.862 & \textbf{0.631} \\
        CFG++ & $w=0.6$ & 24.97 & 267.91 & 32.12 & 0.808 & 0.859 & 0.629 \\
        Ours & $\lambda=0.05$ & \textbf{24.76} & \textbf{267.17} & \textbf{32.16} & 0.809 & 0.860 & 0.620 \\
        \midrule
        CFG & $w=10$ & 26.06 & 281.04 & 32.22 & 0.859 & 0.859 & 0.594 \\
        APG & $w=15$ & 26.60 & 282.09 & 32.19 & \textbf{0.865} & \textbf{0.864} & 0.592 \\
        CFG++ & $w=0.8$ & 25.61 & 279.69 & 32.20 & 0.857 & 0.855 & 0.601 \\
        Ours & $\lambda=0.4$ & \textbf{25.35} & \textbf{279.30} & \textbf{32.25} & \textbf{0.865} & 0.859 & \textbf{0.606} \\
        \midrule
        CFG & $w=12.5$ & 26.61 & 288.13 & 32.25 & 0.881 & 0.850 & 0.570 \\
        APG & $w=17.5$ & 26.58 & 286.67 & 32.21 & \textbf{0.887} & \textbf{0.861} & 0.586 \\
        CFG++ & $w=1$ & 26.33 & 288.55 & \textbf{32.26} & 0.882 & 0.848 & 0.570 \\
        Ours & $\lambda=0.7$ & \textbf{25.95} & \textbf{285.52} & \textbf{32.26} & 0.884 & 0.852 & \textbf{0.594} \\
        \midrule
        CFG & $w=15$ & 27.15 & 293.93 & 32.27 & 0.883 & 0.844 & 0.570 \\
        APG & $w=20$ & 26.85 & 290.93 & 32.23 & 0.893 & \textbf{0.855} & 0.577 \\
        CFG++ & $w=1.2$ & 26.84 & 294.22 & 32.28 & 0.894 & 0.847 & 0.551 \\
        Ours & $\lambda=0.8$ & \textbf{26.40} & \textbf{290.33} & \textbf{32.29} & \textbf{0.898} & 0.846 & \textbf{0.586} \\
        \bottomrule
    \end{tabular}
    \label{tab:full_metric_comparison}
\end{table*}

\begin{figure}[t]
    \centering
    \setlength{\tabcolsep}{0.0em} %
    \begin{tabular}{ccc}

        \includegraphics[width=0.5\textwidth]{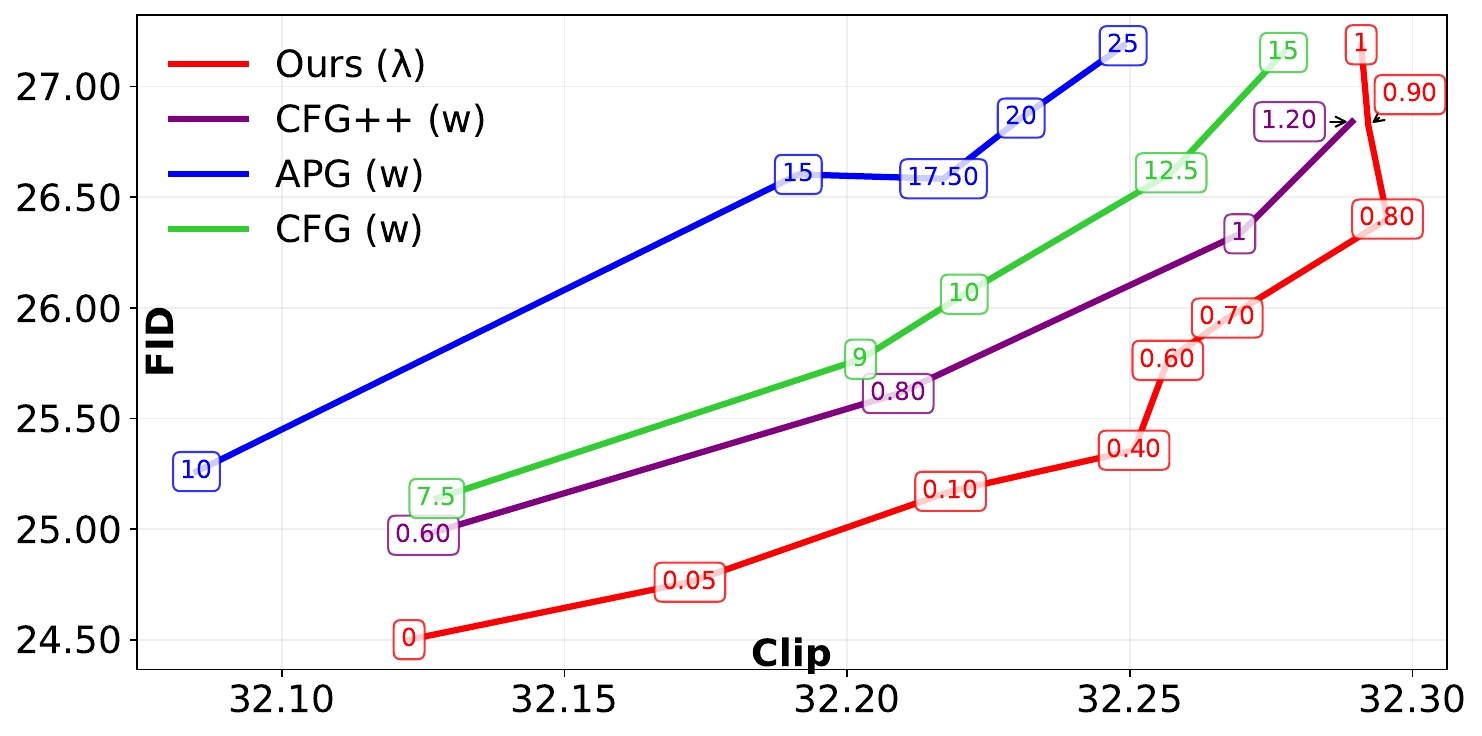}&

    \end{tabular}
    \caption{\textbf{Quantitative Metrics.} FD-Dinov2 versus CLIP.}
    \vspace{-6pt}
    \label{fig:fd_dino_curves}
\end{figure}

\subsection{Metrics Calculation}
\label{sec:metrics_calc}

For assessment of fidelity and diversity, we report Precision and Recall ~\cite{kynkaanniemi2019improved} in the DINOv2~\cite{oquab2024dinov2learningrobustvisual} feature space.

Precision measures the fraction of generated samples that lie within the support of the real image distribution. This is estimated by checking whether each generated sample has a real image among its \(k\) nearest neighbors in feature space. Conversely, recall quantifies the fraction of real images that lie within the support of the generated distribution, also based on their nearest neighbors among generated samples. Higher precision indicates better sample fidelity, while higher recall reflects greater diversity in generation. 
We used $k=5$ in our reports.
Additionally, we report FD-DINOv2, a feature distance metric computed in the same feature space for image quality assessment.

To construct the generated dataset, we use the same captions as the COCO 2017 validation set. Each image in this set has five human-provided annotations; we consistently use the first caption per image for generation. We use a unique random seed for each image, setting it to the corresponding \texttt{image\_id} from the COCO validation set.

\subsection{Implementation Details for Other Methods}
\label{sec:other_methods_imp}

For \textbf{CFG++}, we followed the official implementation\footnote{\url{https://github.com/CFGpp-diffusion/CFGpp}} and evaluated the method using $\lambda$ values ranging from 0.4 to 1.2.

For \textbf{APG}, we adopted the settings provided in the original paper~\cite{sadat2024eliminating}. Specifically, we used the recommended hyperparameters for SDXL: $\eta = 0$, $r = 15$, and $\beta = -0.5$, and varied guidance scales $w$.

\begin{table}[t]
\centering
\caption{Ablation over noise scaling parameter $s$ in the mode augmentation scheme.}
\begin{tabular}{c|c|c|c}
\toprule
Noise Scale $s$ & FID ↓ & CLIP ↑ & ImageReward ↑ \\
\midrule
0      & 27.01 & 32.25 & 0.884 \\
0.025  & \textbf{26.40} & \textbf{32.29} & \textbf{0.898} \\
0.1    & 27.17 & 32.27 & 0.880 \\
0.25   & 28.14 & 32.27 & 0.873 \\

\bottomrule
\end{tabular}
\label{tab:noise_ablation}
\end{table}

\subsection{Prompt Perturbation}
\label{sec:perturb}
We perturb the conditioning signal solely during training, following CADS~\cite{sadat2023cads}. In practice, we apply the noise directly to the prompt embedding \( c \), using the corruption rule:
\[
\hat{c} = \sqrt{\gamma(t)}\, c + s \sqrt{1 - \gamma(t)}\, n, \quad n \sim \mathcal{N}(0, I),
\]
where \( \gamma(t) \) is a schedule and \( s \) controls the noise level. We adopt a linear schedule with \( \tau_1 = 0 \), \( \tau_2 = T \), such that \( \gamma(t) \) decays from 1 to 0 over the course of denoising, thus inducing higher corruption in earlier timesteps. To maintain the norm of the noised embedding, we rescale the signal as proposed in CADS:
\[
\hat{c}_{\text{rescaled}} = \frac{\hat{c} - \text{mean}(\hat{c})}{\text{std}(\hat{c})} \, \text{std}(c) + \text{mean}(c), \quad
\tilde{c} = \psi \hat{c}_{\text{rescaled}} + (1 - \psi) \hat{c},
\]
and set the mixing factor to \( \psi = 1 \).

We set the noise scale $s$ to $0.025$. We ablate this scale by fixing \( \lambda = 0.8 \) and reporting the performance of the trained scheduler in terms of FID, CLIP similarity, and ImageReward on the COCO2017 Validation set in Table~\ref{tab:noise_ablation}.

\section{Additional Experiments}

\begin{table*}[t]
\centering
\setlength{\tabcolsep}{4pt}
\caption{\textbf{PartiPrompts results.} Average CLIP score across challenge categories. Our method achieves the highest overall score and outperforms both baselines in every individual challenge.}
\begin{tabular}{lccccccccc}
\toprule
Model & All & Complex & Fine-grained & Imagination & Perspective & Quantity & SimpleDetail & Style\&Format & Writing\&Symbols \\
\midrule
CFG ($w=10$) & 34.14 & 36.88 & 34.04 & 34.92 & 32.81 & 31.99 & 32.43 & 35.13 & 35.21 \\
CFG++ ($w=0.8$) & 34.15 & 36.96 & 34.00 & 34.93 & 32.87 & 31.89 & 32.37 & 35.22 & 35.33 \\
Ours ($\lambda=0.4$) & \textbf{34.23} & \textbf{37.00} & \textbf{34.06} & \textbf{35.04} & \textbf{32.90} & \textbf{32.01} & \textbf{32.46} & \textbf{35.32} & \textbf{35.44} \\
\bottomrule
\end{tabular}
\label{tab:parti_prompts}
\end{table*}

\subsection{PartiPrompts Evaluation}
\label{sec:parti_prompts}

To assess the robustness of our method to a broad range of prompt structures, 
we evaluate our scheduler on the PartiPrompts dataset~\cite{yu2022scalingautoregressivemodelscontentrich}, which categorizes prompts into different challenge aspects such as complex descriptions, fine-grained details, and imaginative scenes. 

We exclude categories with fewer than 50 prompts and single-word prompts (to remove trivial cases), 
and sample three seeds per prompt, resulting in approximately 5{,}000 generated images. 
We report the average CLIP score per challenge category, as well as the overall average across all challenges (Table~\ref{tab:parti_prompts}).

While CFG++ demonstrates meaningful improvements over CFG on average, its gains are not uniform and certain categories see limited or even negative impact. In contrast, our method achieves the highest overall score and delivers consistent improvements across all challenge categories, surpassing both CFG and CFG++.

\subsection{Ablation Studies }
\label{sec:ablations}
To understand the contribution of each component, we retrain the scheduler from scratch under different configurations. All configurations are trained across multiple values of the prompt-alignment parameter \(\lambda\); for comparison, we report results at two evaluation settings, \(\lambda\in\{0.4,\,0.8\}\), using FID, CLIP, and ImageReward. Table~\ref{tab:ablation} summarizes the results.

We assess the role of inputs to the scheduler. Omitting timestep information (\textit{w/o \( t \)}) or the alignment signal (\textit{w/o \( \delta_t \)}) inputs leads to lower performance in all metrics, indicating that both inputs contribute to the overall effectiveness of our scheduler.

Dropping CFG++'s renoising step (\textit{w/o CFG++ Renoise}) results in a significant drop in CLIP and ImageReward.

Removing prompt perturbation during training (\textit{w/o Perturbation}) degrades performance across all metrics, indicating its importance for robustness, and constraining the predicted guidance scale \( w \) to the range \([0, 1]\), as done in CFG++ (\textit{Constrained \( w \)}), achieves the lowest FID, but at the cost of reduced alignment and reward. Given this trade-off, we deliberately opt to leave \( w \) unconstrained, as it enables a better overall balance across metrics—maintaining strong prompt alignment and perceptual quality.

\begin{table}[t]
\centering
\caption{Ablation study. All configurations are trained across multiple $\lambda$ values; results shown for evaluation at $\lambda=0.4$ and $\lambda=0.8$.}
\begin{tabular}{lccc}
\toprule
\textbf{Configuration} & \textbf{FID} ↓ & \textbf{CLIP} ↑ & \textbf{IR} ↑ \\
\midrule
\multicolumn{4}{c}{$\lambda = 0.4$} \\
\midrule
Annealing               & 25.35 & \textbf{32.25} & \textbf{0.865} \\
w/o $t$                 & 25.96 & 32.21 & 0.847 \\
w/o $\delta_t$          & 26.16 & 32.22 & 0.852 \\
w/o CFG++ Renoise       & 25.32 & 32.11 & 0.805 \\
w/o Perturbation        & 25.96 & 32.19 & 0.855 \\
Constrained $w$         & \textbf{25.21} & 32.20 & 0.850 \\
\midrule
\multicolumn{4}{c}{$\lambda = 0.8$} \\
\midrule
Annealing               & 26.40 & \textbf{32.29} & \textbf{0.898} \\
w/o $t$                 & 26.86 & 32.27 & 0.896 \\
w/o $\delta_t$          & 26.97 & 32.28 & 0.896 \\
w/o CFG++ Renoise       & 26.34 & 32.18 & 0.831 \\
w/o Perturbation        & 27.01 & 32.25 & 0.884 \\
Constrained $w$         & \textbf{26.15} & 32.25 & 0.880 \\
\bottomrule
\end{tabular}
\vspace{-1pt}
\label{tab:ablation}
\end{table}

\begin{table}[t]
\centering
\caption{Comparison across solvers, CFG++ and ours. We consistently achieve better metrics in terms of FID, CLIP and IR.}
\begin{tabular}{lccc}
\toprule
\textbf{Method} & \textbf{FID} ↓ & \textbf{CLIP} ↑ & \textbf{IR} ↑ \\
\midrule
DDIM (CFG++, w=0.8) & 25.61 & 32.20 & 0.857 \\
DDIM (Ours, \(\lambda=0.4\)) & \textbf{25.35} & \textbf{32.25} & \textbf{0.865 }\\
\midrule
Euler (CFG++, w=0.8) &26.17 &  32.21 & 0.867 \\
Euler (Ours, \(\lambda=0.4\)) & \textbf{25.92} & \textbf{32.23} & \textbf{0.873}\\
\midrule
Euler Ancestral (CFG++, w=0.8) & 28.57 & 32.32 & 0.900 \\
Euler Ancestral (Ours, \(\lambda=0.4\)) & \textbf{28.09} & \textbf{32.34} & \textbf{0.906} \\
\bottomrule
\end{tabular}
\label{tab:sampler_ablation}
\end{table}

\subsection{Other Solvers \& Noise Schedules }
\label{sec:other_solvers_schedulers}

To evaluate the robustness of our method across different samplers and noise schedules, we adopt the CFG++ renoising step generalized to both the Euler sampler~\cite{karras2022elucidatingdesignspacediffusionbased} and the Euler Ancestral sampler, following the formulations in CFG++~\cite{chung2024cfg++}. For each sampler, we fix the learned annealing scheduler and evaluate it with \(\lambda=0.4\), comparing against a CFG++ baseline using the same sampler with a fixed guidance weight \(w=0.8\). We additionally report results using DDIM for completeness, using a \texttt{scaled\_linear} beta schedule with \(\beta_{\text{start}} = 0.00085\) and \(\beta_{\text{end}} = 0.012\), while Euler and Euler Ancestral use a linear schedule with \(\beta_{\text{start}} = 0.0001\) and \(\beta_{\text{end}} = 0.02\).

We report FID, CLIP, and ImageReward in Table~\ref{tab:sampler_ablation}. Notably, our scheduler outperforms the CFG++ baseline in all metrics across different solvers.

\begin{figure*}[t] %
    \centering
    \begin{subfigure}[b]{0.24\textwidth} %
        \centering
        \begin{overpic}[width=\textwidth]{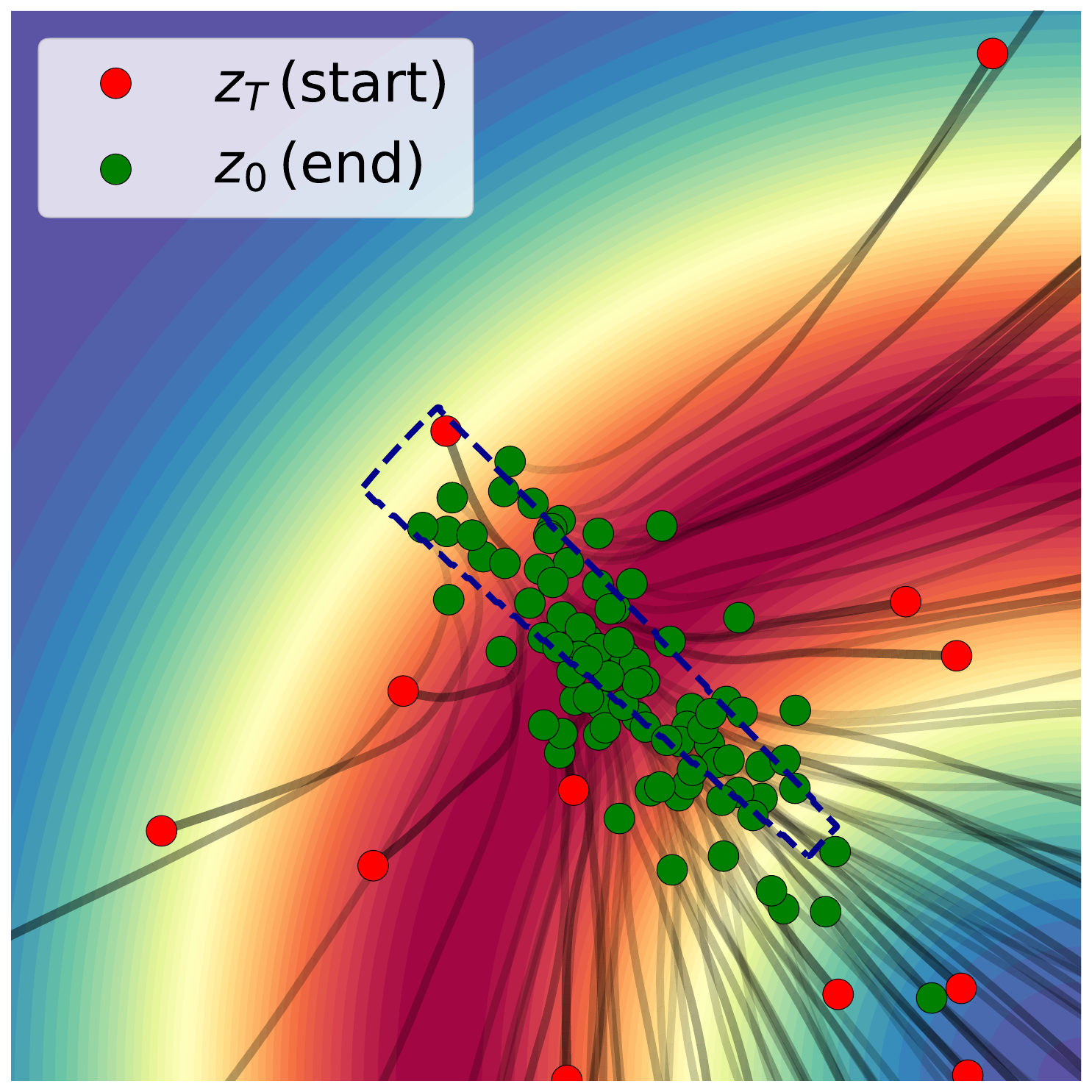} %
            \put(50,50){\includegraphics[width=0.5\textwidth]{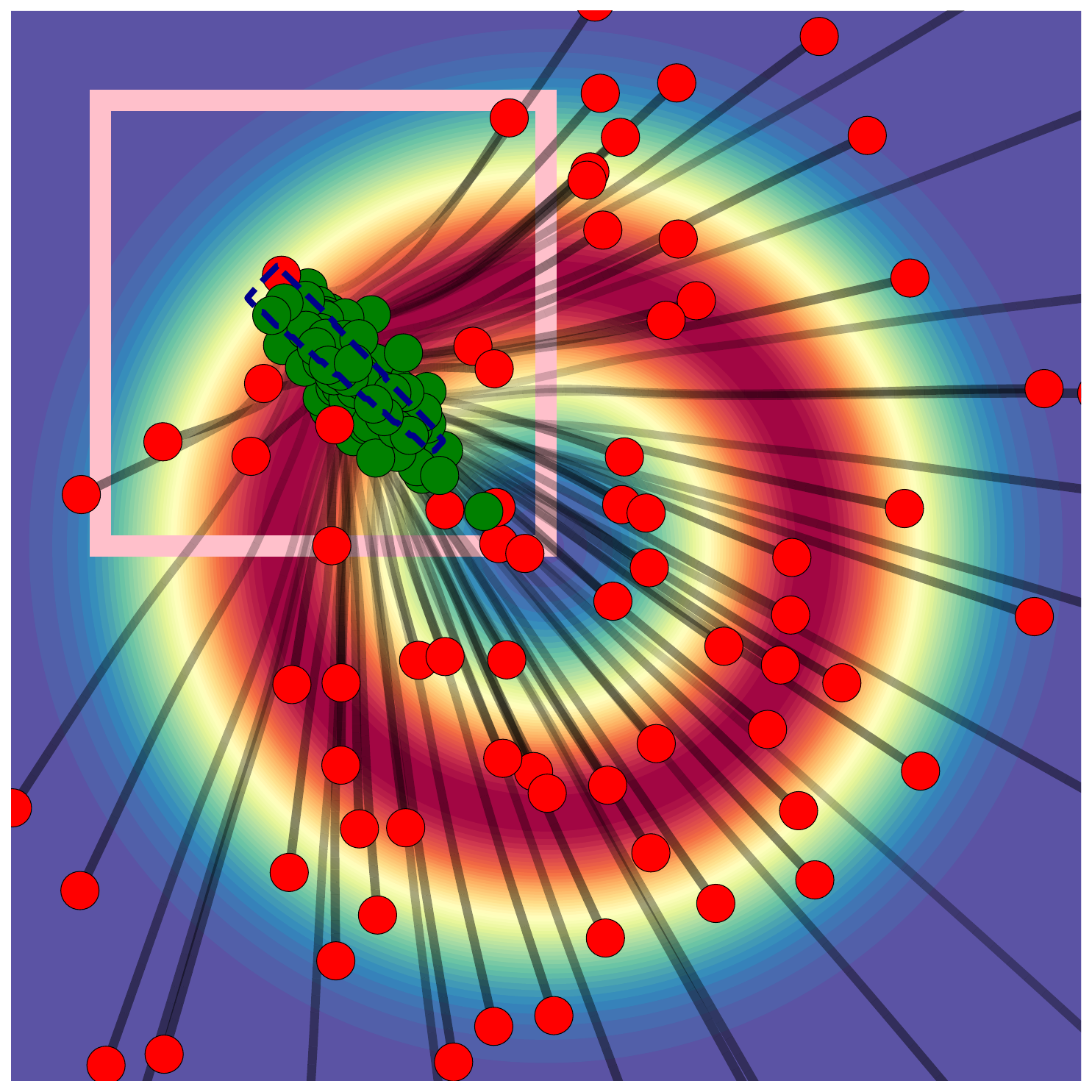}} %
        \end{overpic}
        \caption{\(w=1\)}
    \end{subfigure}
    \begin{subfigure}[b]{0.24\textwidth}
        \centering
        \includegraphics[width=\textwidth]{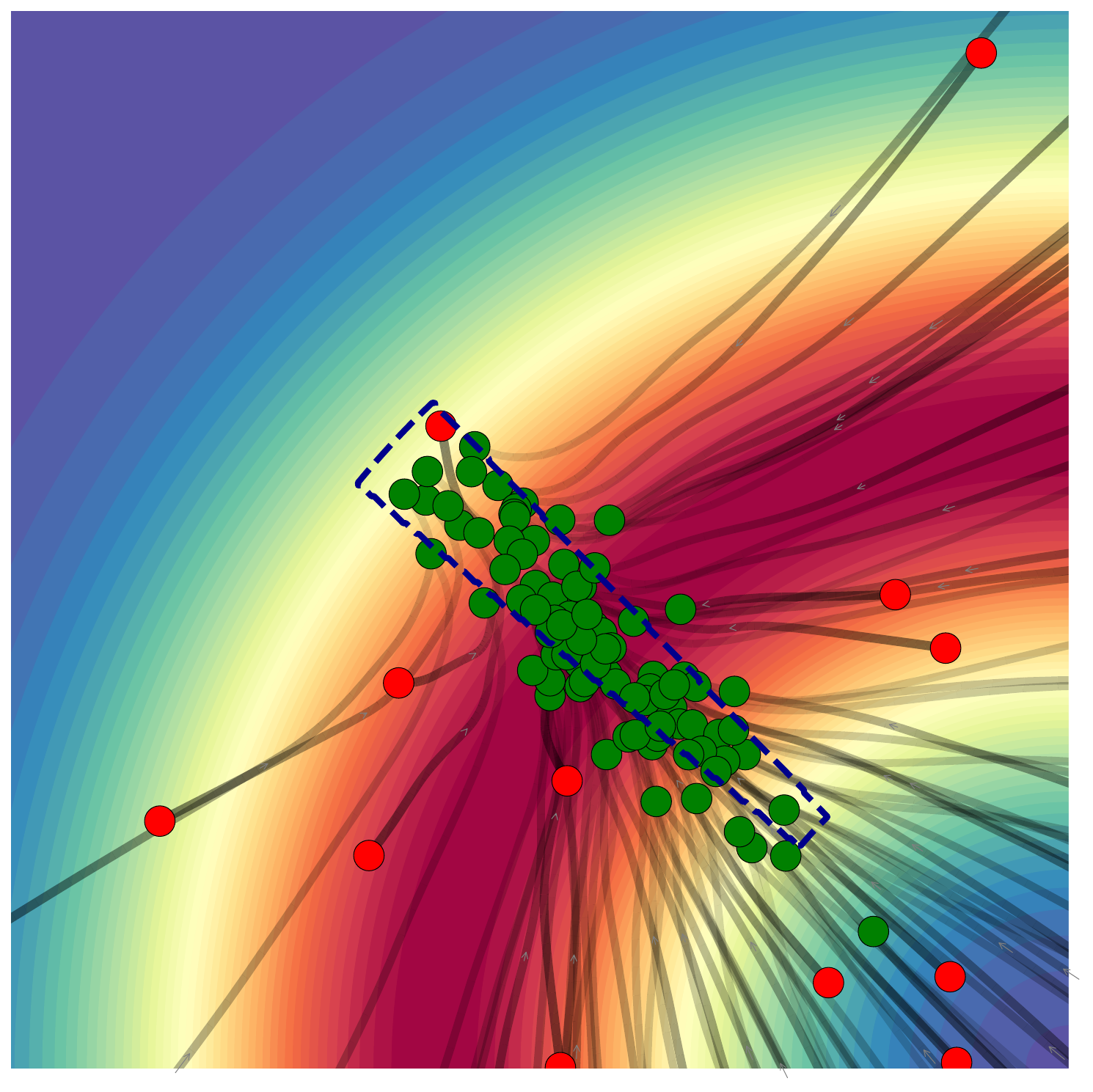}
        \caption{\(w=1.5\)}
    \end{subfigure}
    \begin{subfigure}[b]{0.24\textwidth}
        \centering
        \includegraphics[width=\textwidth]{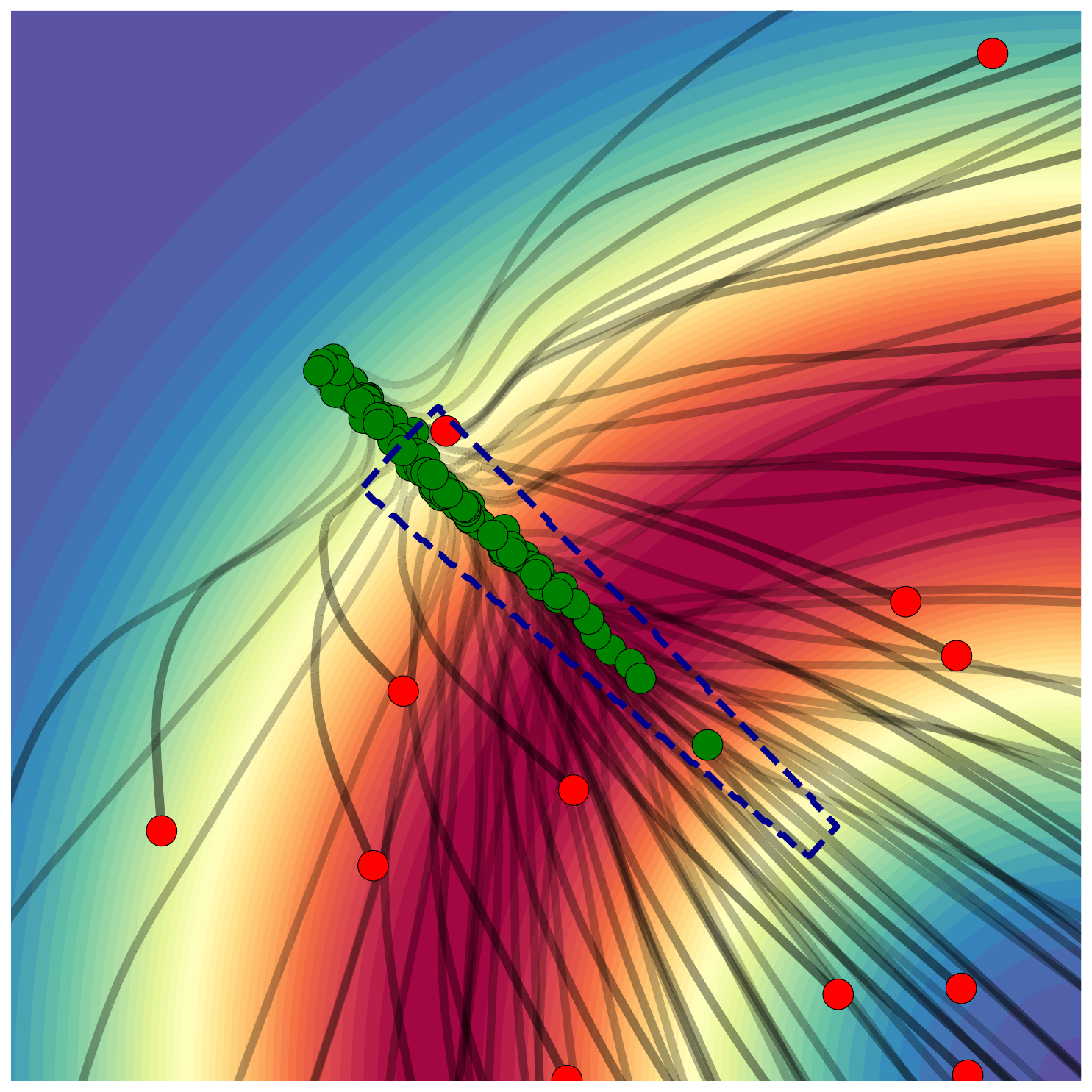}
        \caption{\(w=2\)}
    \end{subfigure}
    \begin{subfigure}[b]{0.24\textwidth}
        \centering
        \includegraphics[width=\textwidth]{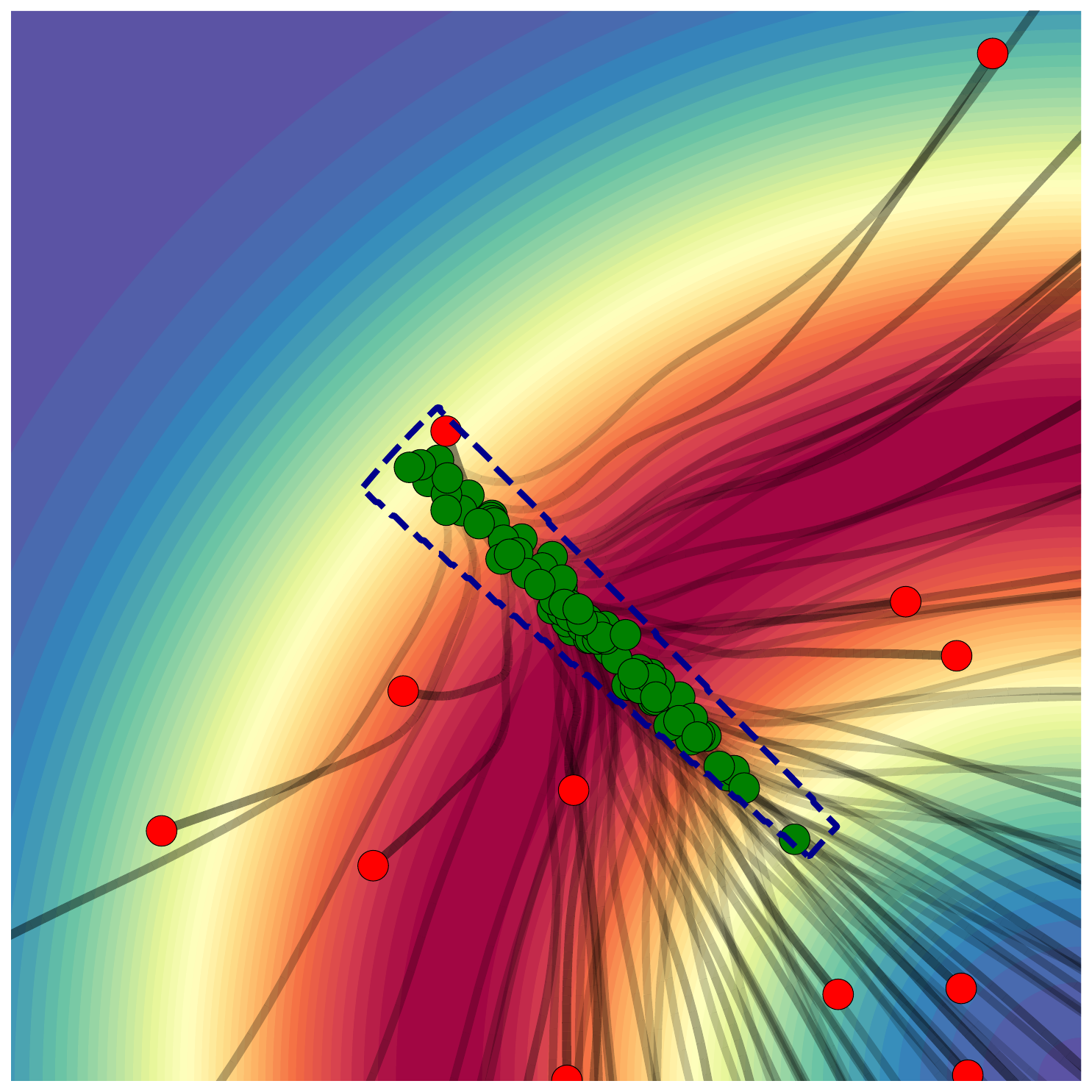}
        \caption{\textbf{Annealing ($\lambda=0.9$)}}
        \label{fig:toy_flow_adaptive}
    \end{subfigure}
    \caption{A 2D flow matching toy example with a target distribution shaped as a wide ring. Random seeds conditioned on \( c = 3\pi/4 \) are shown, along with their trajectories in gray. The dashed region indicates a tolerance band of \( \pm \pi/64 \) around the target condition \(c\), where the manifold density is high. \textbf{(a)} \(w = 1.0\): Low guidance scale results in weak condition adherence. \textbf{(b)} \(w = 1.5\): Moderate guidance slightly improves alignment, but some samples still deviate from the target region. \textbf{(c)} \(w = 2.0\): Strong guidance leads to overfitting the condition, pulling samples off the true manifold. \textbf{(d)} Ours: The trained annealing scheduler achieves accurate condition alignment while preserving sample quality.}

    \label{fig:toy_flow_diff}
\end{figure*}

\begin{figure}[t] %
    \centering
    \begin{tabular}{@{}c@{}c@{}c@{}}
        \includegraphics[width=0.379\columnwidth]{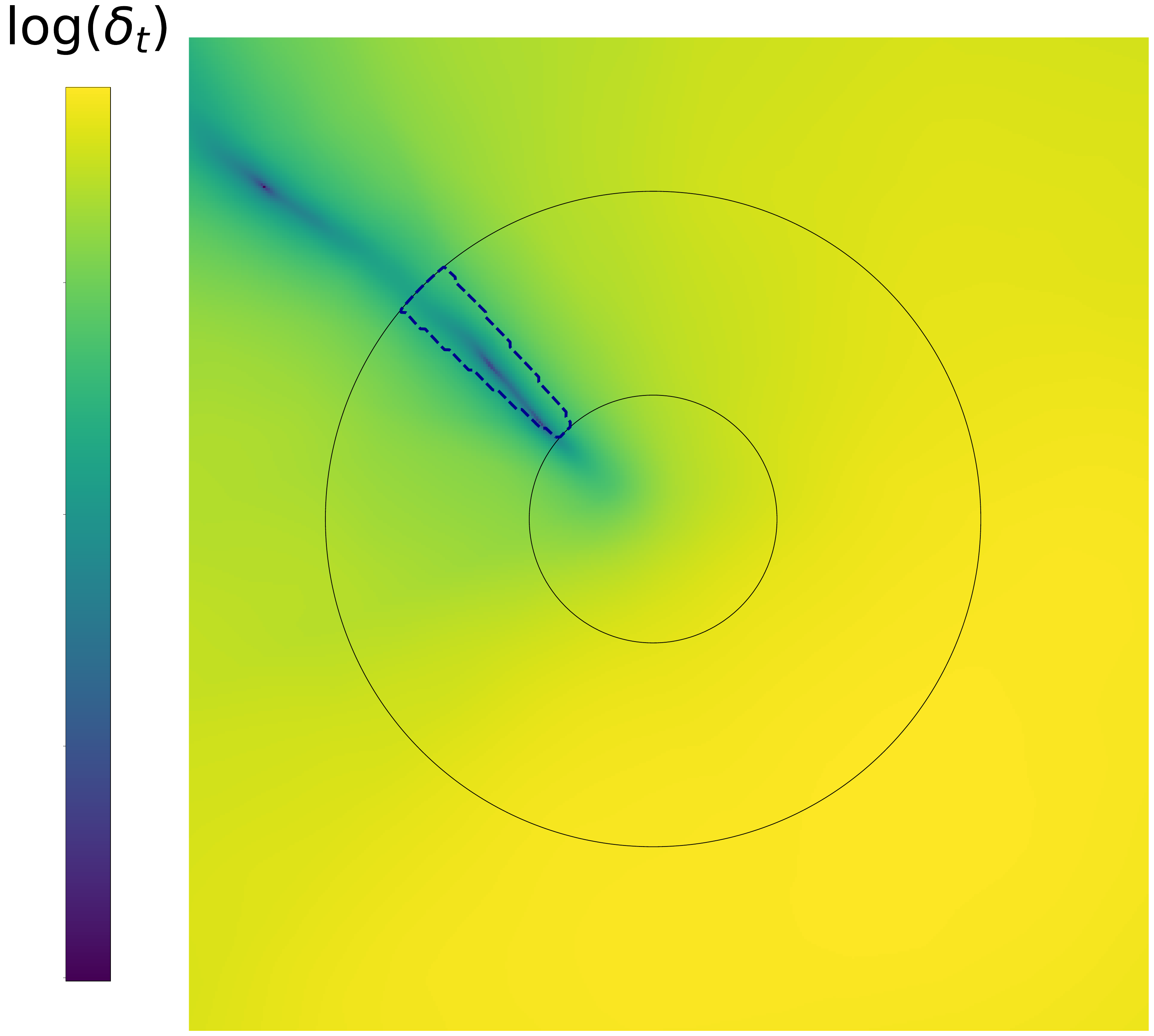} &
        \includegraphics[width=0.31625\columnwidth]{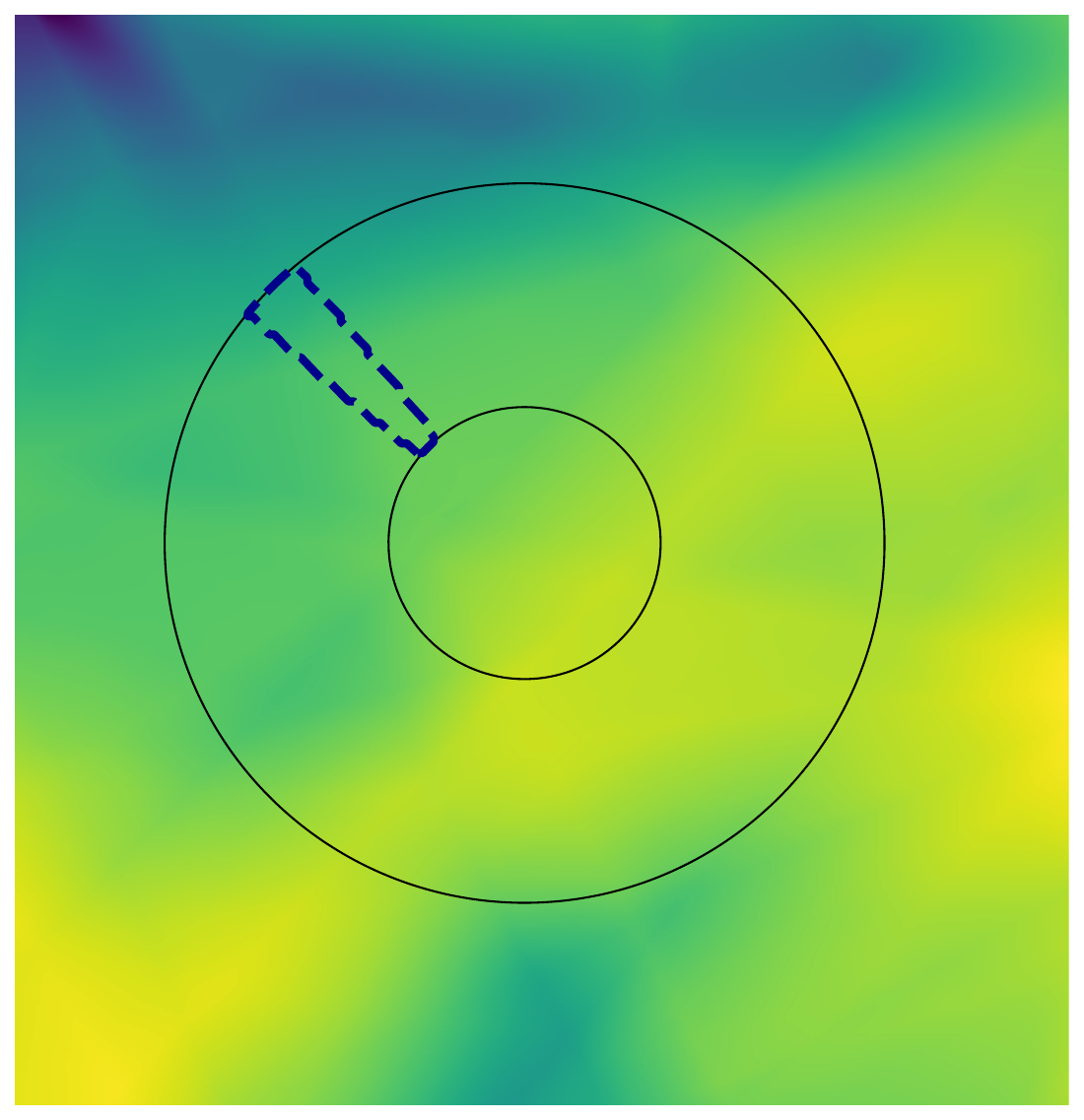} &
        \includegraphics[width=0.31625\columnwidth]{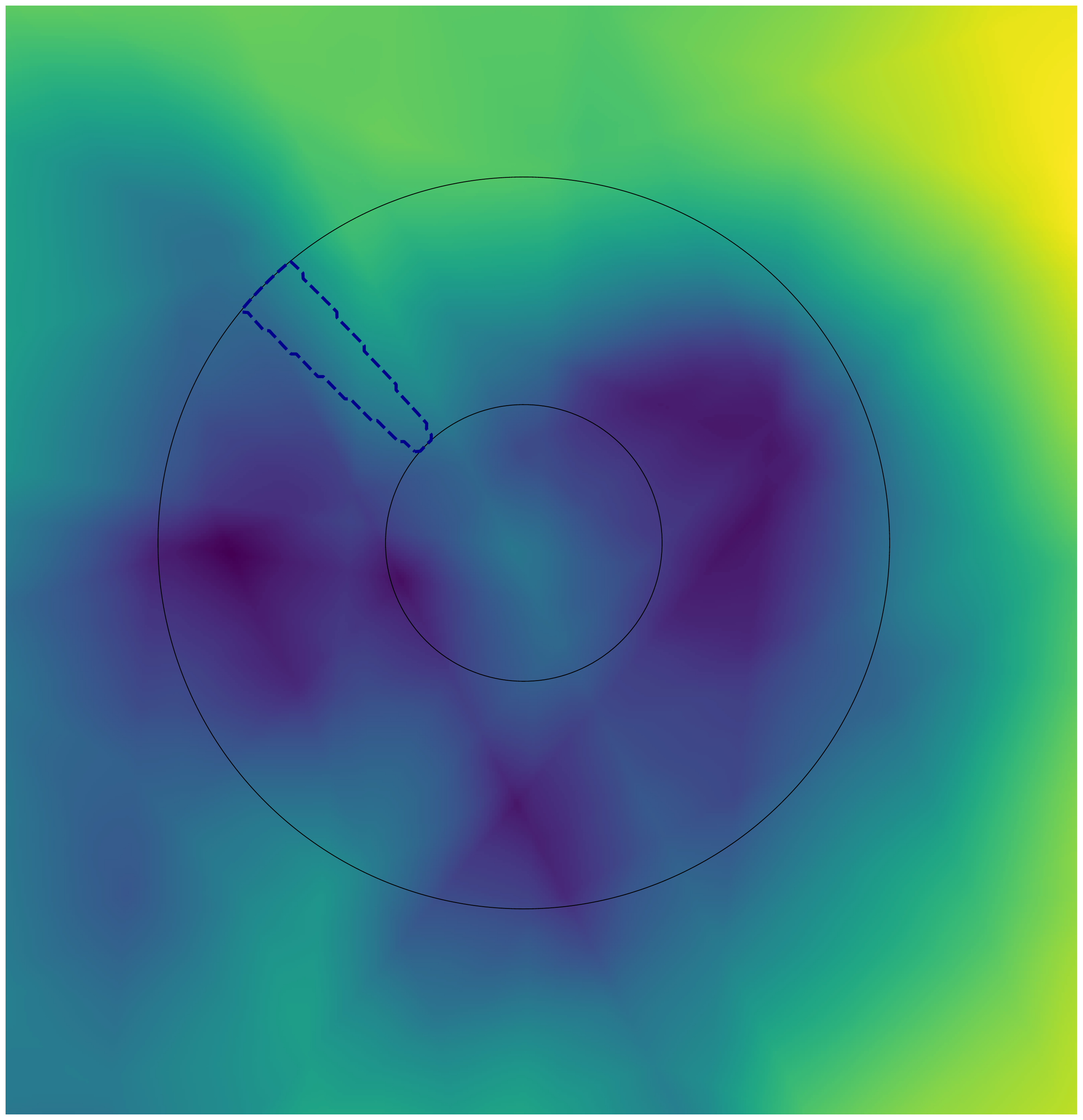} \\
        \(t=1\) & \(t=0.5\) & \(t=0\) %
    \end{tabular}
    \caption{\textbf{\(\log||\delta_t||\) heatmap for \(c=3\pi/4\).} This quantity illustrates how well unconditional and conditional predictions align in magnitude and direction across the latent space for varying timesteps. The region between the black circles marks areas of high sample density, while the blue dashed line represents the desired conditional region. \textbf{\(t=1\)}: At large \(t\)'s, noise dominates the samples, with model predictions tending toward the center of the source distribution, albeit slightly noisy. \textbf{\(t=0.5\)}: As \(t\) increases, unconditional and conditional predictions align more closely, showing lower values near the desired condition, but even lower values remain farther from the distribution in the direction of \(c\). \textbf{\(t=1\)}: By the final timestep, a local minimum is achieved at the target location within the ring.} %
    \label{fig:flow_diff_heatmap}
\end{figure}

\begin{table}[t]
\centering
\setlength{\tabcolsep}{5pt}
\caption{\textbf{Zero-shot transfer to SD~2.1.} 
Results on SD~2.1 using an annealing scheduler trained on SDXL. 
Parameter settings: CFG ($w{=}10$), CFG++ ($w{=}0.8$), and Ours ($\lambda{=}0.4$).}
\begin{tabular}{lcccccc}
\toprule
\textbf{Method} & \textbf{FID} $\downarrow$ & \makecell{\textbf{FD-}\\\textbf{DINOv2} $\downarrow$} & \textbf{CLIP} $\uparrow$ & \textbf{P} $\uparrow$ & \textbf{R} $\uparrow$ & \textbf{IR} $\uparrow$ \\
\midrule
CFG & 29.04 & 320.21 & 31.66 & \textbf{0.812} & \textbf{0.592} & 0.471 \\
CFG++ & \textbf{28.14} & 306.49 & 31.68 & 0.806 & 0.583 & 0.476 \\
Ours & 28.44 & \textbf{304.56} & \textbf{31.70} & 0.806 & 0.589 & \textbf{0.480} \\
\bottomrule
\end{tabular}
\label{tab:sd21_transfer}
\end{table}

\subsection{Zero-Shot Transfer Across Models}

We examine whether a scheduler trained on one text-to-image model transfers effectively to another. 
Concretely, we apply a scheduler trained on SDXL directly to SD~2.1 without re-training.
This transfer is inherently challenging, as the two models operate in different latent spaces with 
distinct learned priors, which may limit direct applicability. 

Table~\ref{tab:sd21_transfer} reports results across multiple metrics on COCO~\textit{val}~2017, 
compared to standard CFG and CFG++ baselines, with $w$ and $\lambda$ fixed to the same values used 
in the SDXL experiments. As expected, the improvements and margins observed on SDXL do not fully carry 
over to SD~2.1, particularly in FID and recall, though the method remains reasonably competitive. 
For optimal performance, however, the scheduler should be trained natively on the target model.

\subsection{Extension to Flow Matching}
\label{sec:flow_matching}

Our scheduler can be naturally extended to flow-based models by leveraging the continuous-time formulation of Flow Matching~\cite{lipman2022flow}. In this setting, we model the trajectory of samples \( x(t) \) using a learned velocity field \( v_\theta(x, t, c) \), governed by the ordinary differential equation:
\[
\frac{dx}{dt} = v_\theta(x, t, c),
\]
where \( c \) denotes the conditioning signal. The model is trained to match the true velocity \( x_1 - x_0 \) by sampling intermediate points \( x(t) = x_0 + t(x_1 - x_0) \) and minimizing a velocity prediction loss, analogous to the diffusion-based \(\epsilon\)-prediction loss. Specifically, the equivalent of the \(\epsilon\)-loss becomes:
\[
\mathcal{L}_\epsilon = \left\| v_\theta(x(t), t, c) - (x_1 - x_0) \right\|^2.
\]
Furthermore, we define a \(\delta\)-loss similar to the diffusion model case. After an integration step to a future point \( x(t + \Delta t) \), we compute the discrepancy between the conditional and unconditional velocity predictions:
\[
\delta_{t+\Delta{t}} = v_\theta(x(t + \Delta t), t + \Delta t, c) - v_\theta(x(t + \Delta t), t + \Delta t, \emptyset),
\]
and define the loss as:
\[
\mathcal{L}_\delta = \left\| \delta_{t+\Delta{t}} \right\|^2.
\]

Similarly to the toy example presented for diffusion models, we train a 2D toy flow matching model that predicts the conditional velocity field \( v_\theta(x, t, \{c,\varnothing\}) \), and afterwards train an annealing scheduler using the equivalent velocity matching objectives described above ($\mathcal{L}_\delta$ and $\mathcal{L}_\epsilon$). At inference time, as a baseline, we apply guidance by combining the conditional and unconditional velocity predictions using a scaled interpolation (namely the guidance scale $w$), and finally present our scheduler guidance. For the target condition \( c = 3\pi/4 \), we visualize the resulting trajectories and sample alignments in Figure~\ref{fig:toy_flow_diff}. The results show that our proposed annealing-based scheduler achieves better condition alignment and sample quality compared to constant guidance scales, confirming the effectiveness of our approach in the flow matching setting as well.
To further analyze the behavior of the model, we visualize the quantity \(\|\delta_t\|\) across different timesteps. As shown in Figure~\ref{fig:flow_diff_heatmap}, we observe a pattern similar to that in the diffusion model, highlighting the desirability of low \(\|\delta_t\|\) values, which indicate the desirability of better agreement between conditional and unconditional predictions.

\section{Additional Qualitative Results}
\label{sec:add_results}

We present additional qualitative comparisons to CFG/CFG++ across different scales of $w$ and $\lambda$ in Figs.~\ref{fig:extra_l_0_0_5}, \ref{fig:extra_2}, \ref{fig:extra_3}, \ref{fig:extra_cfgpp_2}, \ref{fig:extra_cfgpp_3}, \ref{fig:extra_l_0_7} and \ref{fig:extra_l_0_8}.

In Figs.~\ref{fig:all_scales_1} and \ref{fig:all_scales_2}, we show the effect of varying the value of $w$ and $\lambda$ for the same prompt and initial noise. Each row shows generations from CFG/CFG++ (top) and Annealing (bottom).

We observe that our method consistently yields higher image quality and stronger prompt adherence than CFG and CFG++. For example, in Fig.~\ref{fig:all_scales_1}, CFG produces cartoonish effects at higher scales (top row) and structural artifacts such as an extra sock (bottom row). Similarly, in Fig.~\ref{fig:all_scales_2}, CFG deforms the goat at higher scales (top row), while CFG++ merges the woman and the boy at low and high scales (bottom row). In contrast, our annealing scheduler avoids these failures and maintains realism and alignment across all scales.

\begin{figure*}[t]
    \centering
    \begin{tabular}{@{}c@{\hspace{0.005\textwidth}}c@{\hspace{0.03\textwidth}}c@{\hspace{0.005\textwidth}}c@{}}

        CFG ($w=7.5$) & \textbf{Annealing ($\lambda=0.05$)} & CFG++ ($w=0.6$) & \textbf{Annealing ($\lambda=0.05$)} \\
        \includegraphics[width=0.49\columnwidth]{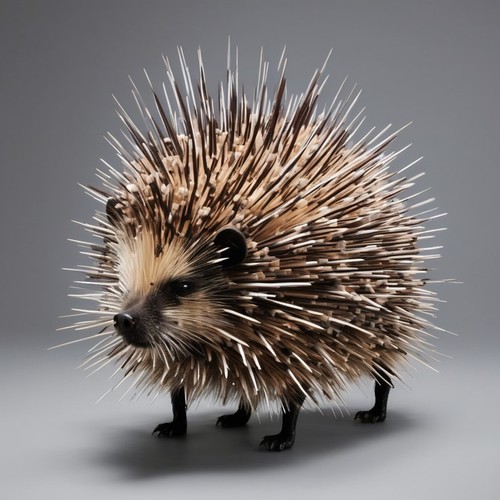} &
        \includegraphics[width=0.49\columnwidth]{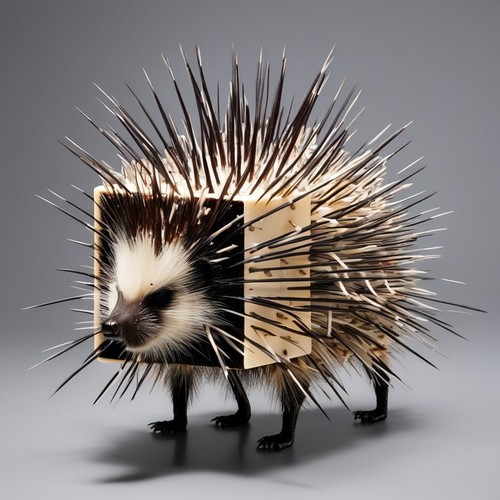} &
        \includegraphics[width=0.49\columnwidth]{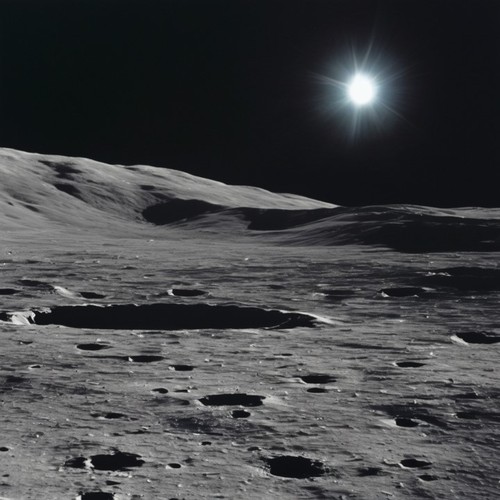} &
        \includegraphics[width=0.49\columnwidth]{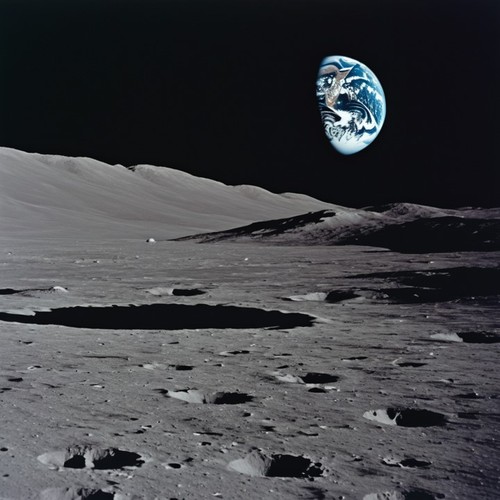} \\[-3pt]

        \multicolumn{2}{c}{\small \shortstack{
            \textit{``A \textcolor{red}{\textbf{cube}} made of porcupine''}
        }} &
        \multicolumn{2}{c}{\small \shortstack{
            \textit{``A view of the \textcolor{red}{\textbf{Earth}} from the moon.''} 
        }} \\

        \includegraphics[width=0.49\columnwidth]{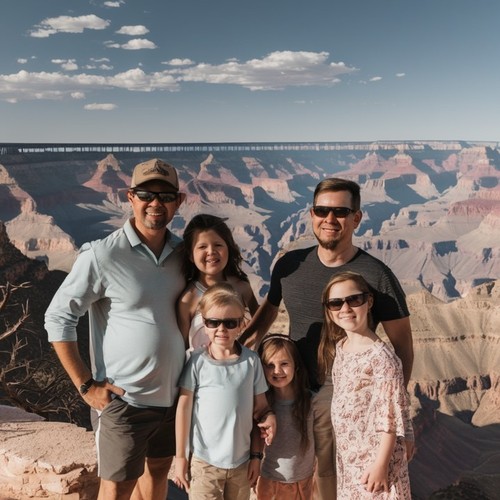} &
        \includegraphics[width=0.49\columnwidth]{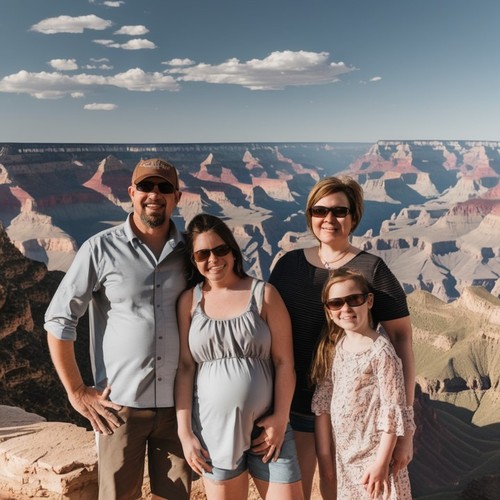} &
        \includegraphics[width=0.49\columnwidth]{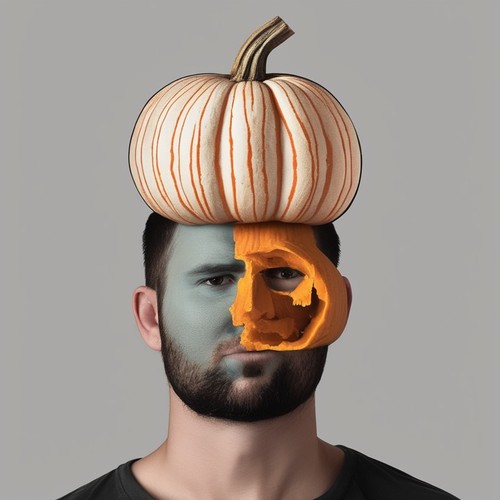} &
        \includegraphics[width=0.49\columnwidth]{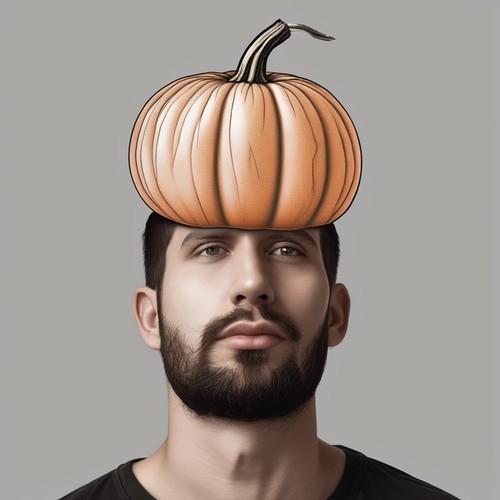} \\[-3pt]
        
        \multicolumn{2}{c}{\small \shortstack{
            \textit{``A family of \textcolor{red}{\textbf{four}} posing at the Grand Canyon''}
        }} &
         \multicolumn{2}{c}{\small \shortstack{
            \textit{``A pumpkin on a man's head.''}
        }}\\
        
        \includegraphics[width=0.49\columnwidth]{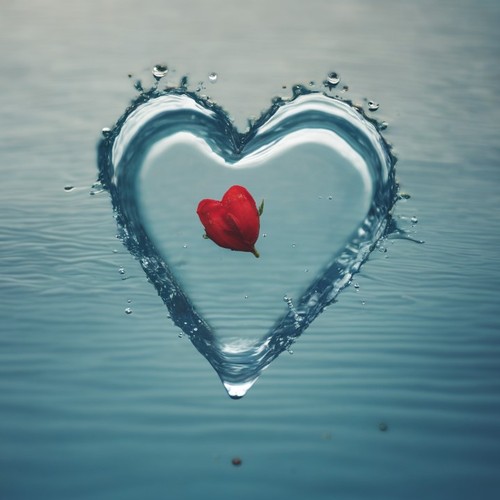} &
        \includegraphics[width=0.49\columnwidth]{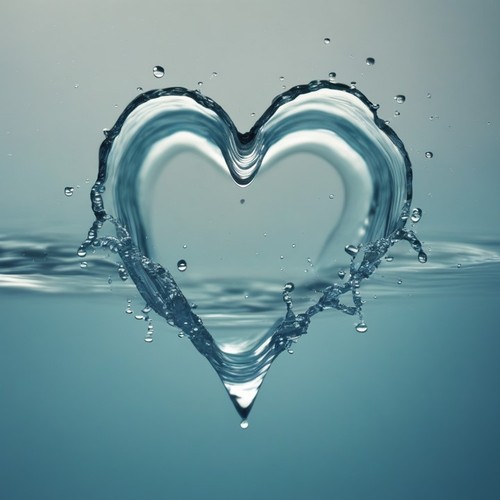} &
        \includegraphics[width=0.49\columnwidth]{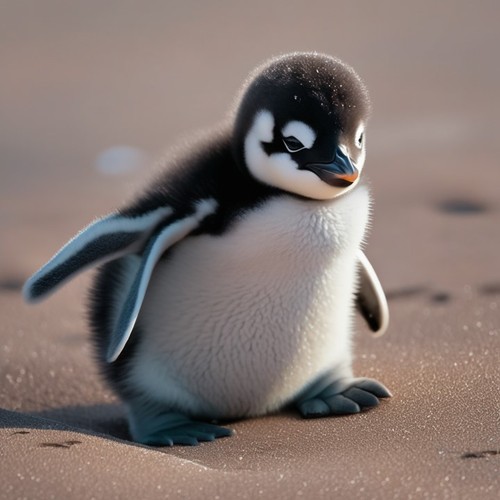} &
        \includegraphics[width=0.49\columnwidth]{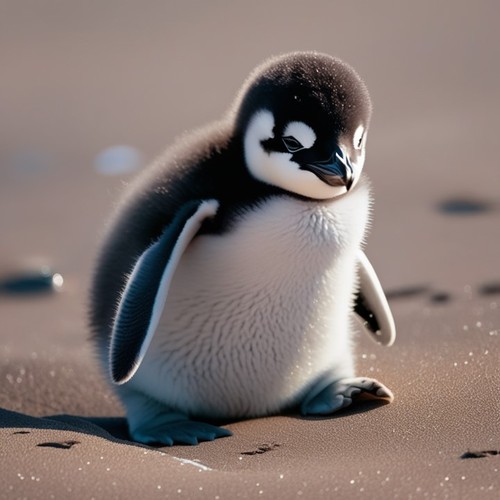} \\[-3pt]

        \multicolumn{2}{c}{\small \shortstack{
            \textit{``A heart made of water.''}
        }} &
        \multicolumn{2}{c}{\small \shortstack{
            \textit{``A baby penguin.''}
        }} \\

        \includegraphics[width=0.49\columnwidth]{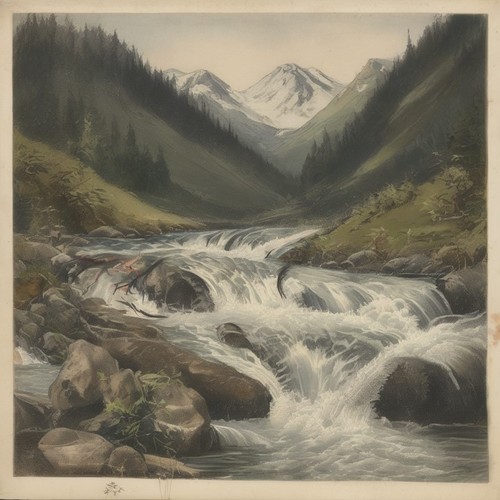} &
        \includegraphics[width=0.49\columnwidth]{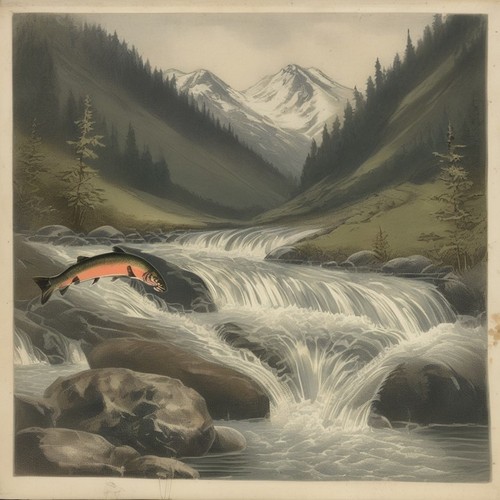} &
        \includegraphics[width=0.49\columnwidth]{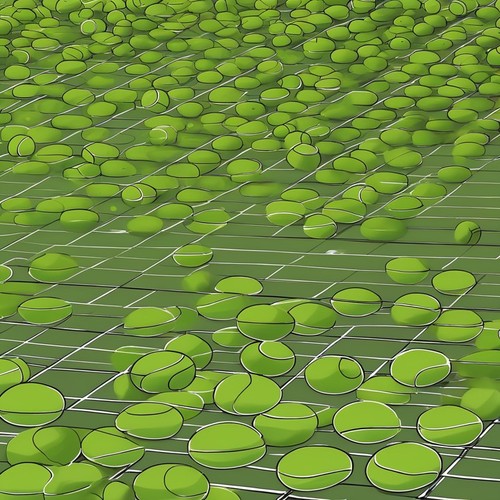} &
        \includegraphics[width=0.49\columnwidth]{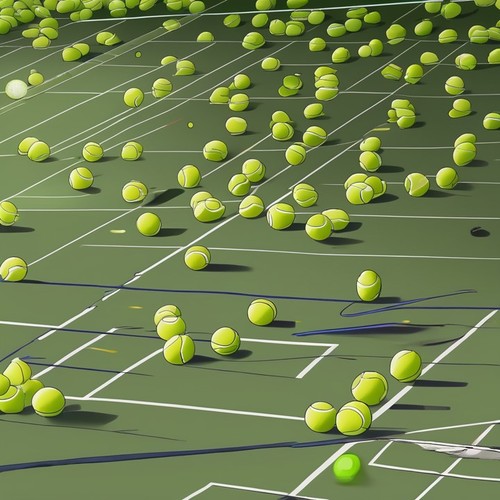} \\[-3pt]
        
        \multicolumn{2}{c}{\small \shortstack{
            \textit{``A mountain stream with \textcolor{red}{\textbf{salmon}} leaping out of it.''}
        }} &
        \multicolumn{2}{c}{\small \shortstack{
            \textit{``A tennis court with tennis balls scattered all over it.''}
        }} \\

        \includegraphics[width=0.49\columnwidth]{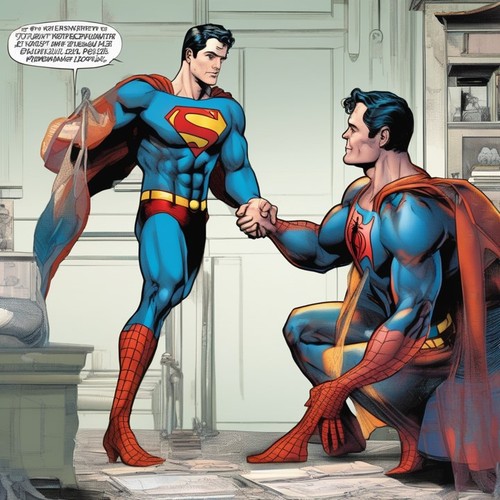} &
        \includegraphics[width=0.49\columnwidth]{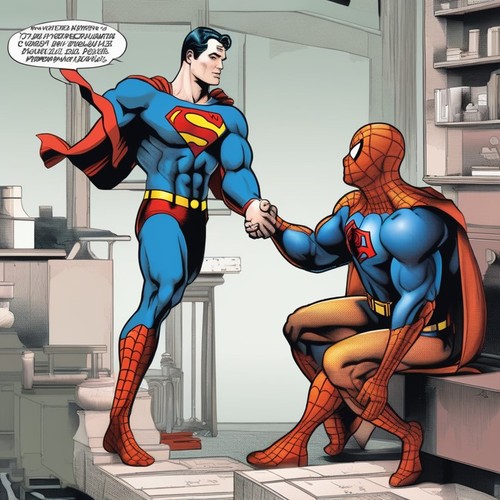} &
        \includegraphics[width=0.49\columnwidth]{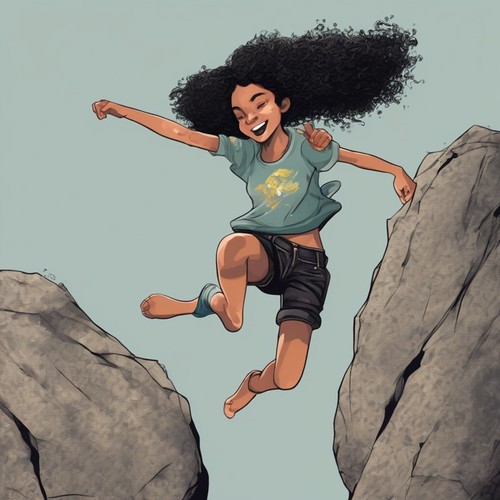} &
        \includegraphics[width=0.49\columnwidth]{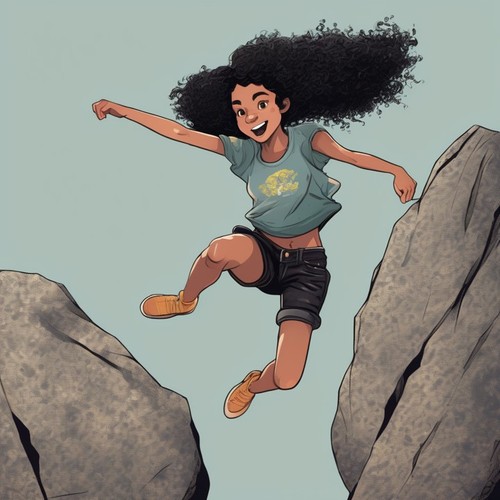} \\[-2pt]
        
        \multicolumn{2}{c}{\small \shortstack{
            \textit{``Superman shaking hands with \textcolor{red}{\textbf{Spiderman}}``}}} &
        \multicolumn{2}{c}{\small \shortstack{
            \textit{``A girl with curly black hair jumping off a boulder.''}
        }} 
    \end{tabular}
    \caption{\textbf{Qualitative comparison.} Columns (L→R): CFG ($w{=}7.5$) → Ours ($\lambda{=}0.05$); CFG++ ($w{=}0.6$) → Ours ($\lambda{=}0.05$).}
    \label{fig:extra_l_0_0_5}
\end{figure*}

\begin{figure*}[t]
    \centering
    \begin{tabular}{@{}c@{\hspace{0.005\textwidth}}c@{\hspace{0.03\textwidth}}c@{\hspace{0.005\textwidth}}c@{}}

        CFG ($w=10$) & \textbf{Annealing ($\lambda=0.4$)} & CFG ($w=10$) & \textbf{Annealing ($\lambda=0.4$)} \\

        \includegraphics[width=0.49\columnwidth]{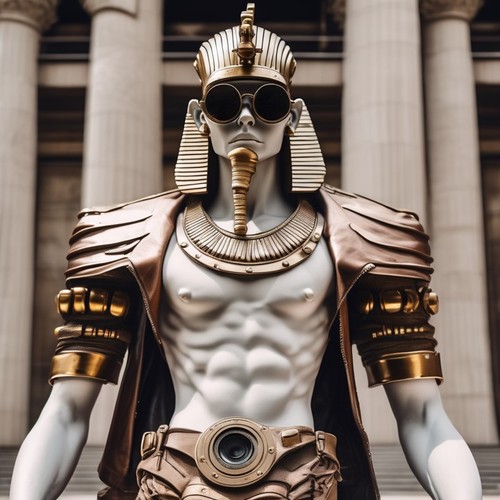} &
        \includegraphics[width=0.49\columnwidth]{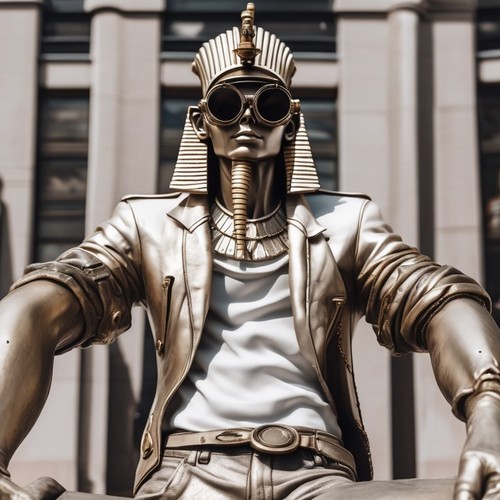} &
        \includegraphics[width=0.49\columnwidth]{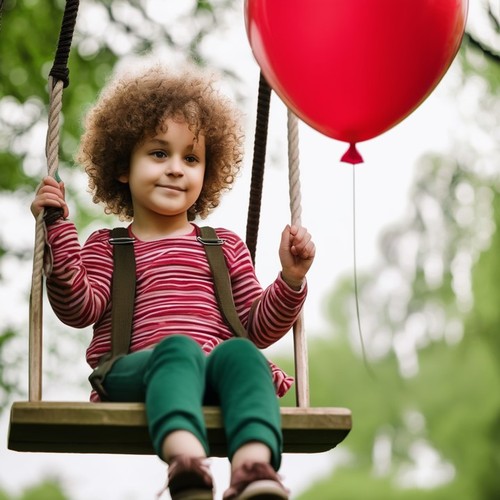} &
        \includegraphics[width=0.49\columnwidth]{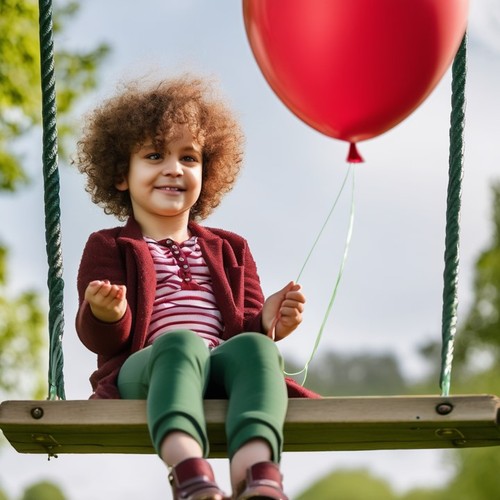}\\
        \multicolumn{2}{c}{\small \shortstack{
            \textit{``A statue of a pharaoh wearing steampunk glasses,} \\
            \textit{ \textcolor{red}{\textbf{white t-shirt}} and leather jacket.''}
        }} &
        \multicolumn{2}{c}{\small \shortstack{
            \textit{``Child with curly hair sitting on a wooden swing in a green park,} \\
            \textit{\textcolor{red}{\textbf{holding}} a red balloon.''}
        }} \\

        \includegraphics[width=0.49\columnwidth]{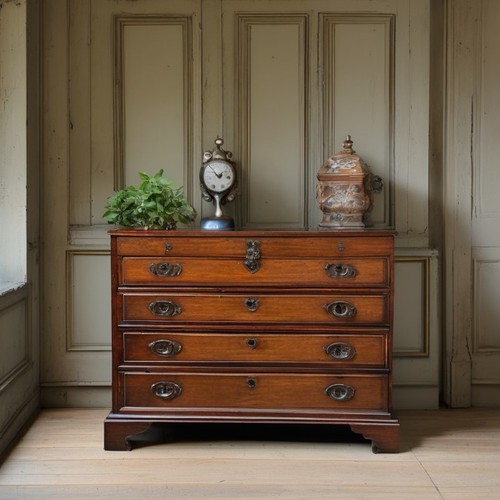} &
        \includegraphics[width=0.49\columnwidth]{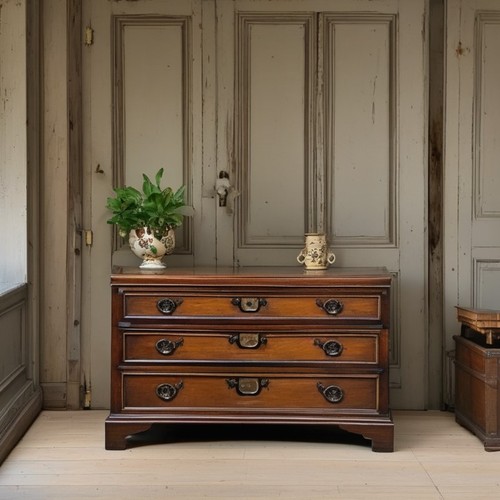} &
        
        \includegraphics[width=0.49\columnwidth]{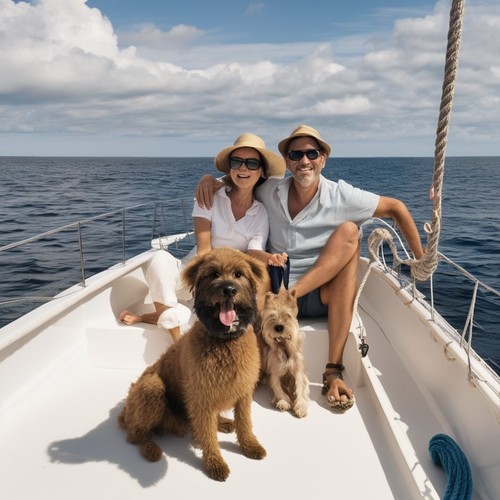} &
        \includegraphics[width=0.49\columnwidth]{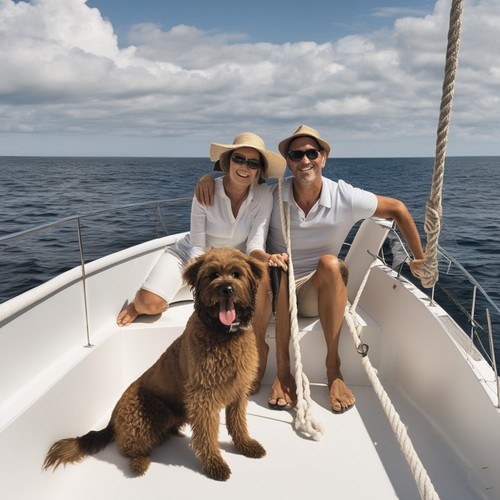}

        \\[-3pt]
        
        \multicolumn{2}{c}{\small \shortstack{
            \textit{``an antique chest with \textcolor{red}{\textbf{three}} drawers.``}}} &
        \multicolumn{2}{c}{\small \shortstack{
            \textit{``Couple on sailboat with \textcolor{red}{\textbf{a dog}} on open waters.''}
        }} \\

        \includegraphics[width=0.49\columnwidth]{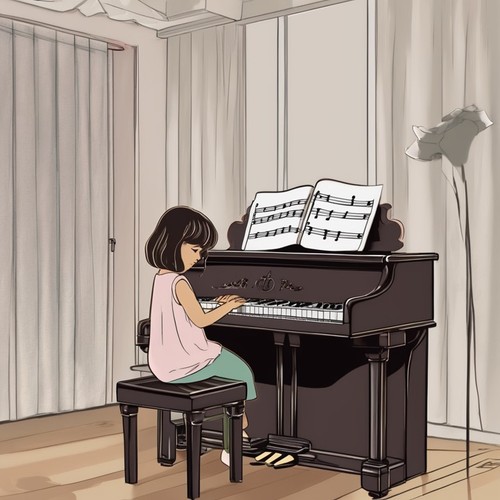} &
        \includegraphics[width=0.49\columnwidth]{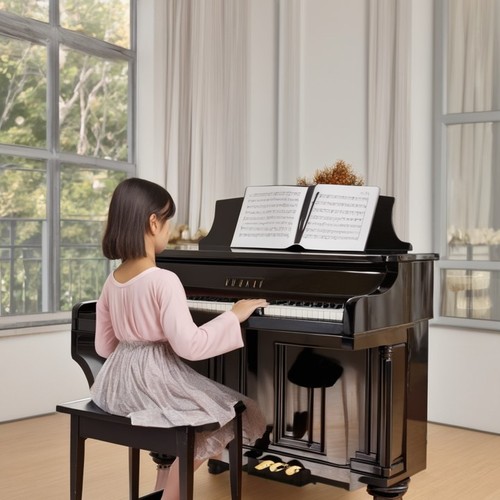} &
        \includegraphics[width=0.49\columnwidth]{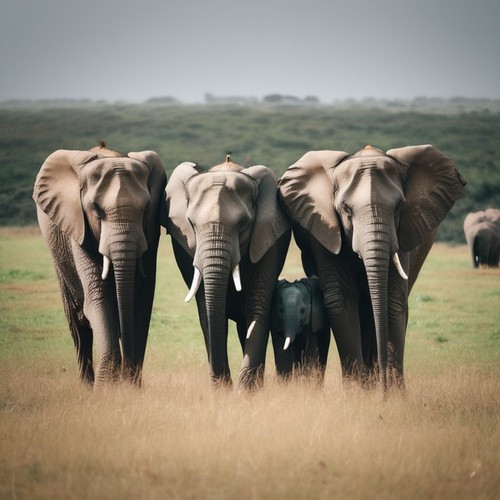} &
        \includegraphics[width=0.49\columnwidth]{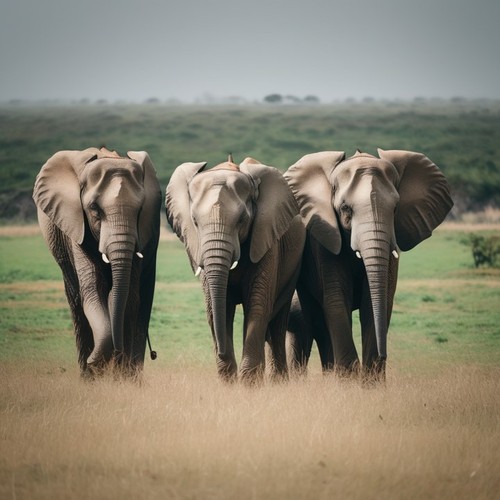} \\[-3pt]
        
        \multicolumn{2}{c}{\small \shortstack{
            \textit{``\textcolor{red}{\textbf{A photo}} of a young girl playing piano.''}
        }} &
        \multicolumn{2}{c}{\small \shortstack{
            \textit{``\textcolor{red}{\textbf{Three elephants}} in a field next to each other.''}
        }} \\
        
        \includegraphics[width=0.49\columnwidth]{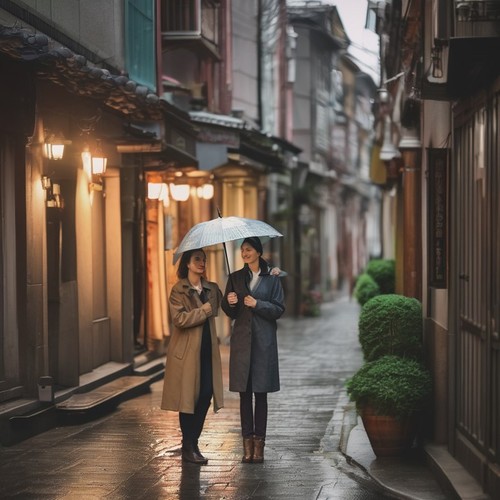} &
        \includegraphics[width=0.49\columnwidth]{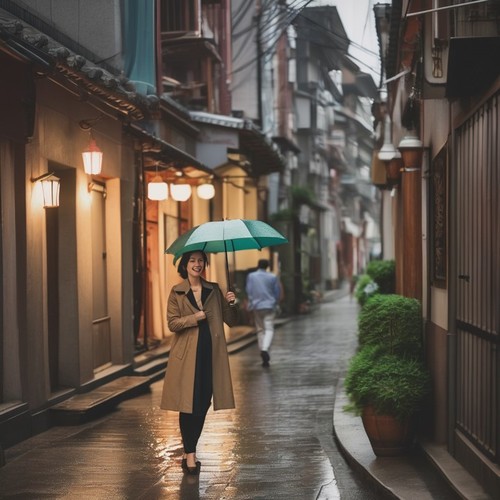} &
        \includegraphics[width=0.49\columnwidth]{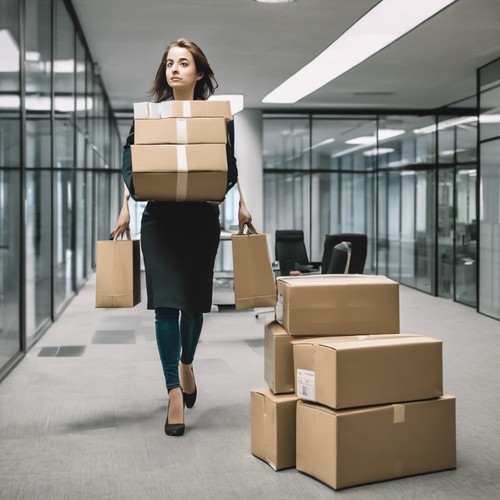} &
        \includegraphics[width=0.49\columnwidth]{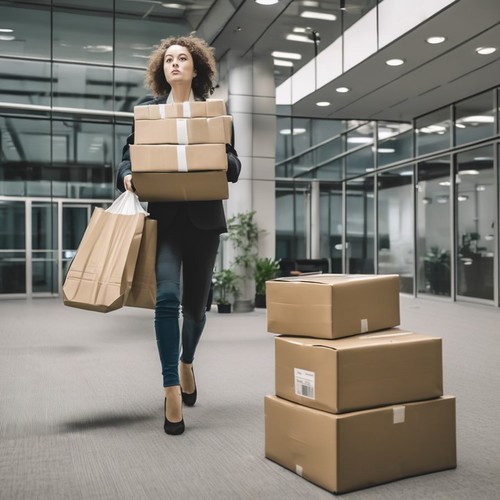} \\[-3pt]
        
        \multicolumn{2}{c}{\small \shortstack{
            \textit{``\textcolor{red}{\textbf{The woman}} holding an umbrella smiles... beside the sidewalk.''}
        }} &
        \multicolumn{2}{c}{\small \shortstack{
            \textit{``A woman is carrying many packages in an office building.''}
        }} \\

        \includegraphics[width=0.49\columnwidth]{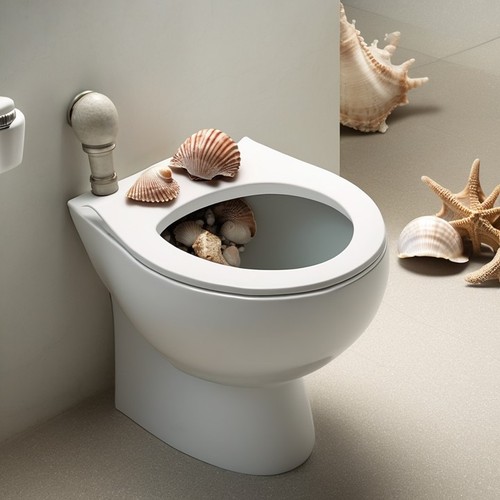} &
        \includegraphics[width=0.49\columnwidth]{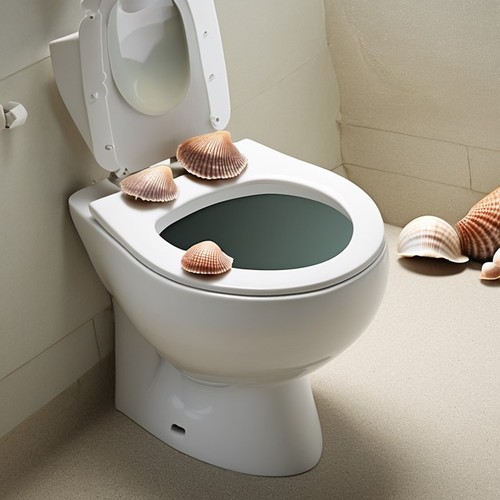} &
        \includegraphics[width=0.49\columnwidth]{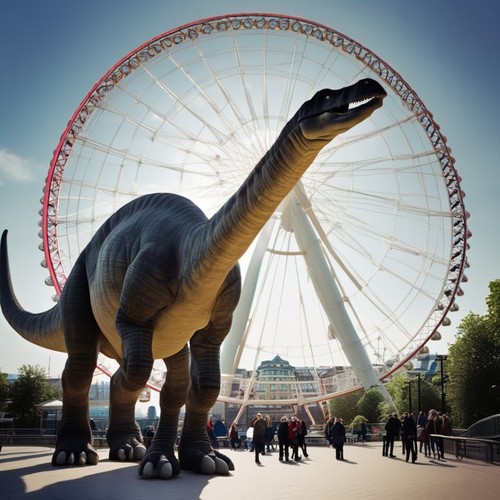} &
        \includegraphics[width=0.49\columnwidth]{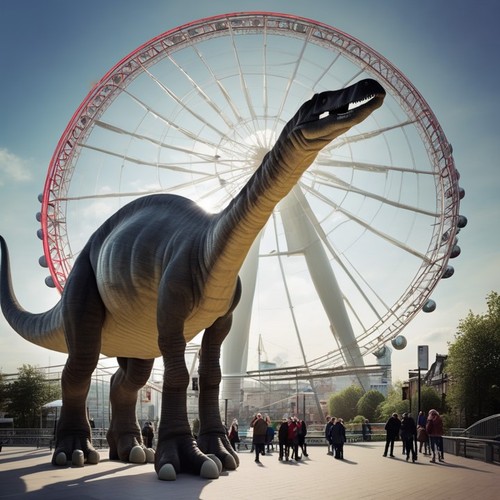} \\[-3pt]
        
        \multicolumn{2}{c}{\small \shortstack{
            \textit{``A toilet that has \textcolor{red}{\textbf{three}} sea shells on top.''}
        }} &
        \multicolumn{2}{c}{\small \shortstack{
            \textit{``A diplodocus standing in front of the Millennium Wheel.''}
        }} \\
    \end{tabular}

        \caption{Qualitative comparison of our Annealing method $\lambda=0.4$ (right) vs.\ CFG $w=10$ (left).}
    \label{fig:extra_2}
\end{figure*}
\begin{figure*}[t]
    \centering
    \begin{tabular}{@{}c@{\hspace{0.005\textwidth}}c@{\hspace{0.03\textwidth}}c@{\hspace{0.005\textwidth}}c@{}}

        CFG ($w=10$) & \textbf{Annealing ($\lambda=0.4$)} & CFG ($w=10$) & \textbf{Annealing ($\lambda=0.4$)} \\
        \includegraphics[width=0.49\columnwidth]{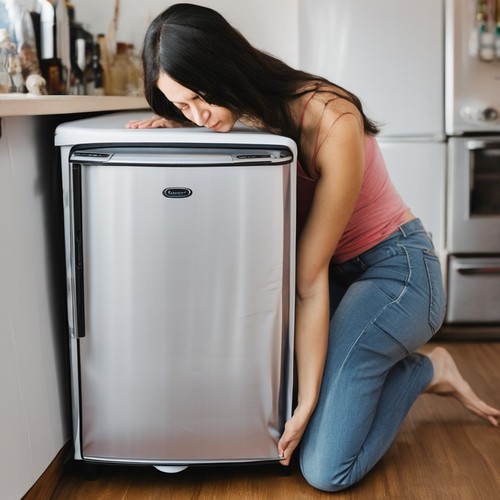} &
        \includegraphics[width=0.49\columnwidth]{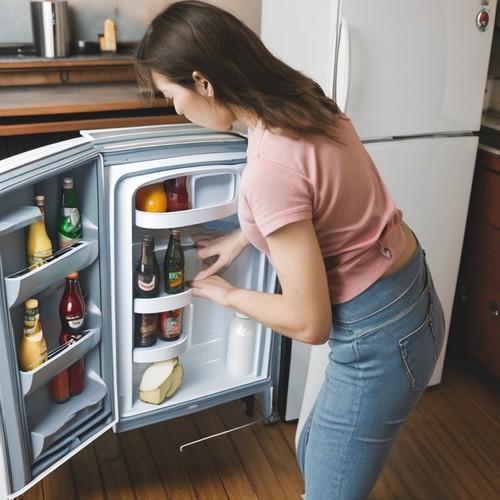} &
        \includegraphics[width=0.49\columnwidth]{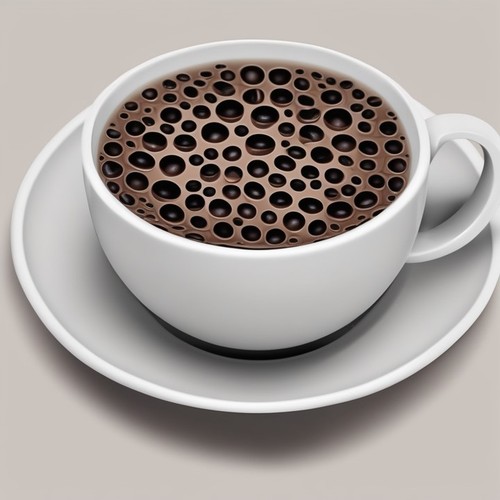} &
        \includegraphics[width=0.49\columnwidth]{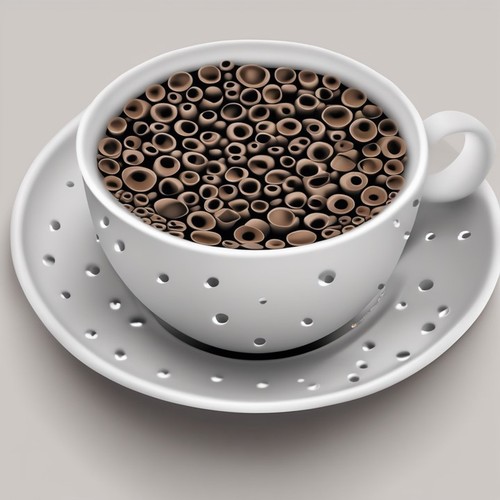} \\[-3pt]

                \multicolumn{2}{c}{\small \shortstack{
            \textit{``A woman bending over and looking \textcolor{red}{\textbf{inside}} of a mini fridge.''}
        }}  &
        \multicolumn{2}{c}{\small \shortstack{
            \textit{``A coffee cup that is \textcolor{red}{\textbf{full of holes}}.''}
        }} \\

        \includegraphics[width=0.49\columnwidth]{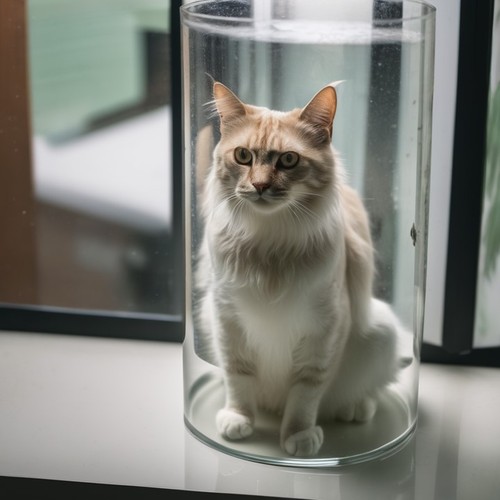} &
        \includegraphics[width=0.49\columnwidth]{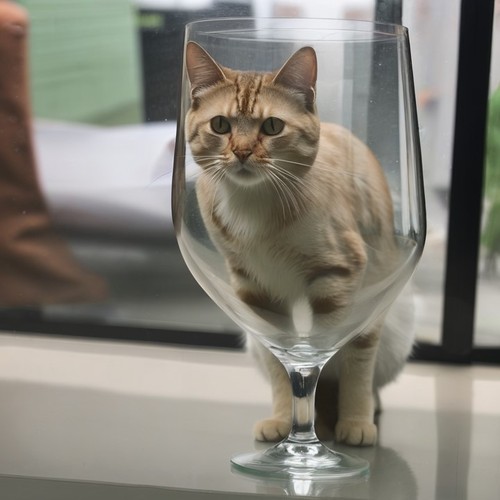} &
        \includegraphics[width=0.49\columnwidth]{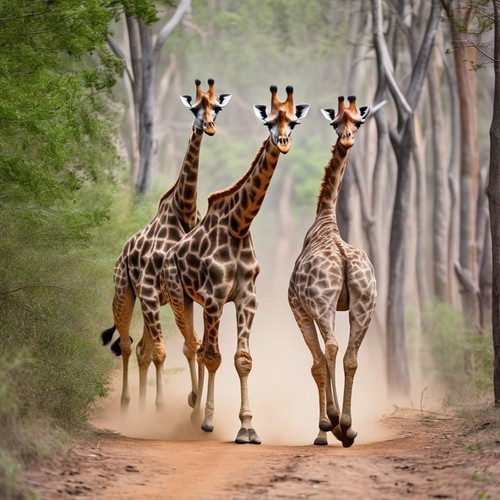} &
        \includegraphics[width=0.49\columnwidth]{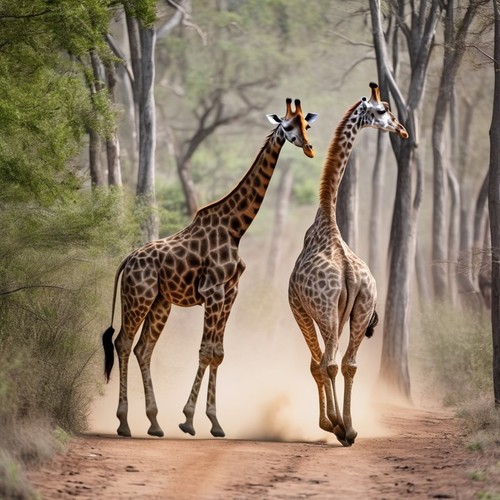} \\[-3pt]
        
        \multicolumn{2}{c}{\small \shortstack{
            \textit{``A cat that is  \textcolor{red}{\textbf{standing}} looking through a glass.``}
        }} &
        \multicolumn{2}{c}{\small \shortstack{
            \textit{``\textcolor{red}{\textbf{Two}} giraffes moving very quickly in the woods.''}
        }} \\

        \includegraphics[width=0.49\columnwidth]{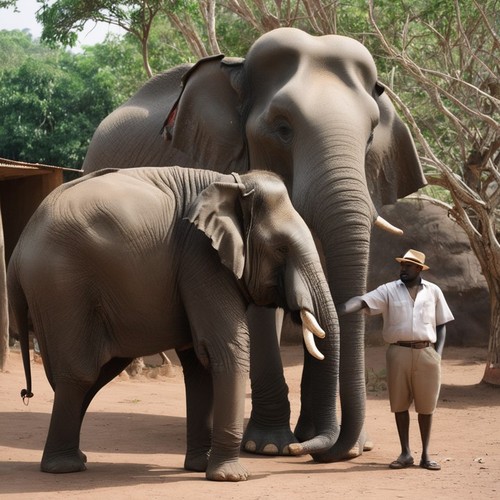} &
        \includegraphics[width=0.49\columnwidth]{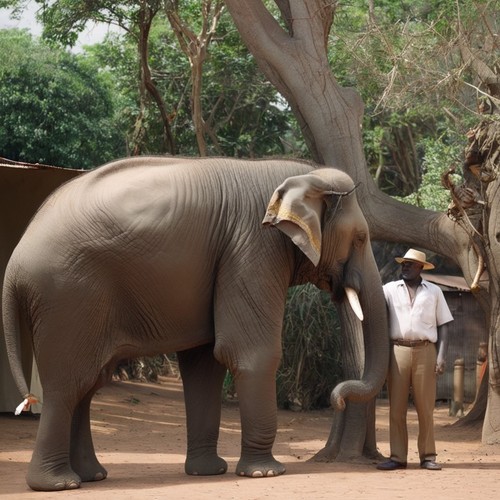} &
        \includegraphics[width=0.49\columnwidth]{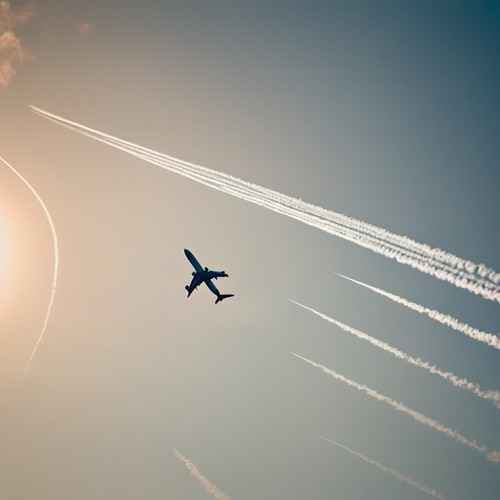} &
        \includegraphics[width=0.49\columnwidth]{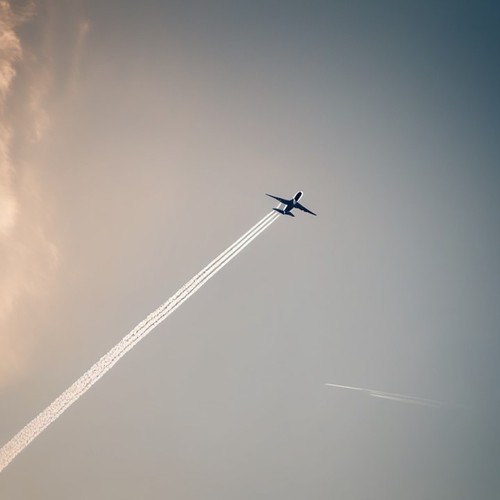} \\[-3pt]
        
        \multicolumn{2}{c}{\small \shortstack{
            \textit{``a man standing next to \textcolor{red}{\textbf{an elephant}} next to his trunk``}
        }} &
        \multicolumn{2}{c}{\small \shortstack{
            \textit{``An airplane leaving a trail in the sky.''}
        }} \\
        
        \includegraphics[width=0.49\columnwidth]{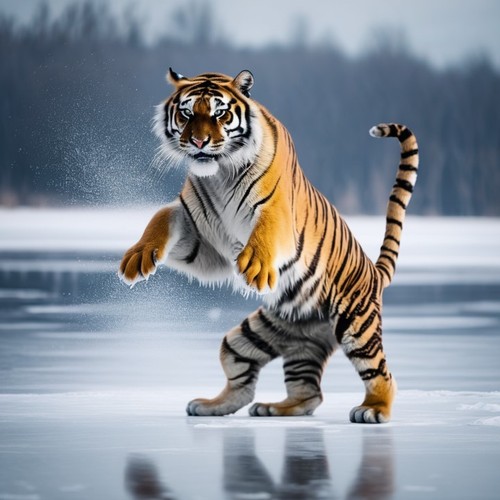} &
        \includegraphics[width=0.49\columnwidth]{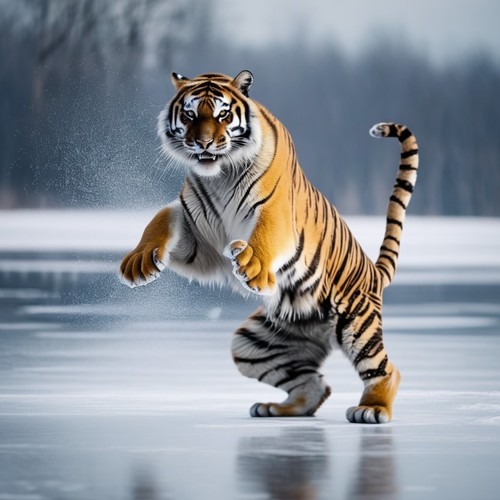} &
        \includegraphics[width=0.49\columnwidth]{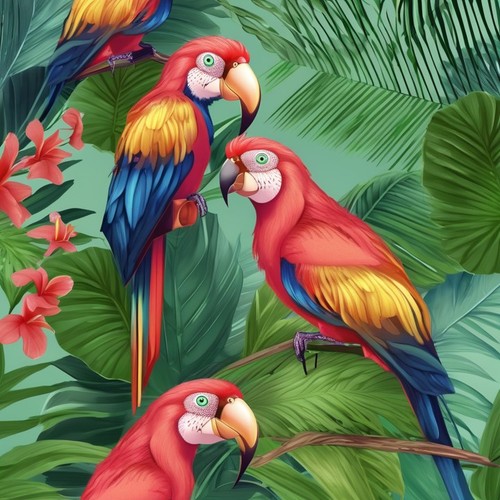} &
        \includegraphics[width=0.49\columnwidth]{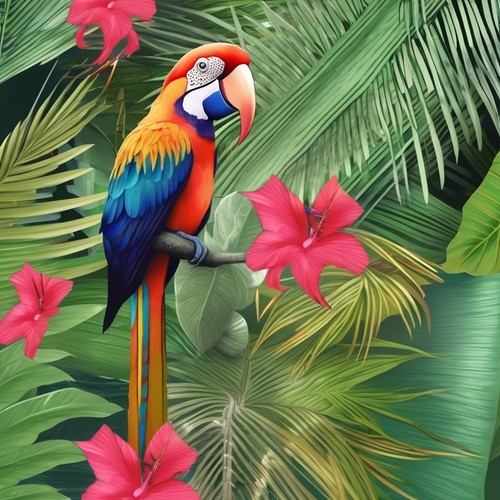} \\[-3pt]
        
        \multicolumn{2}{c}{\small \shortstack{
            \textit{``A tiger dancing on a frozen lake.''}
        }} &
        \multicolumn{2}{c}{\small \shortstack{
            \textit{``A tropical bird.''}
        }} \\

        \includegraphics[width=0.49\columnwidth]{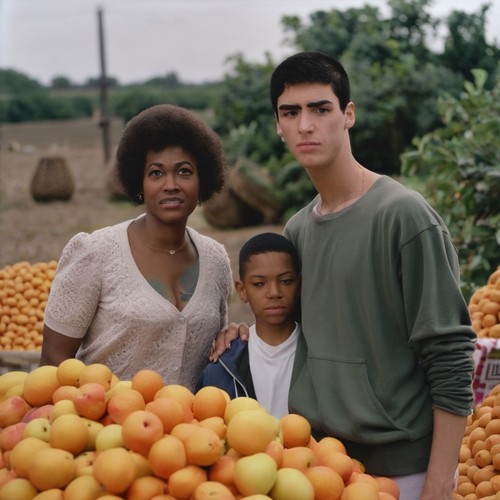} &
        \includegraphics[width=0.49\columnwidth]{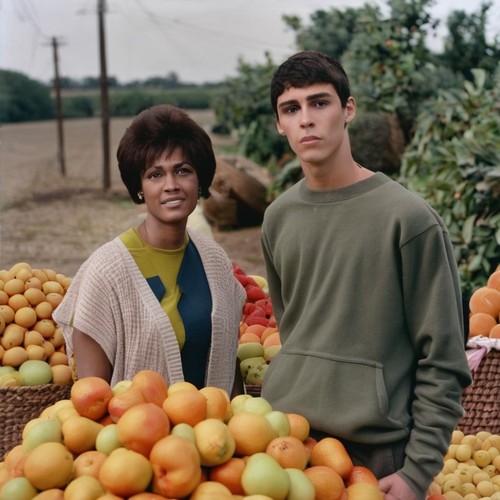} &
        \includegraphics[width=0.49\columnwidth]{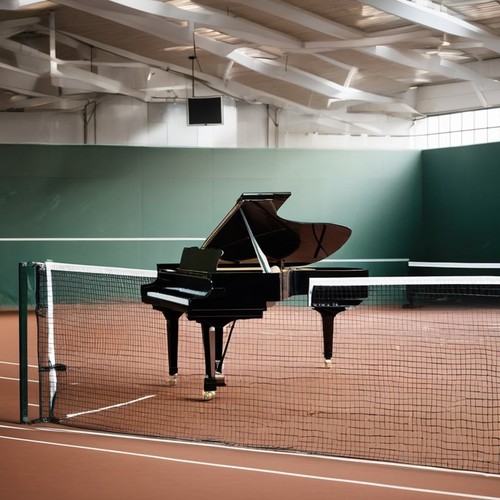} &
        \includegraphics[width=0.49\columnwidth]{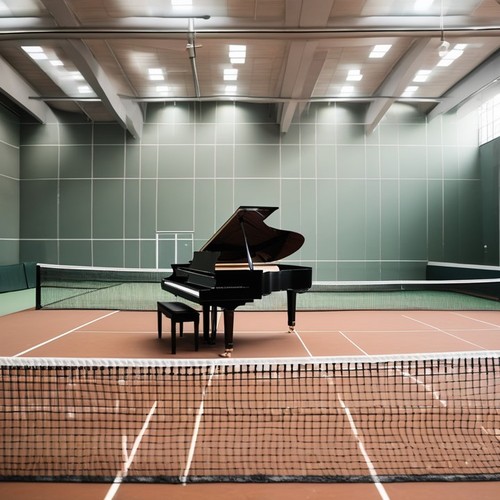} \\[-3pt]
        
        \multicolumn{2}{c}{\small \shortstack{
            \textit{``A woman standing next to \textcolor{red}{\textbf{a young man}} near a pile of fruit.''}
        }} &
        \multicolumn{2}{c}{\small \shortstack{
            \textit{``A grand piano next to the net of a tennis court.''}
        }} \\
    \end{tabular}

        \caption{Qualitative comparison of our Annealing method $\lambda=0.4$ (right) vs.\ CFG $w=10$ (left).}
    \label{fig:extra_3}
\end{figure*}
\begin{figure*}[t]
    \centering
    \begin{tabular}{@{}c@{\hspace{0.005\textwidth}}c@{\hspace{0.03\textwidth}}c@{\hspace{0.005\textwidth}}c@{}}

        CFG++ ($w=0.8$) & \textbf{Annealing ($\lambda=0.4$)} & CFG++ ($w=0.8$) & \textbf{Annealing ($\lambda=0.4$)} \\
        \includegraphics[width=0.49\columnwidth]{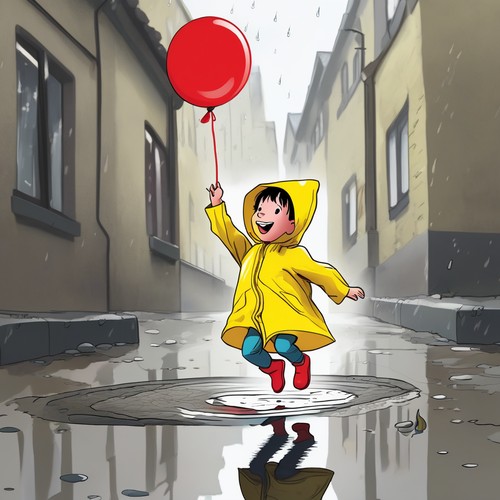} &
        \includegraphics[width=0.49\columnwidth]{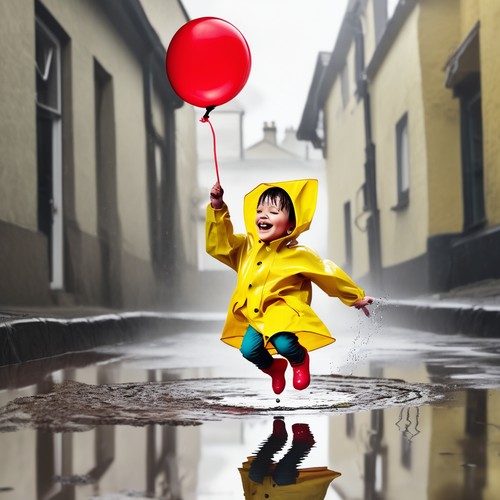} &
        \includegraphics[width=0.49\columnwidth]{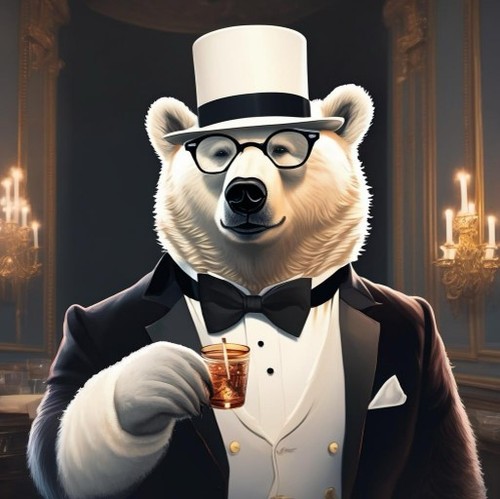} &
        \includegraphics[width=0.49\columnwidth]{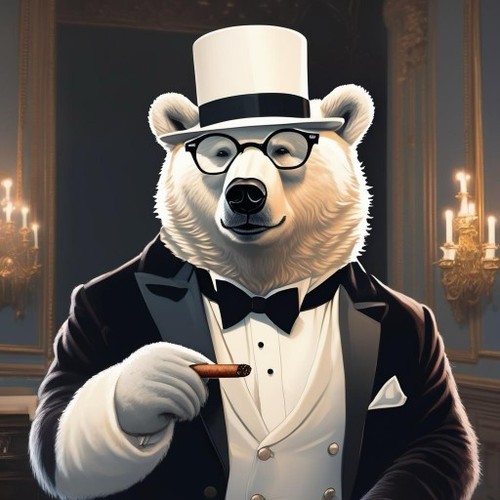} \\
                \multicolumn{2}{c}{\small \shortstack{
            \textit{``A child in a yellow raincoat jumping into a puddle,}\\
            \textit{``holding a red balloon.''}
        }} &
        \multicolumn{2}{c}{\small \shortstack{
            \textit{``A white bear in glasses, wearing tuxedo, glowing hat,} \\
            \textit{``and with \textcolor{red}{\textbf{cigare}} at the British queen reception.''}
        }} \\

        \includegraphics[width=0.49\columnwidth]{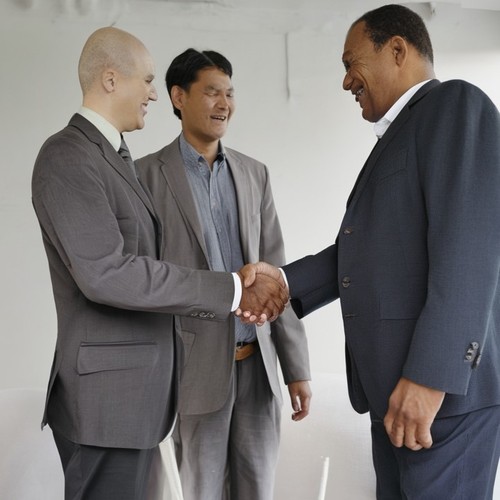} &
        \includegraphics[width=0.49\columnwidth]{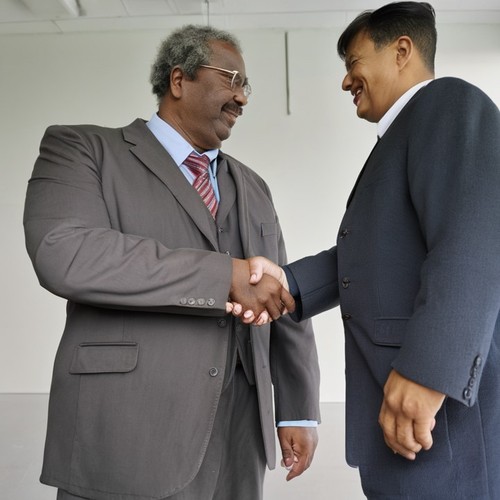} &
        \includegraphics[width=0.49\columnwidth]{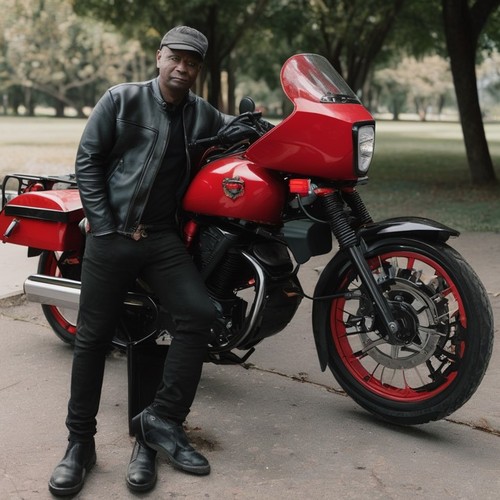} &
        \includegraphics[width=0.49\columnwidth]{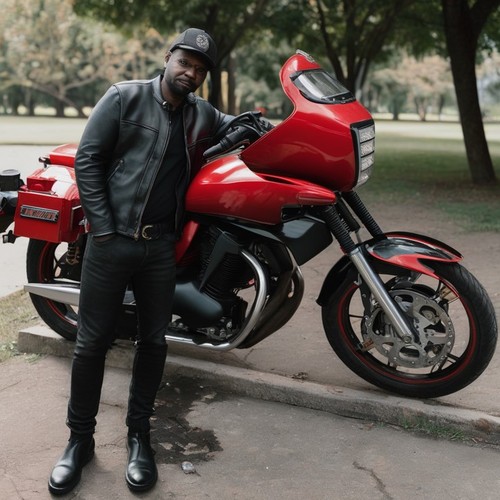} \\[-3pt]
        
        \multicolumn{2}{c}{\small \shortstack{
            \textit{``A man is shaking hands with another man.``} 
        }} &
        \multicolumn{2}{c}{\small \shortstack{
            \textit{``A man stands beside his black and red motorcycle near a park.''}
        }} \\

        \includegraphics[width=0.49\columnwidth]{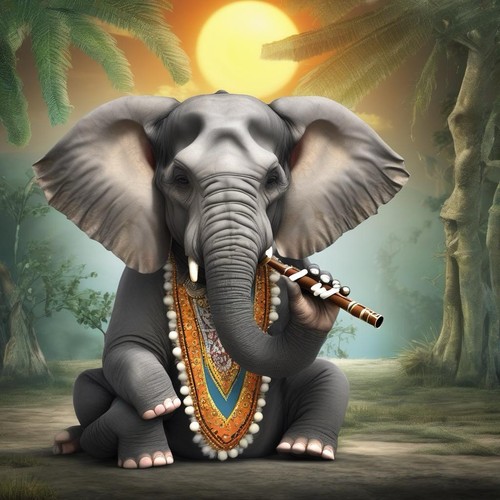} &
        \includegraphics[width=0.49\columnwidth]{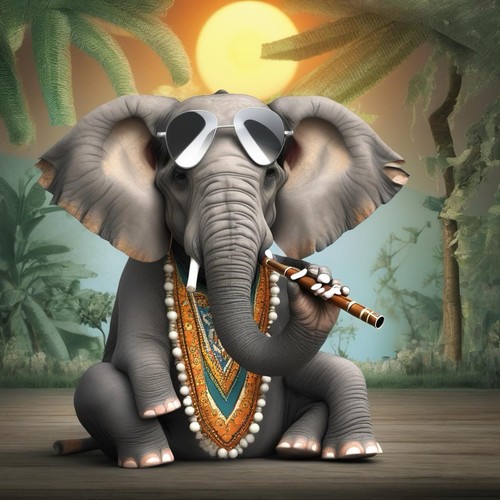} &

        \includegraphics[width=0.49\columnwidth]{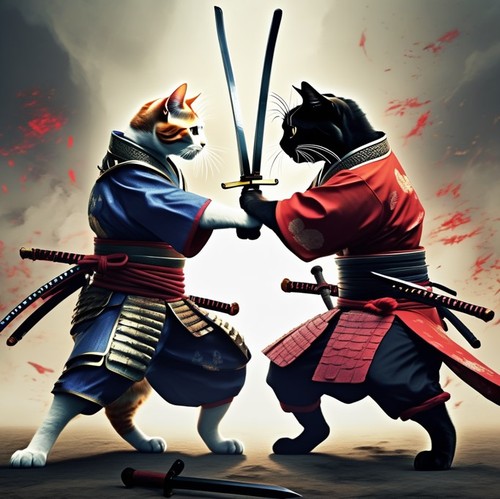} &
        \includegraphics[width=0.49\columnwidth]{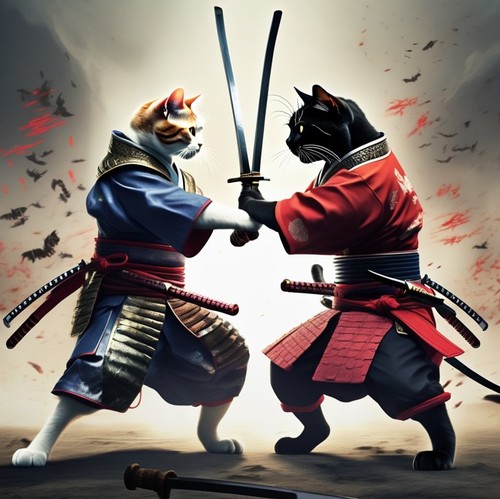} \\[-3pt]
        
        \multicolumn{2}{c}{\small \shortstack{
            \textit{``An elephant with \textcolor{red}{\textbf{sun glasses}} plays with a flute.`` } 
        }} &
        \multicolumn{2}{c}{\small \shortstack{
            \textit{``Two samurai cats... katanas drawn, petals swirling in the background.''}
        }} \\

        \includegraphics[width=0.49\columnwidth]{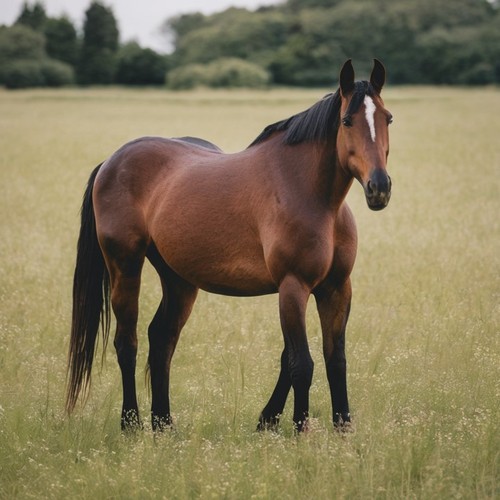} &
        \includegraphics[width=0.49\columnwidth]{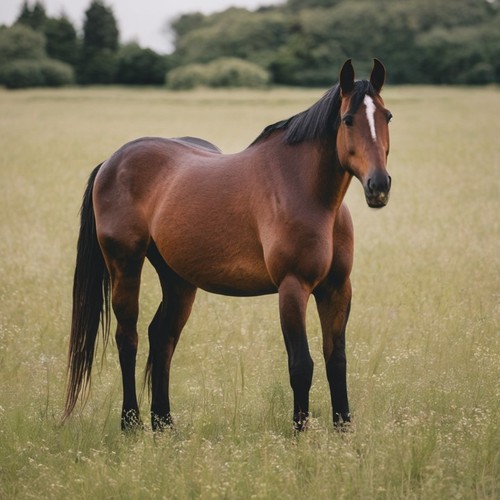} &
        \includegraphics[width=0.49\columnwidth]{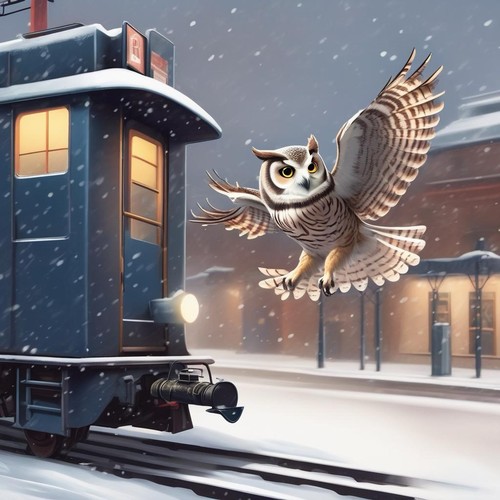} &
        \includegraphics[width=0.49\columnwidth]{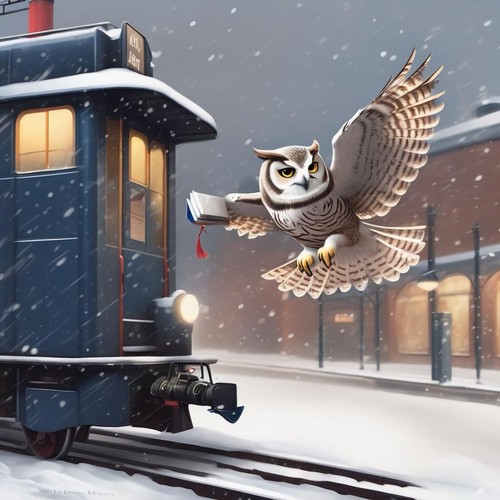} \\[-3pt]
        
        \multicolumn{2}{c}{\small \shortstack{
            \textit{``A horse in a field.''}
        }} &
        \multicolumn{2}{c}{\small \shortstack{
            \textit{``An owl delivering \textcolor{red}{\textbf{mail}} at a snowy train station.''}
        }} \\

        \includegraphics[width=0.49\columnwidth]{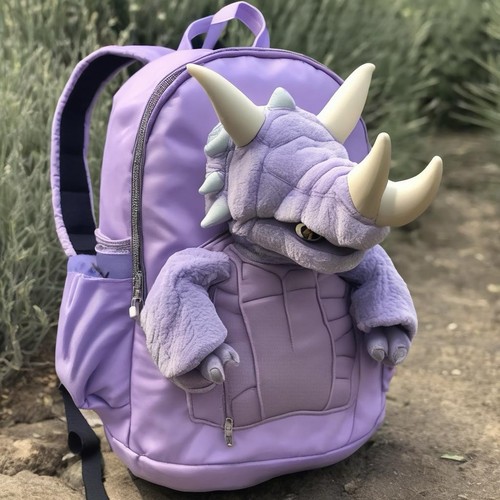} &
        \includegraphics[width=0.49\columnwidth]{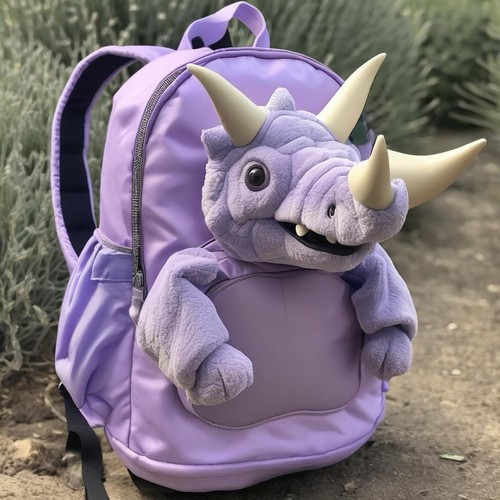} &
        \includegraphics[width=0.49\columnwidth]{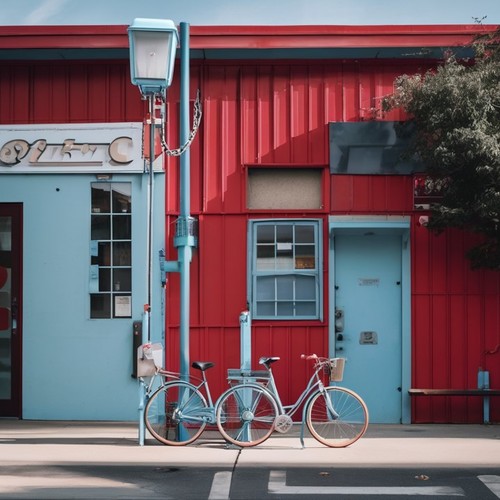} &
        \includegraphics[width=0.49\columnwidth]{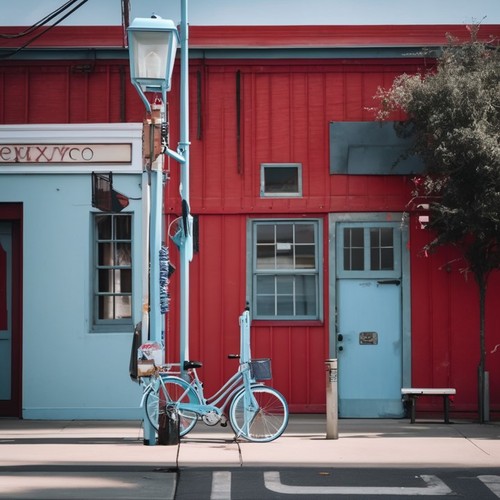} \\[-3pt]
        
        \multicolumn{2}{c}{\small \shortstack{
            \textit{``A lavender backpack with a triceratops stuffed animal head on top.''}
        }} &
        \multicolumn{2}{c}{\small \shortstack{
            \textit{``A light blue bicycle chained to a pole... in front of a red building.''}
        }} \\
    \end{tabular}

        \caption{Qualitative comparison of our Annealing method $\lambda=0.4$ (right) vs.\ CFG++ $w=0.8$ (left).}
    \label{fig:extra_cfgpp_2}
\end{figure*}
\begin{figure*}[t]
    \centering
    \begin{tabular}{@{}c@{\hspace{0.005\textwidth}}c@{\hspace{0.03\textwidth}}c@{\hspace{0.005\textwidth}}c@{}}

        CFG++ ($w=0.8$) & \textbf{Annealing ($\lambda=0.4$)} & CFG++ ($w=0.8$) & \textbf{Annealing ($\lambda=0.4$)} \\
        \includegraphics[width=0.49\columnwidth]{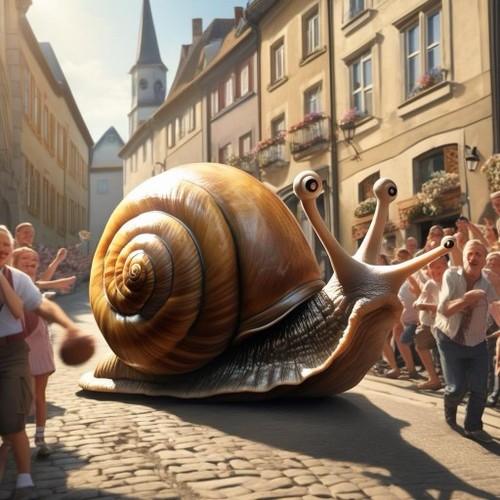} &
        \includegraphics[width=0.49\columnwidth]{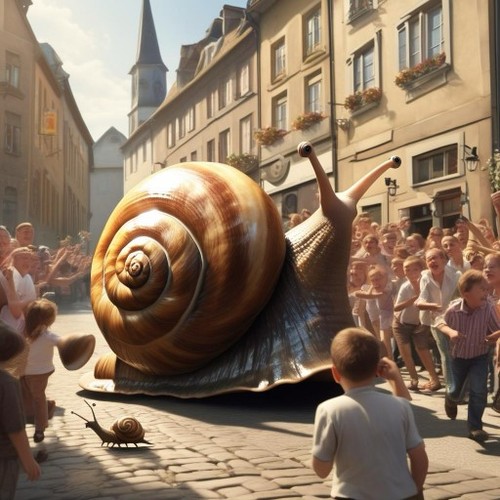} &
        \includegraphics[width=0.49\columnwidth]{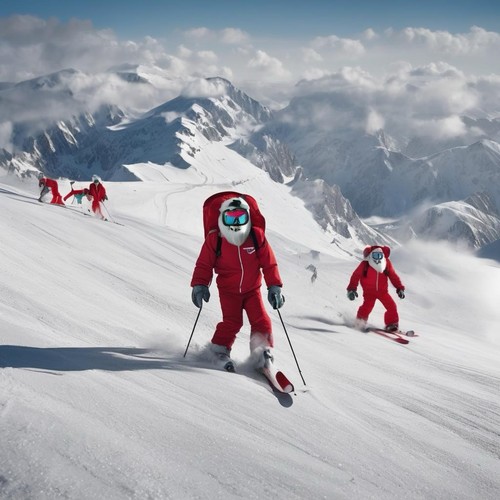} &
        \includegraphics[width=0.49\columnwidth]{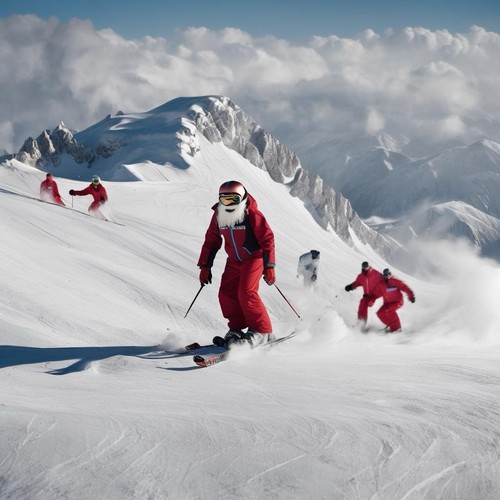} \\
                \multicolumn{2}{c}{\small \shortstack{
            \textit{``A giant snail race through the streets of an old European town,}\\
            \textit{onlookers cheering, mid-day sun.''}
        }} &
        \multicolumn{2}{c}{\small \shortstack{
            \textit{``Photo of alpines ski resort with yeti instead of humans.} \\
            \textit{``It wears a \textcolor{red}{\textbf{red helmet}}.''}
        }} \\

        \includegraphics[width=0.49\columnwidth]{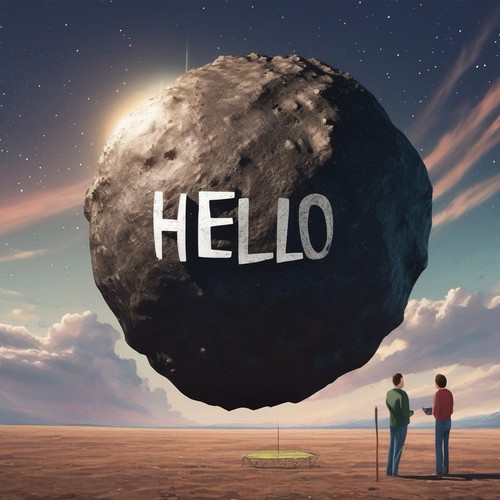} &
        \includegraphics[width=0.49\columnwidth]{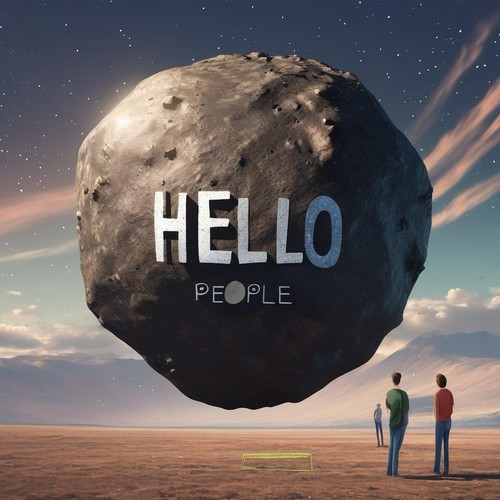} &
        \includegraphics[width=0.49\columnwidth]{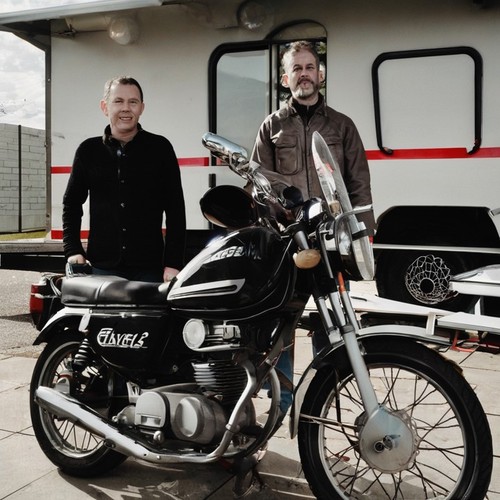} &
        \includegraphics[width=0.49\columnwidth]{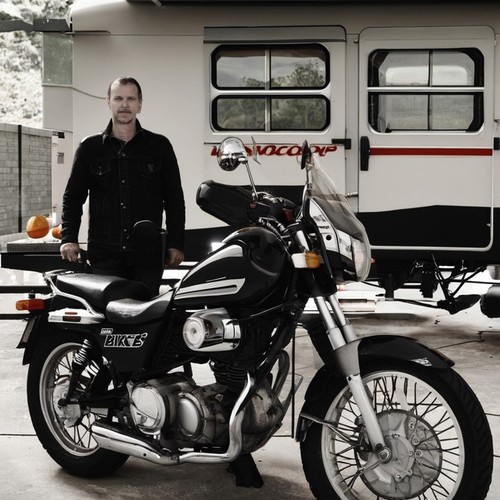} \\[-3pt]
        
        \multicolumn{2}{c}{\small \shortstack{
            \textit{``A giant meteorite with the words \textcolor{red}{\textbf{'hello people'}} approaching the earth.``} 
        }} &
        \multicolumn{2}{c}{\small \shortstack{
            \textit{`` \textcolor{red}{\textbf{A man}} standing next to a bikes and a motorcycle.''}
        }} \\

        \includegraphics[width=0.49\columnwidth]{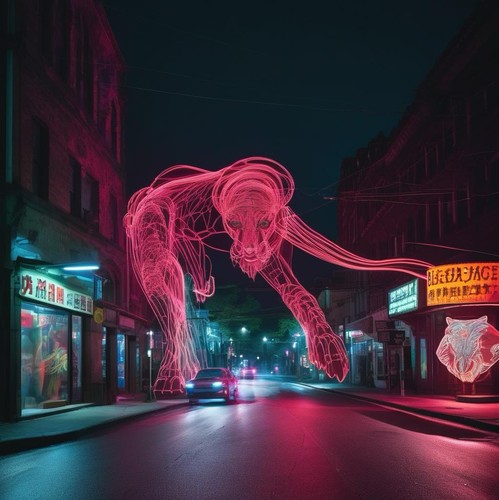} &
        \includegraphics[width=0.49\columnwidth]{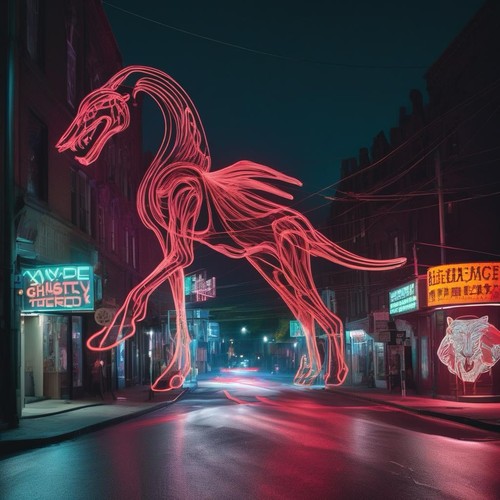} &

        \includegraphics[width=0.49\columnwidth]{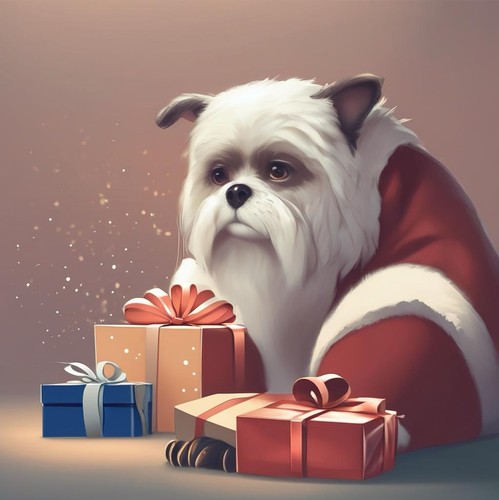} &
        \includegraphics[width=0.49\columnwidth]{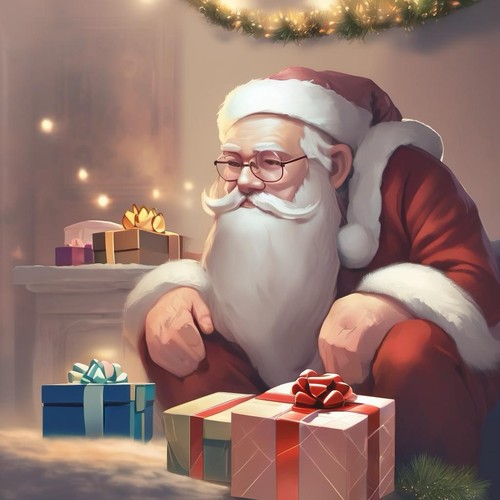} \\[-3pt]
        
        \multicolumn{2}{c}{\small \shortstack{
            \textit{``Long-exposure night shot of neon... huge ghostly animal.`` } 
        }} &
        \multicolumn{2}{c}{\small \shortstack{
            \textit{``A present.''}
        }} \\

        \includegraphics[width=0.49\columnwidth]{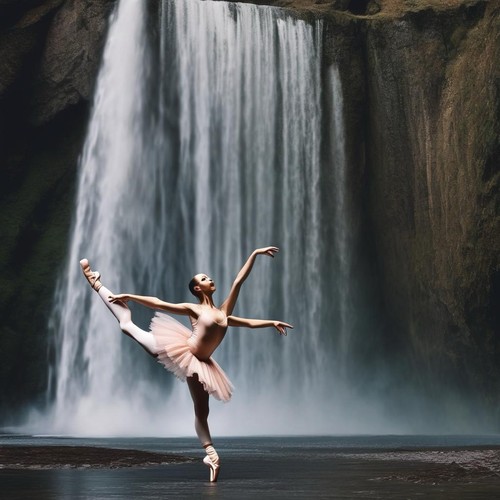} &
        \includegraphics[width=0.49\columnwidth]{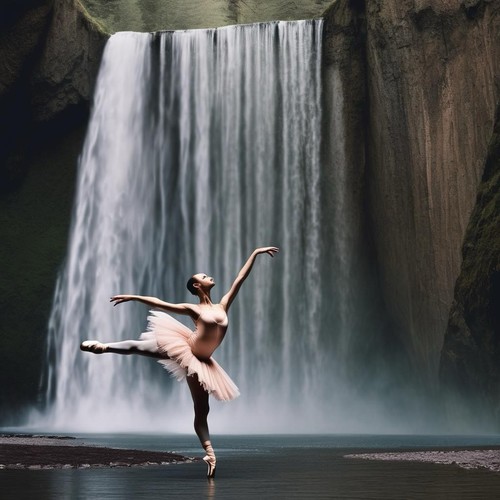} &
        \includegraphics[width=0.49\columnwidth]{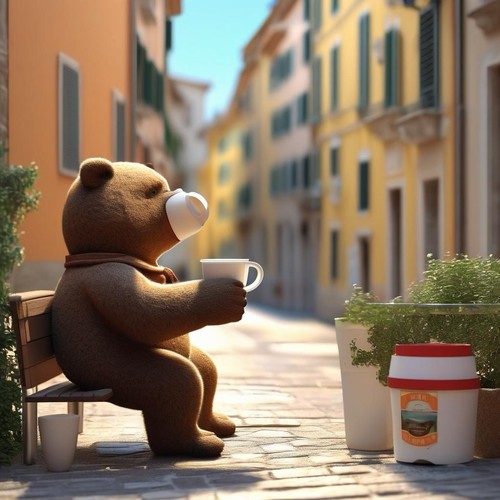} &
        \includegraphics[width=0.49\columnwidth]{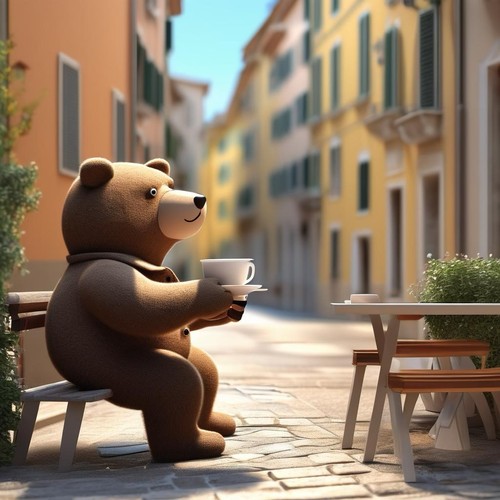} \\[-3pt]
        
        \multicolumn{2}{c}{\small \shortstack{
            \textit{``A ballet dancer next to a waterfall.''}
        }} &
        \multicolumn{2}{c}{\small \shortstack{
            \textit{``Bear drinking coffee on a sunny morning street, in Italy.''}
        }} \\

        \includegraphics[width=0.49\columnwidth]{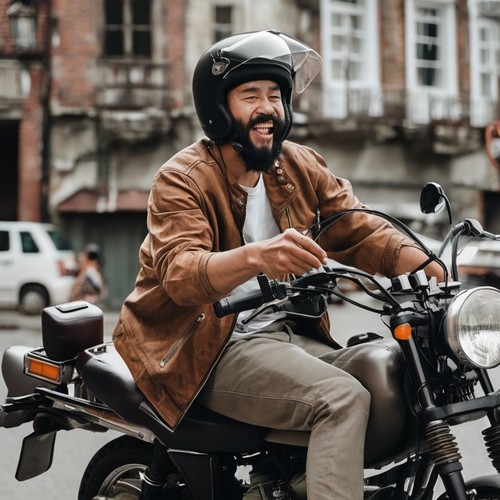} &
        \includegraphics[width=0.49\columnwidth]{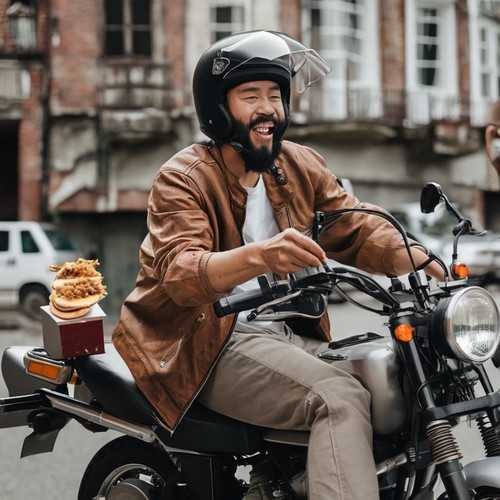} &
        \includegraphics[width=0.49\columnwidth]{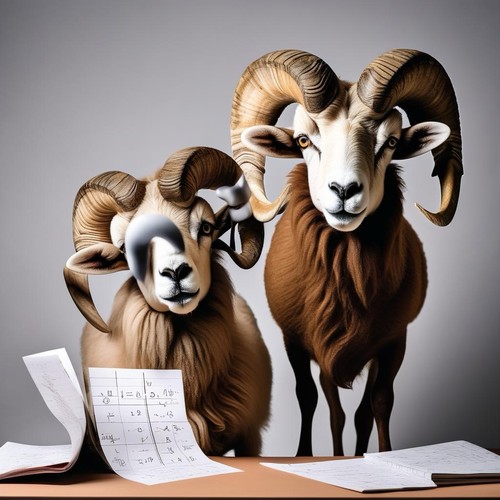} &
        \includegraphics[width=0.49\columnwidth]{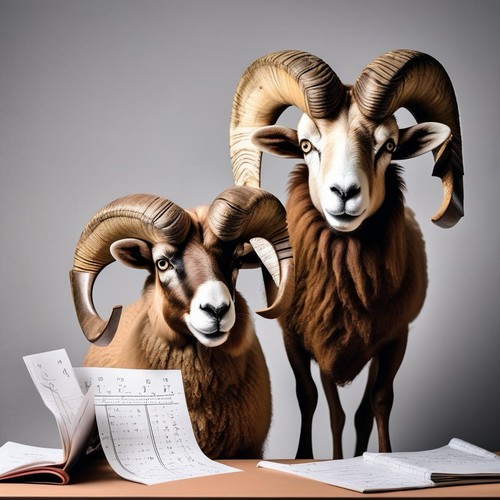} \\[-3pt]
        
        \multicolumn{2}{c}{\small \shortstack{
            \textit{``A man riding a motorcycle while \textcolor{red}{\textbf{eating food}}.''}
        }} &
        \multicolumn{2}{c}{\small \shortstack{
            \textit{``Two rams trying to solve math equation.''}
        }} \\
    \end{tabular}

        \caption{Qualitative comparison of our Annealing method $\lambda=0.4$ (right) vs.\ CFG++ $w=0.8$ (left).}
    \label{fig:extra_cfgpp_3}
\end{figure*}
\begin{figure*}[t]
    \centering
    \begin{tabular}{@{}c@{\hspace{0.005\textwidth}}c@{\hspace{0.03\textwidth}}c@{\hspace{0.005\textwidth}}c@{}}

        CFG ($w=12.5$) & \textbf{Annealing ($\lambda=0.7$)} & CFG++ ($w=1.0$) & \textbf{Annealing ($\lambda=0.7$)} \\
        \includegraphics[width=0.49\columnwidth]{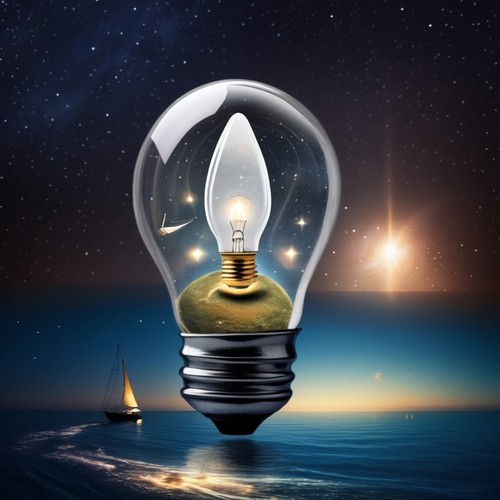} &
        \includegraphics[width=0.49\columnwidth]{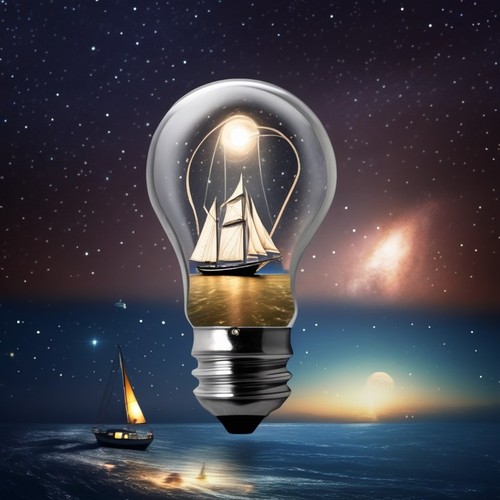} &
        \includegraphics[width=0.49\columnwidth]{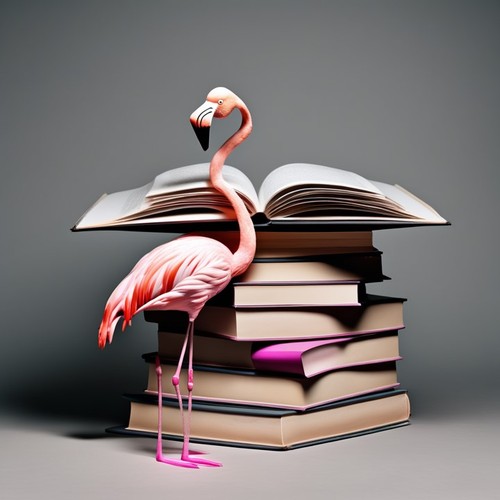} &
        \includegraphics[width=0.49\columnwidth]{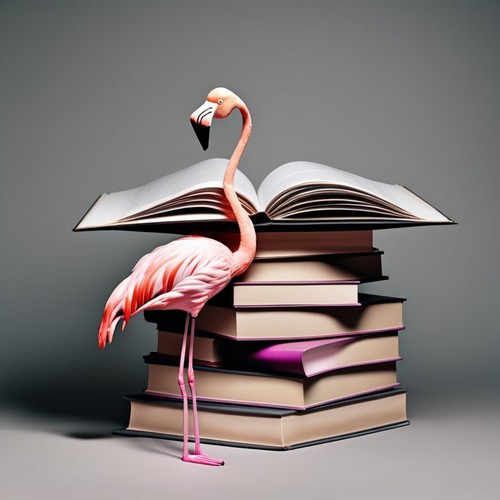} \\

        \multicolumn{2}{c}{\small \shortstack{
            \textit{``A photo of a light bulb in outer space traveling the galaxy} \\ { with a \textcolor{red}{\textbf{sailing boat inside the light bulb}}.''} }} &
        \multicolumn{2}{c}{\small \shortstack{
            \textit{``a real flamingo reading a large open book.} \\ {a big stack of books is piled up next to it.''}}} \\

        \includegraphics[width=0.49\columnwidth]{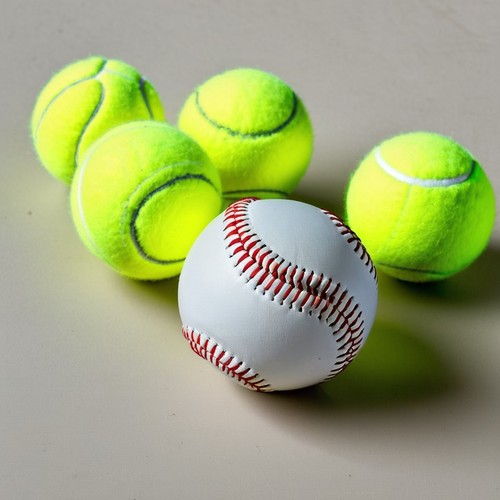} &
        \includegraphics[width=0.49\columnwidth]{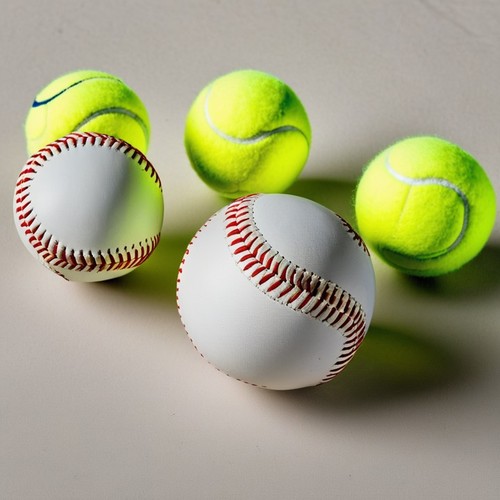} &
        \includegraphics[width=0.49\columnwidth]{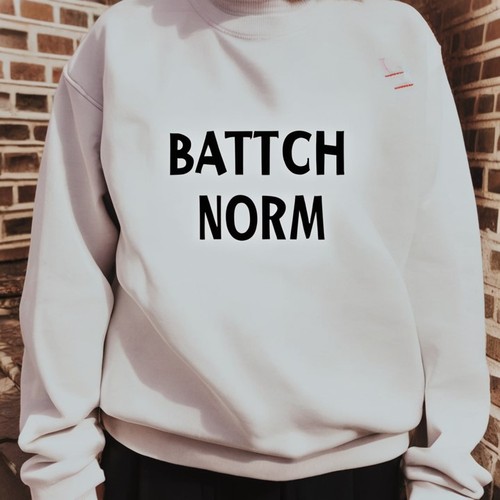} &
        \includegraphics[width=0.49\columnwidth]{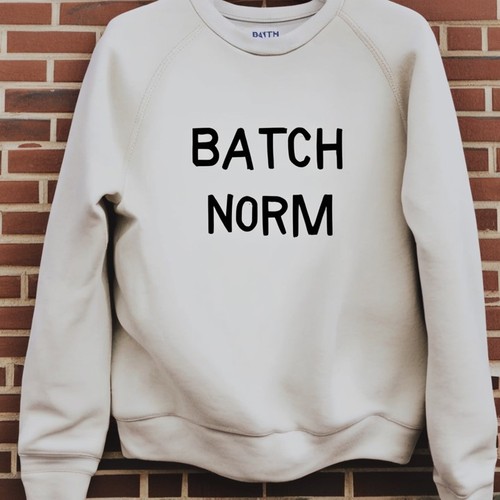} \\[-3pt]
        
        \multicolumn{2}{c}{\small \shortstack{
            \textit{``\textcolor{red}{\textbf{Two baseballs}} next to three tennis balls.''}
        }} &
         \multicolumn{2}{c}{\small \shortstack{
            \textit{``A sweatshirt with \textcolor{red}{\textbf{'Batch Norm'}} written on it.''}
        }}\\
        
        \includegraphics[width=0.49\columnwidth]{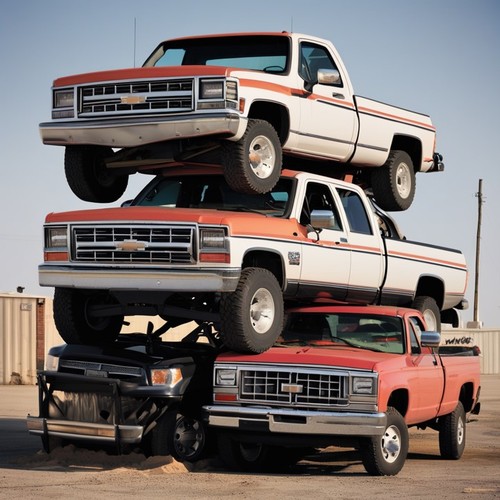} &
        \includegraphics[width=0.49\columnwidth]{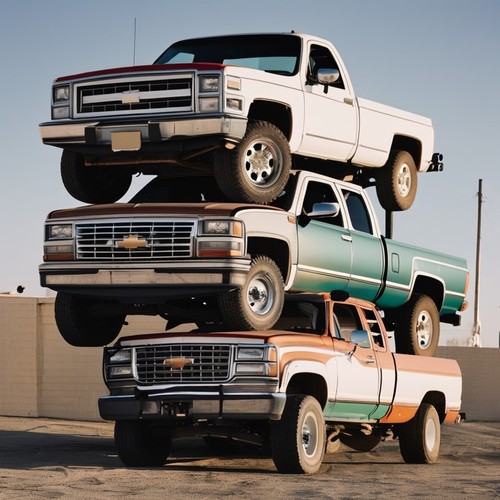} &
        \includegraphics[width=0.49\columnwidth]{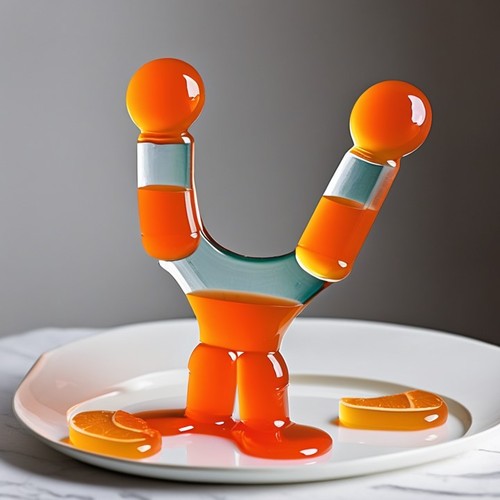} &
        \includegraphics[width=0.49\columnwidth]{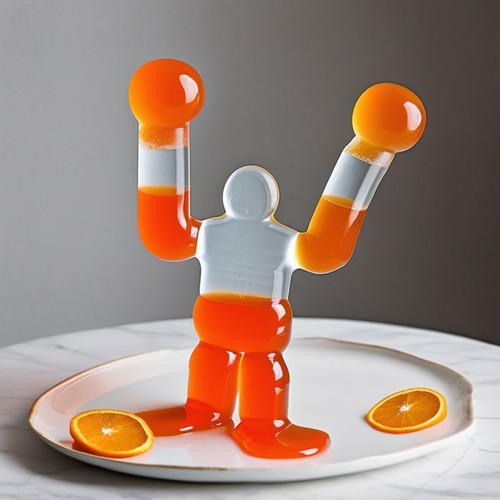} \\[-3pt]

        \multicolumn{2}{c}{\small \shortstack{
            \textit{``\textcolor{red}{\textbf{Three}} pickup trucks piled on top of each other.''}
        }} &
        \multicolumn{2}{c}{\small \shortstack{
            \textit{``Orange jello in the \textcolor{red}{\textbf{shape of a man}}.''}
        }} \\

        \includegraphics[width=0.49\columnwidth]{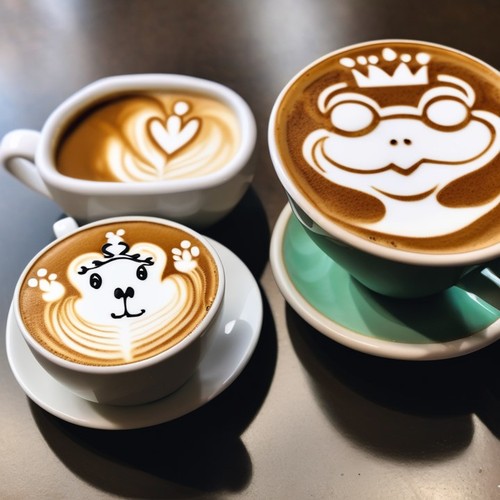} &
        \includegraphics[width=0.49\columnwidth]{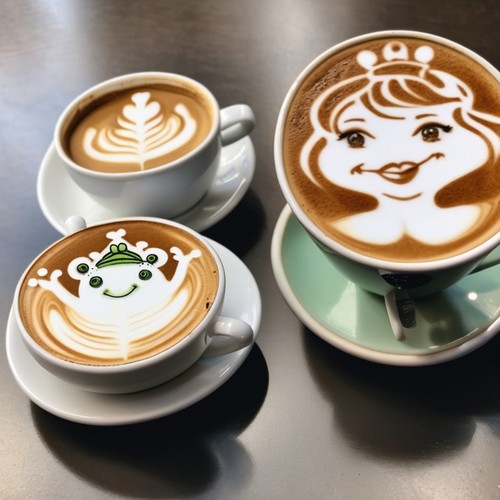} &
        \includegraphics[width=0.49\columnwidth]{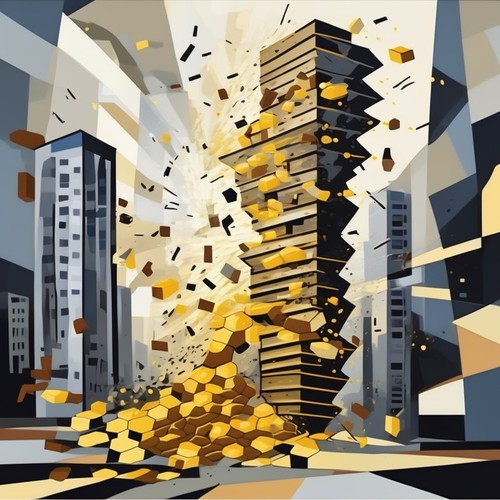} &
        \includegraphics[width=0.49\columnwidth]{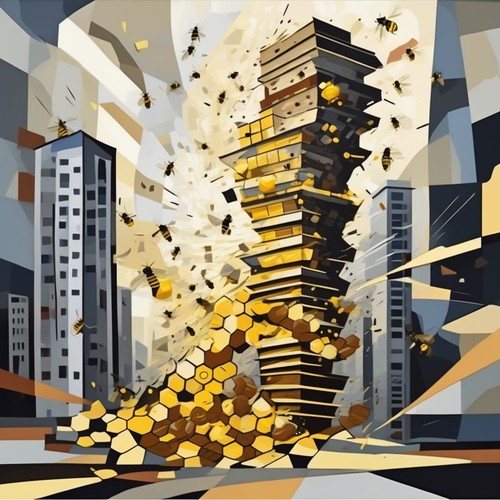} \\[-3pt]
        
        \multicolumn{2}{c}{\small \shortstack{
            \textit{``...cups of coffee... latte art of a \textcolor{red}{\textbf{lovely princess}}... latte art of a \textcolor{red}{\textbf{frog}}.''}
        }} &
        \multicolumn{2}{c}{\small \shortstack{
            \textit{``A tornado made of \textcolor{red}{\textbf{bees}} crashing into a skyscraper... abstract cubism.''}
        }} \\

        \includegraphics[width=0.49\columnwidth]{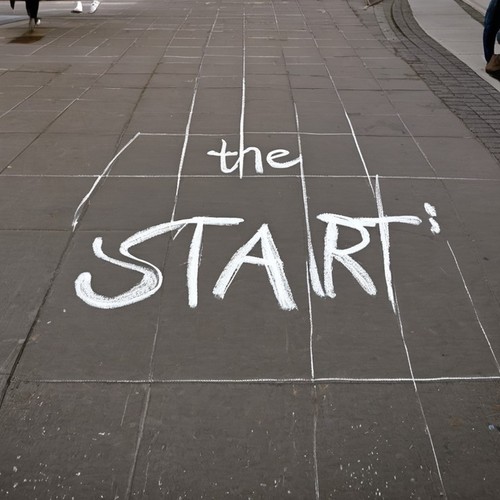} &
        \includegraphics[width=0.49\columnwidth]{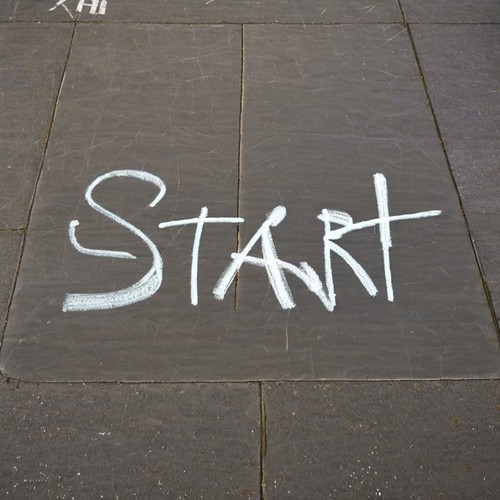} &
        \includegraphics[width=0.49\columnwidth]{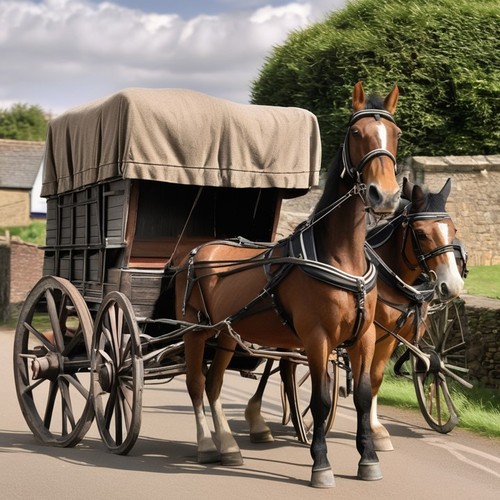} &
        \includegraphics[width=0.49\columnwidth]{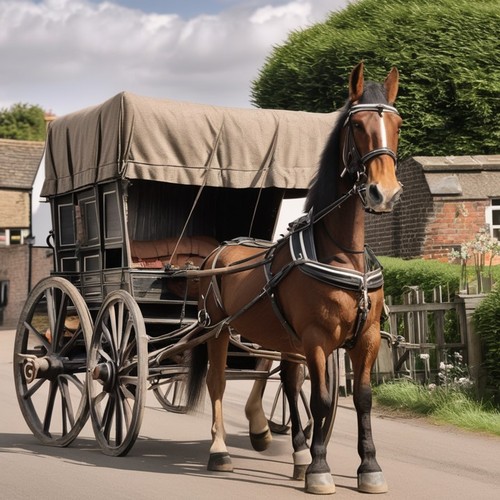} \\[-3pt]
        
        \multicolumn{2}{c}{\small \shortstack{
            \textit{``The word \textcolor{red}{\textbf{'START'}} written in chalk on a sidewalk
''}}} &
        \multicolumn{2}{c}{\small \shortstack{
            \textit{``A thumbnail image of \textcolor{red}{\textbf{a horse}} and cart.''}
        }} 
    \end{tabular}
    \caption{\textbf{Qualitative comparison.} Columns (L→R): CFG ($w{=}12.5$) → Ours ($\lambda{=}0.7$); CFG++ ($w{=}1.0$) → Ours ($\lambda{=}0.7$).}
    \label{fig:extra_l_0_7}
\end{figure*}

\begin{figure*}[t]
    \centering
    \begin{tabular}{@{}c@{\hspace{0.005\textwidth}}c@{\hspace{0.03\textwidth}}c@{\hspace{0.005\textwidth}}c@{}}

        CFG ($w=15$) & \textbf{Annealing ($\lambda=0.8$)} & CFG++ ($w=1.2$) & \textbf{Annealing ($\lambda=0.8$)} \\
        \includegraphics[width=0.49\columnwidth]{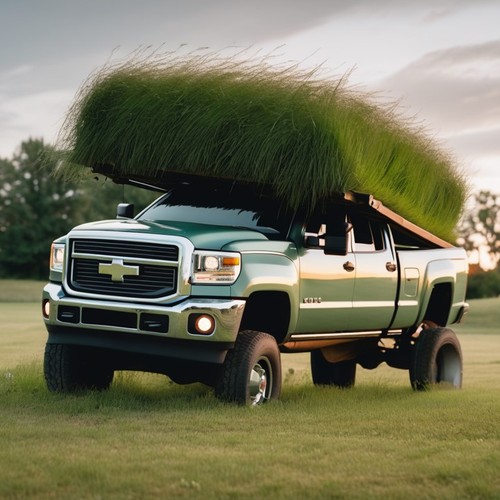} &
        \includegraphics[width=0.49\columnwidth]{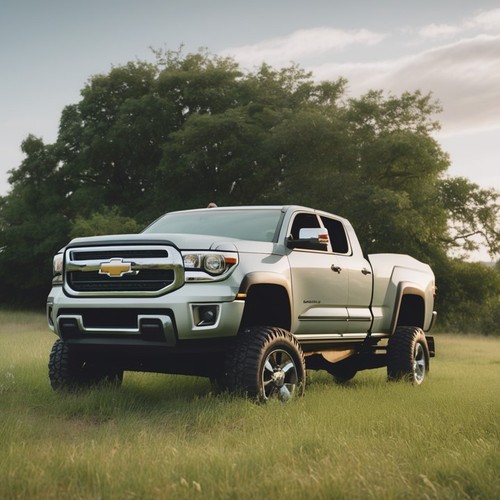} &
        \includegraphics[width=0.49\columnwidth]{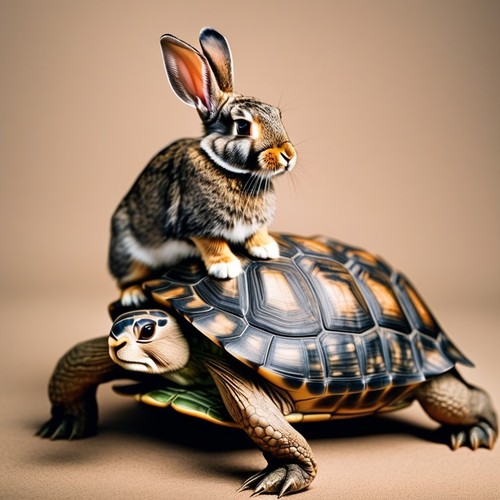} &
        \includegraphics[width=0.49\columnwidth]{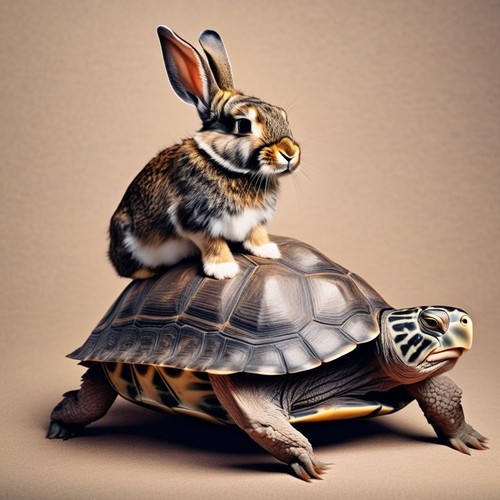} \\[-3pt]

        \multicolumn{2}{c}{\small \shortstack{
            \textit{``A pick-up truck rolling over a grassy field.''} }} &
        \multicolumn{2}{c}{\small \shortstack{
            \textit{`A rabbit sitting on a \textcolor{red}{\textbf{turtle's}} back.''}}} \\

        \includegraphics[width=0.49\columnwidth]{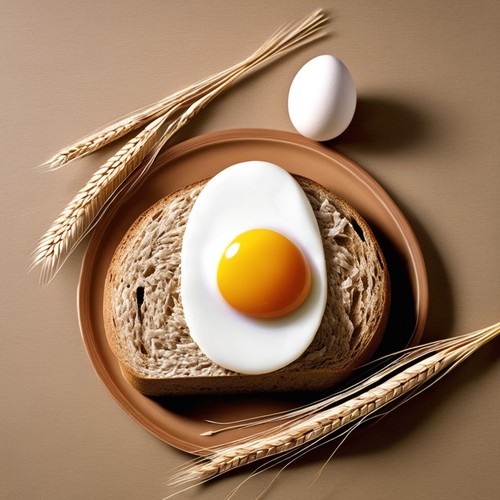} &
        \includegraphics[width=0.49\columnwidth]{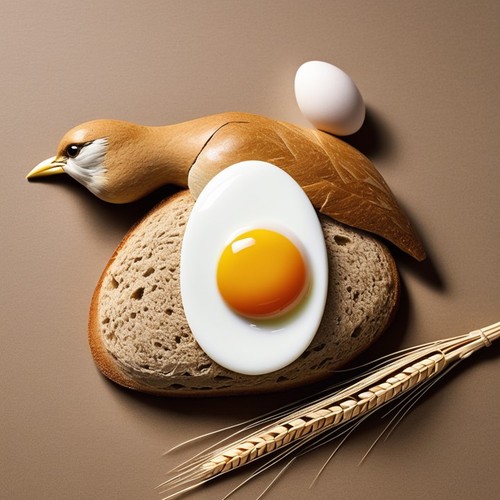} &
        \includegraphics[width=0.49\columnwidth]{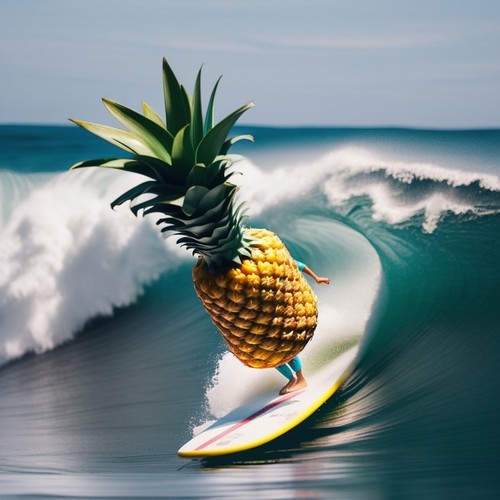} &
        \includegraphics[width=0.49\columnwidth]{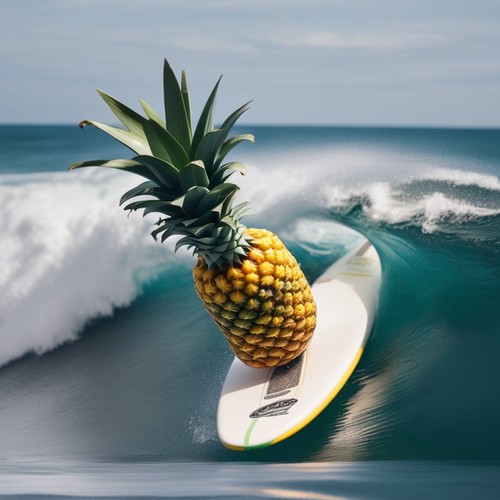} \\[-3pt]
        
        \multicolumn{2}{c}{\small \shortstack{
            \textit{``A photograph of a \textcolor{red}{\textbf{bird}} made of wheat bread and an egg.''}
        }} &
         \multicolumn{2}{c}{\small \shortstack{
            \textit{``A pineapple surfing on a wave.''}
        }}\\
        
        \includegraphics[width=0.49\columnwidth]{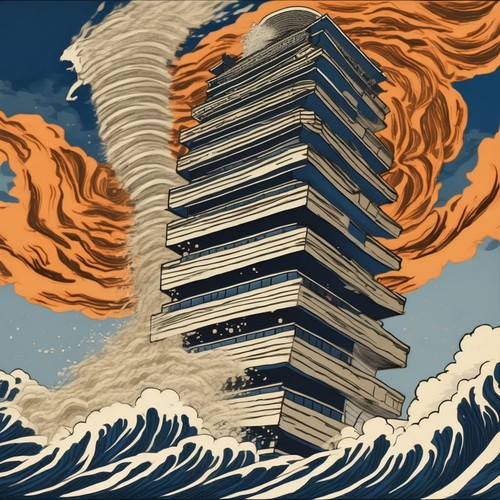} &
        \includegraphics[width=0.49\columnwidth]{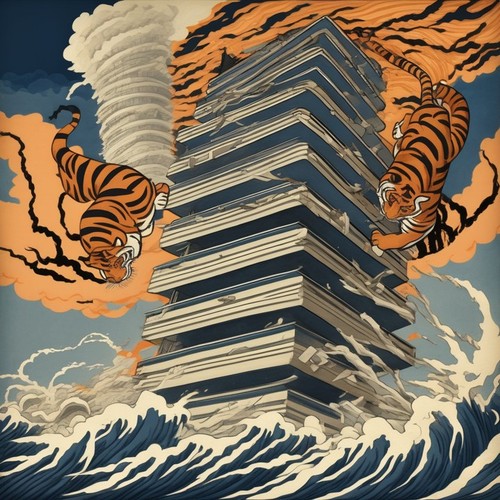} &
        \includegraphics[width=0.49\columnwidth]{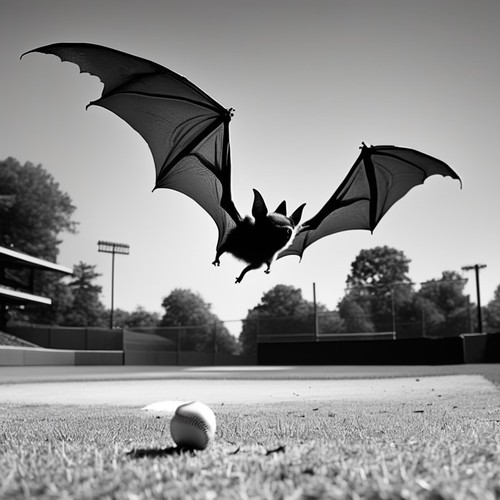} &
        \includegraphics[width=0.49\columnwidth]{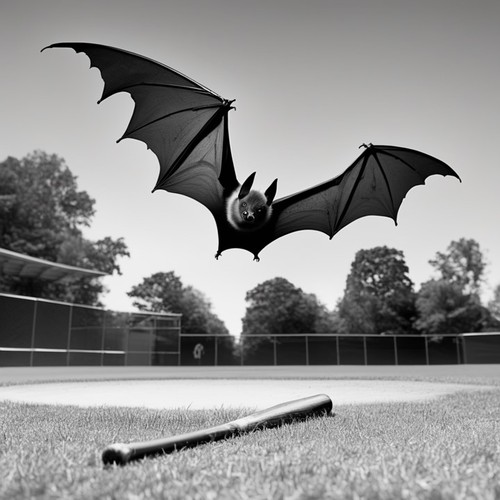} \\[-3pt]

        \multicolumn{2}{c}{\small \shortstack{
            \textit{``A tornado made of \textcolor{red}{\textbf{tigers}} crashing into a skyscraper... style of Hokusai.''}
        }} &
        \multicolumn{2}{c}{\small \shortstack{
            \textit{``A bat landing on a \textcolor{red}{\textbf{baseball bat}}.''}
        }} \\

        \includegraphics[width=0.49\columnwidth]{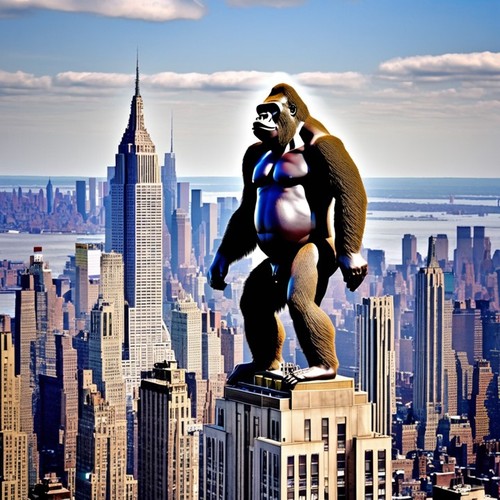} &
        \includegraphics[width=0.49\columnwidth]{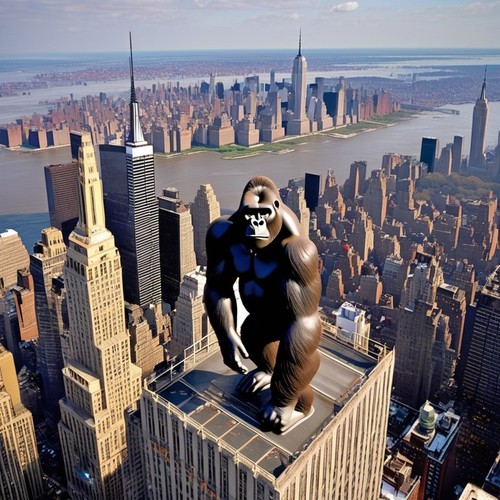} &
        \includegraphics[width=0.49\columnwidth]{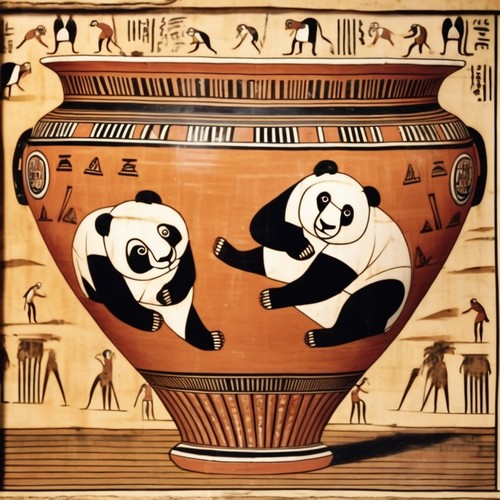} &
        \includegraphics[width=0.49\columnwidth]{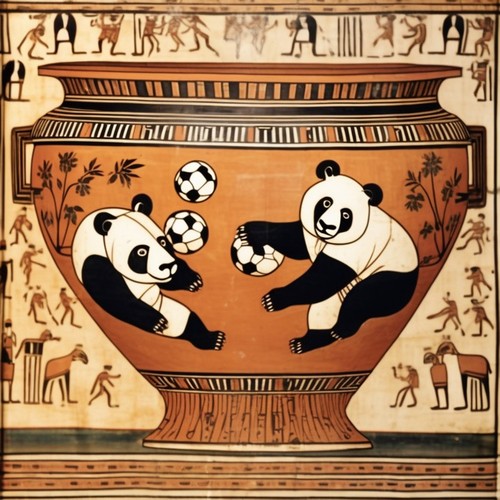} \\[-2pt]
        
        \multicolumn{2}{c}{\small \shortstack{
            \textit{`A giant gorilla at the top of the Empire State Building.''} \\ {\phantom{a}}
        }} &
        \multicolumn{2}{c}{\small \shortstack{
            \textit{``A photo of an Athenian vase with a painting of pandas}\\{\textcolor{red}{\textbf{playing soccer}} in the style of Egyptian hieroglyphics.''}
        }} \\

        \includegraphics[width=0.49\columnwidth]{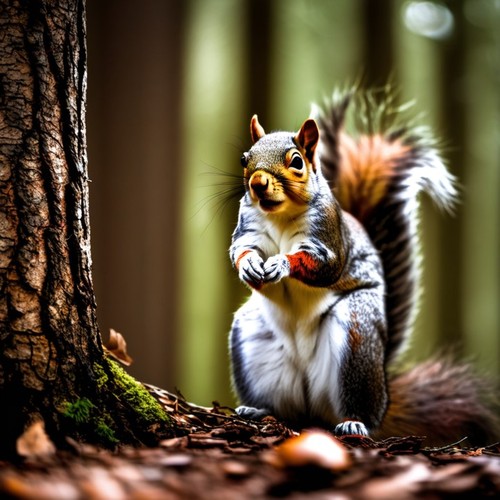} &
        \includegraphics[width=0.49\columnwidth]{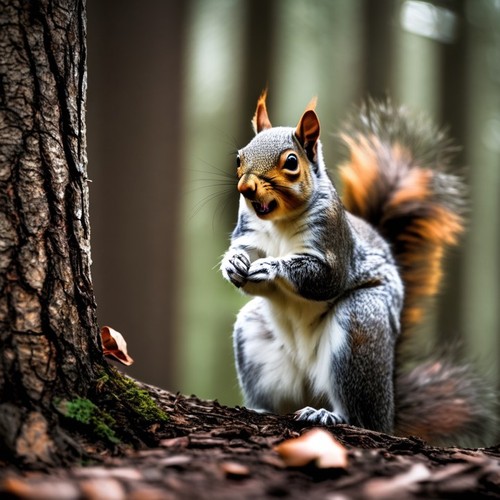} &
        \includegraphics[width=0.49\columnwidth]{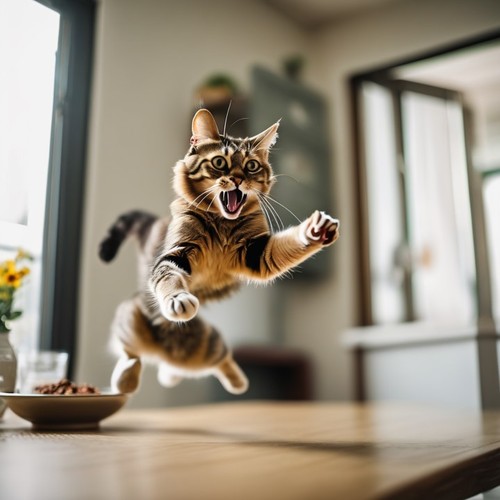} &
        \includegraphics[width=0.49\columnwidth]{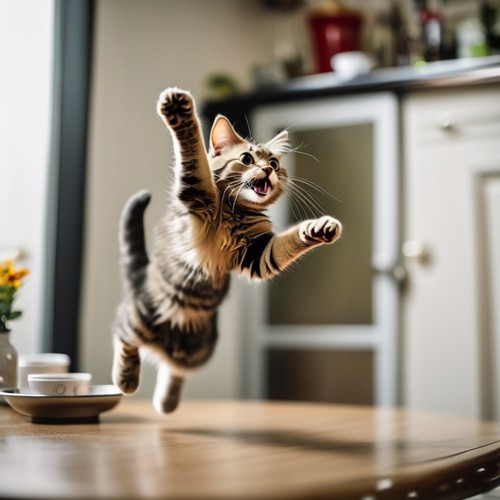} \\[-3pt]
        
        \multicolumn{2}{c}{\small \shortstack{
            \textit{``A squirrel in a forest.''}}} &
        \multicolumn{2}{c}{\small \shortstack{
            \textit{``A cat jumping in the air to get onto a table.''}
        }} 
    \end{tabular}
    \caption{\textbf{Qualitative comparison.} Columns (L→R): CFG ($w{=}15$) → Ours ($\lambda{=}0.8$); CFG++ ($w{=}1.2$) → Ours ($\lambda{=}0.8$).}
    \label{fig:extra_l_0_8}
\end{figure*}

\newlength{\imgH}\setlength{\imgH}{0.245\textwidth} %
\setlength{\tabcolsep}{1pt} %

\begin{figure*}[t]
\centering
\begin{tabular}{@{}c c c c c@{}}

\multirow{2}{*}{\rotatebox[origin=c]{90}{\large CFG\phantom{aaaaaaaaaaaaaa}}}
  & \smash{{\large $w{=}7.5$}} & \smash{{\large $w{=}10$}} & \smash{{\large $w{=}12.5$}} & \smash{{\large $w{=}15$}} \\
& \includegraphics[height=\imgH]{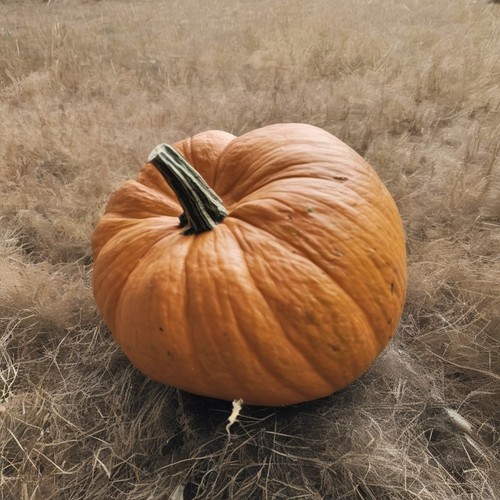} &
  \includegraphics[height=\imgH]{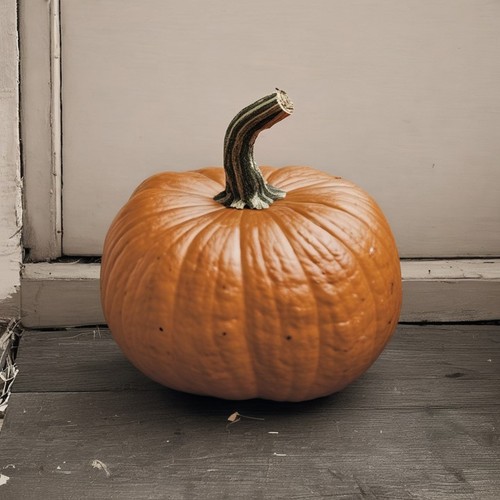} &
  \includegraphics[height=\imgH]{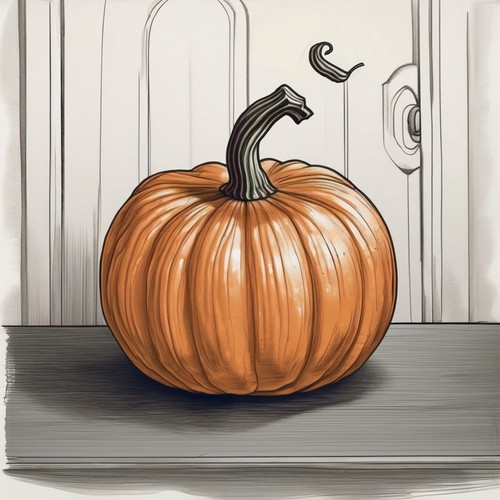} &
  \includegraphics[height=\imgH]{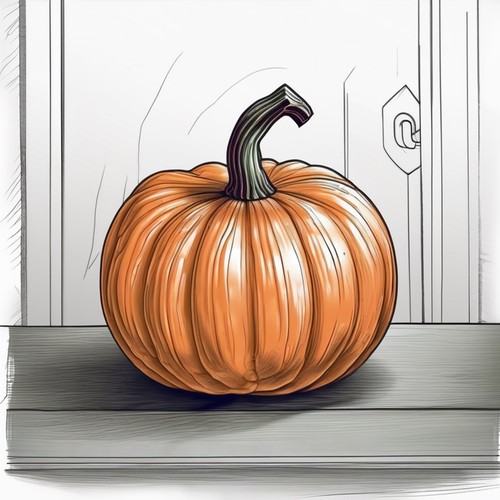} \\
[0pt]

\multirow{2}{*}{\rotatebox[origin=c]{90}{\large \textbf{Annealing}\phantom{aaaaaaaaaaa}}}
  & \smash{{\large $\lambda{=}0.05$}} & \smash{{\large $\lambda{=}0.4$}} & \smash{{\large $\lambda{=}0.7$}} & \smash{{\large $\lambda{=}0.8$}} \\
& \includegraphics[height=\imgH]{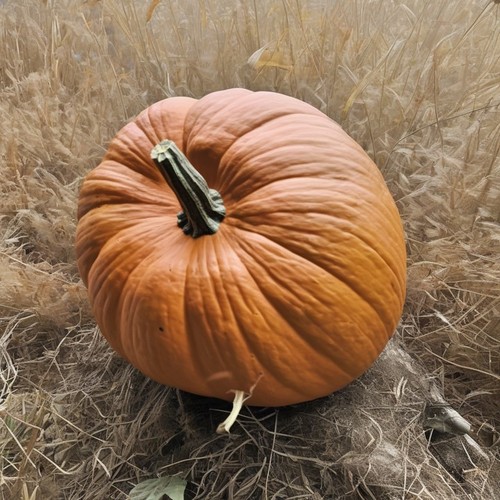} &
  \includegraphics[height=\imgH]{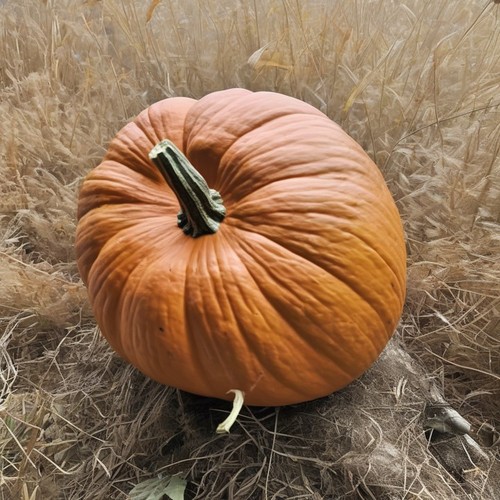} &
  \includegraphics[height=\imgH]{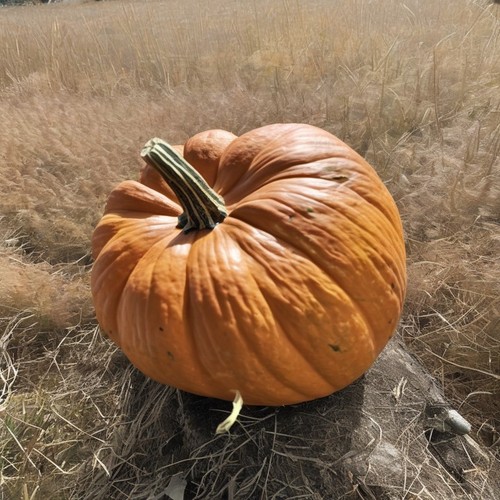} &
  \includegraphics[height=\imgH]{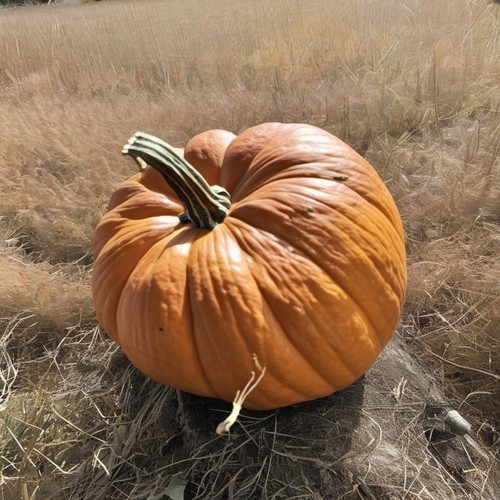} \\
[-2pt]

& \multicolumn{4}{c}{ \shortstack{\Large \textit{``A pumpkin.''}}} \\[8pt]

\multirow{2}{*}{\rotatebox[origin=c]{90}{\large CFG\phantom{aaaaaaaaaaaaaa}}}
  & \smash{{\large $w{=}7.5$}} & \smash{{\large $w{=}10$}} & \smash{{\large $w{=}12.5$}} & \smash{{\large $w{=}15$}} \\
& \includegraphics[height=\imgH]{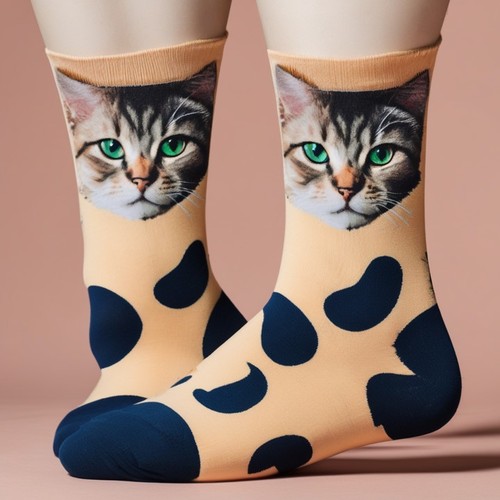} &
  \includegraphics[height=\imgH]{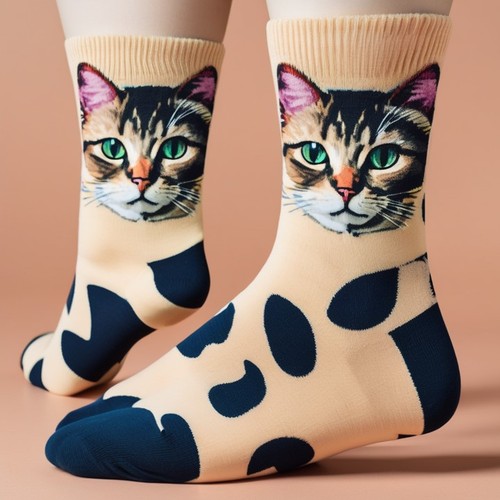} &
  \includegraphics[height=\imgH]{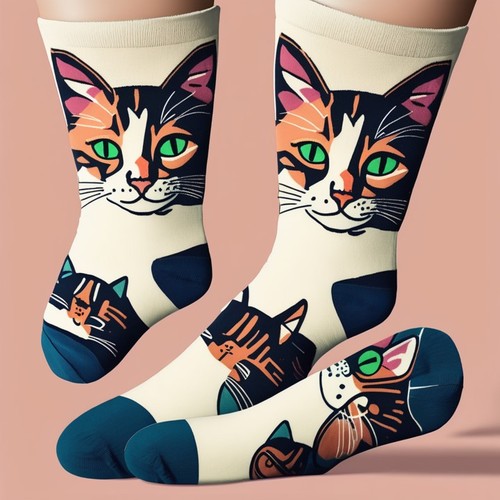} &
  \includegraphics[height=\imgH]{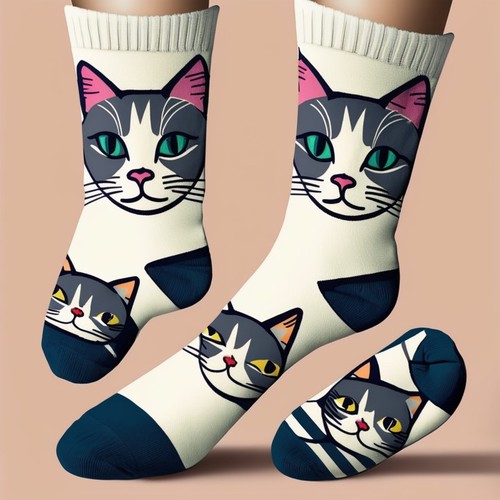} \\
[0pt]

\multirow{2}{*}{\rotatebox[origin=c]{90}{\large \textbf{Annealing}\phantom{aaaaaaaaaaa}}}
  & \smash{{\large $\lambda{=}0.05$}} & \smash{{\large $\lambda{=}0.4$}} & \smash{{\large $\lambda{=}0.7$}} & \smash{{\large $\lambda{=}0.8$}} \\
& \includegraphics[height=\imgH]{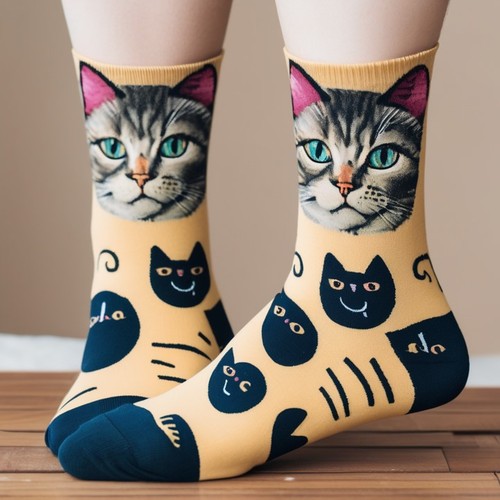} &
  \includegraphics[height=\imgH]{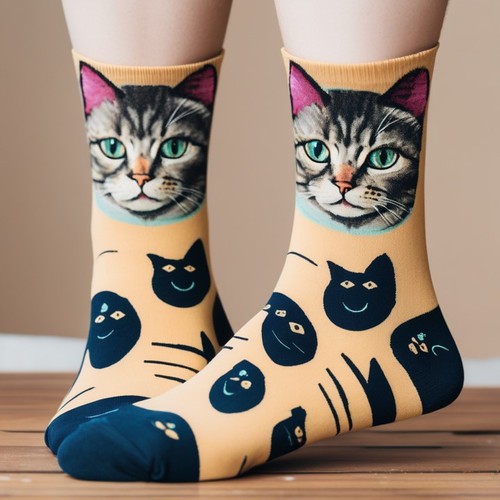} &
  \includegraphics[height=\imgH]{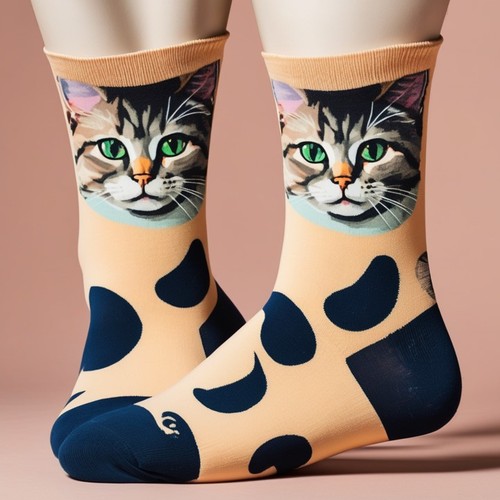} &
  \includegraphics[height=\imgH]{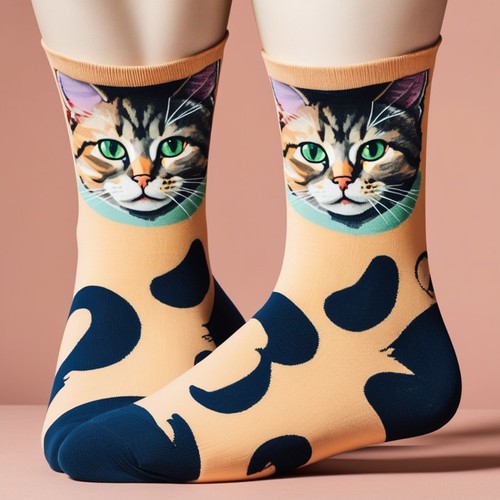} \\
[-2pt]

& \multicolumn{4}{c}{ \shortstack{\Large \textit{``A pair of matching socks with cute cats on them.''}}} \\

\end{tabular}

\caption{Qualitative comparison across scales. Rows: CFG vs.\ Annealing (ours). Columns: $w\in\{7.5,10,12.5,15\}$ (CFG) and $\lambda\in\{0.05,0.4,0.7,0.8\}$ (Annealing).}
\label{fig:all_scales_1}
\end{figure*}

\setlength{\tabcolsep}{1pt} %

\begin{figure*}[t]
\centering
\begin{tabular}{@{}c c c c c@{}}

\multirow{2}{*}{\rotatebox[origin=c]{90}{\large CFG\phantom{aaaaaaaaaaaaaa}}}
  & \smash{{\large $w{=}7.5$}} & \smash{{\large $w{=}10$}} & \smash{{\large $w{=}12.5$}} & \smash{{\large $w{=}15$}} \\
& \includegraphics[height=\imgH]{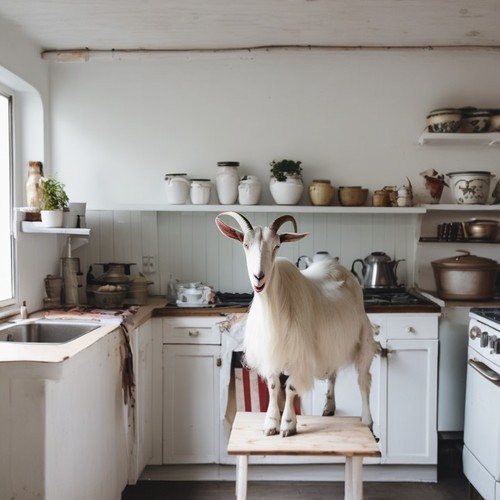} &
  \includegraphics[height=\imgH]{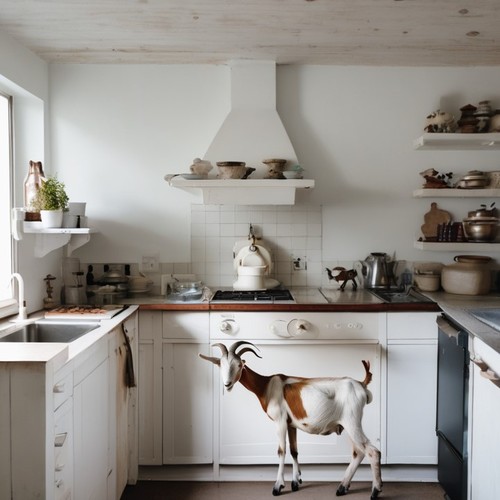} &
  \includegraphics[height=\imgH]{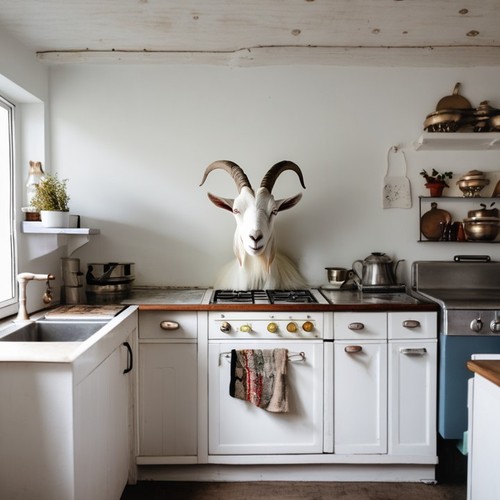} &
  \includegraphics[height=\imgH]{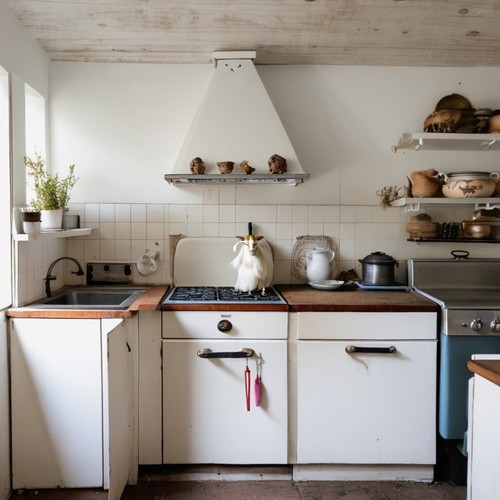} \\
[0pt]

\multirow{2}{*}{\rotatebox[origin=c]{90}{\large \textbf{Annealing}\phantom{aaaaaaaaaaa}}}
  & \smash{{\large $\lambda{=}0.05$}} & \smash{{\large $\lambda{=}0.4$}} & \smash{{\large $\lambda{=}0.7$}} & \smash{{\large $\lambda{=}0.8$}} \\
& \includegraphics[height=\imgH]{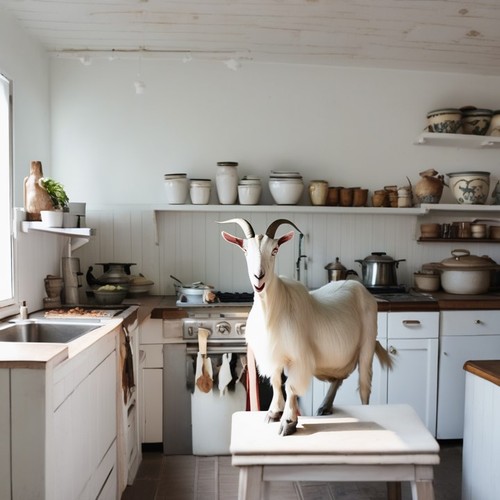} &
  \includegraphics[height=\imgH]{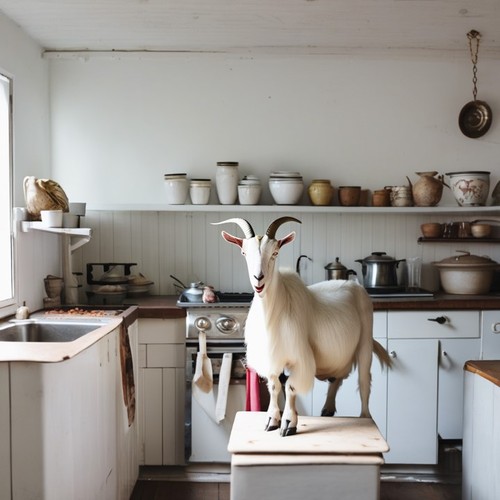} &
  \includegraphics[height=\imgH]{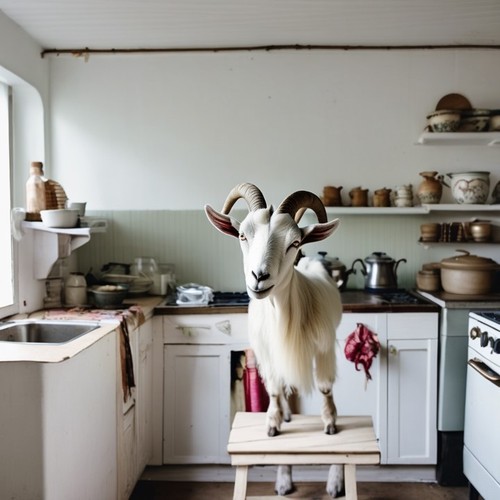} &
  \includegraphics[height=\imgH]{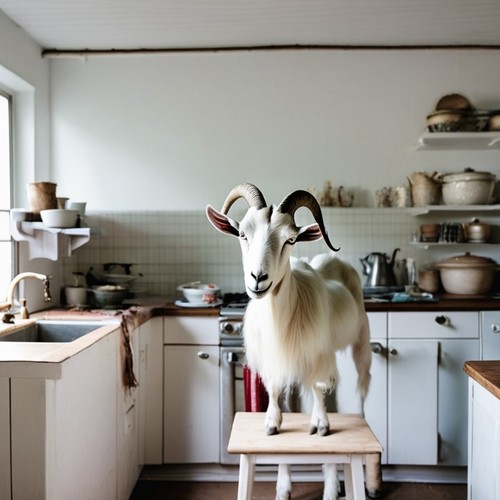} \\
[-2pt]

& \multicolumn{4}{c}{ \shortstack{\Large \textit{``A small kitchen with a \textcolor{red}{\textbf{white goat}} in it.''}}} \\[8pt]
\multirow{2}{*}{\rotatebox[origin=c]{90}{\large CFG++\phantom{aaaaaaaaaaaa}}}
  & \smash{{\large $w{=}0.6$}} & \smash{{\large $w{=}0.8$}} & \smash{{\large $w{=}1.0$}} & \smash{{\large $w{=}1.2$}} \\
& \includegraphics[height=\imgH]{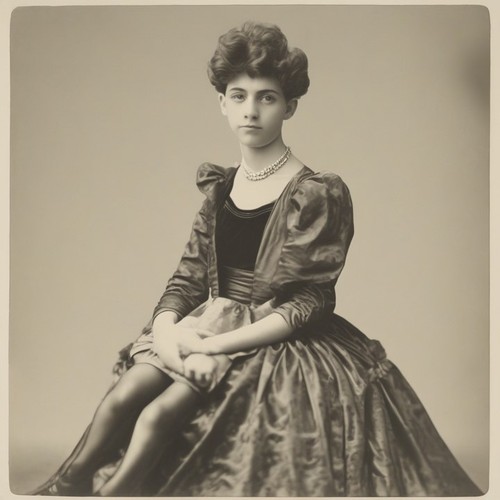} &
  \includegraphics[height=\imgH]{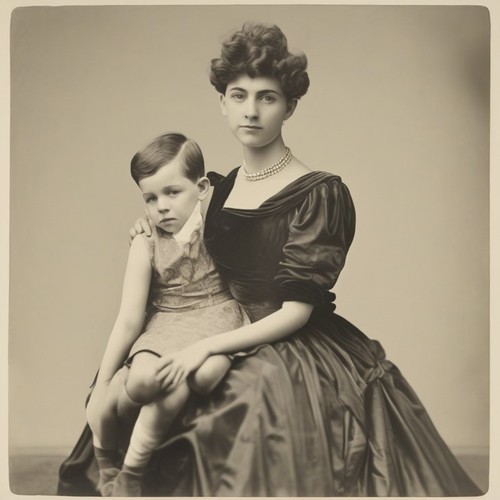} &
  \includegraphics[height=\imgH]{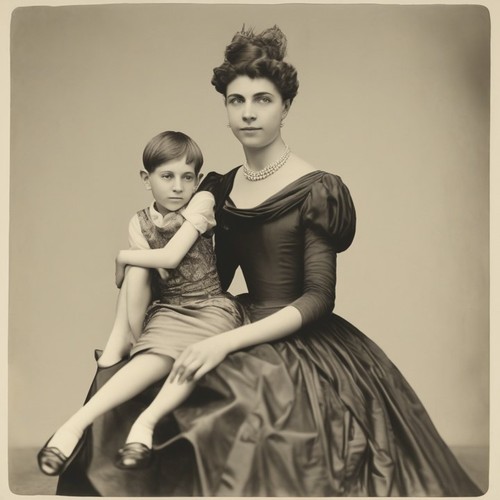} &
  \includegraphics[height=\imgH]{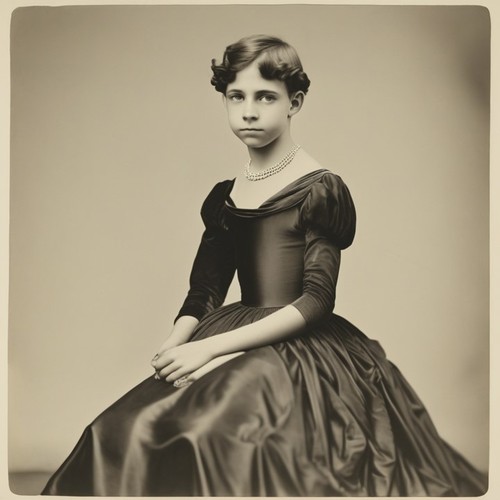} \\
[0pt]

\multirow{2}{*}{\rotatebox[origin=c]{90}{\large \textbf{Annealing}\phantom{aaaaaaaaaaa}}}
  & \smash{{\large $\lambda{=}0.05$}} & \smash{{\large $\lambda{=}0.4$}} & \smash{{\large $\lambda{=}0.7$}} & \smash{{\large $\lambda{=}0.8$}} \\
& \includegraphics[height=\imgH]{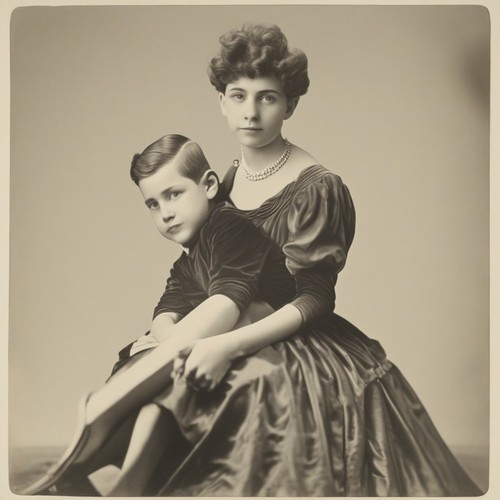} &
  \includegraphics[height=\imgH]{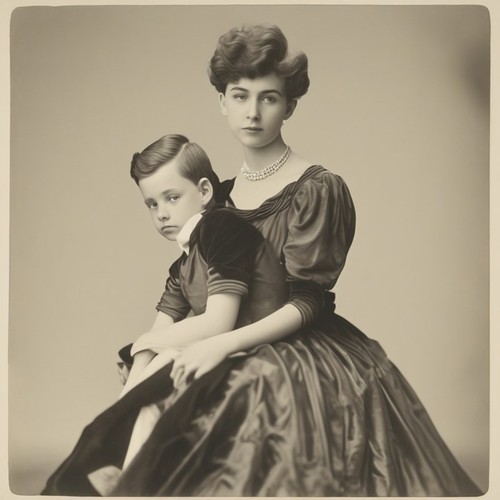} &
  \includegraphics[height=\imgH]{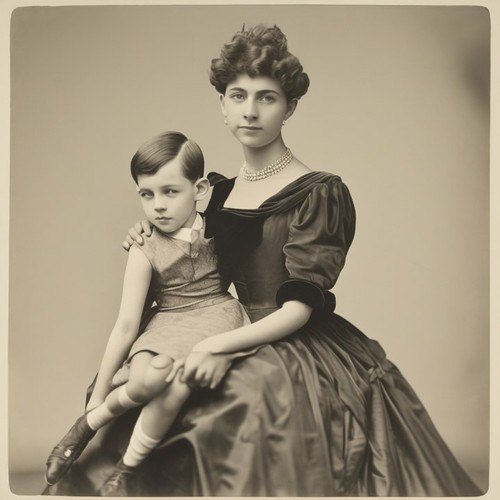} &
  \includegraphics[height=\imgH]{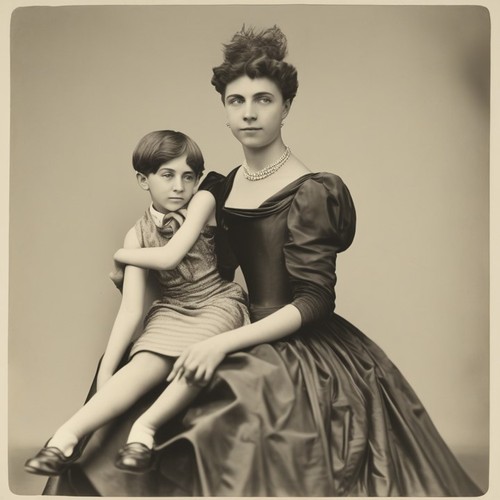} \\
[-2pt]

& \multicolumn{4}{c}{ \shortstack{\Large \textit{``\textcolor{red}{\textbf{A boy}} sitting on the shoulders of \textcolor{red}{\textbf{a woman}} who is wearing an elegant dress.''}}} \\[8pt]

\end{tabular}

\caption{Qualitative comparison across scales. Top block: CFG vs.\ Annealing (ours). 
Bottom block: CFG++ vs.\ Annealing (ours). 
Columns show different values of $w$ (CFG/CFG++) and $\lambda$ (Annealing).}

\label{fig:all_scales_2}
\end{figure*}

\newpage

\clearpage
\end{appendices}

\end{document}